\documentclass[english,aps,prb,amsmath,amssymb,superscriptaddress,showpacs,superscriptaddress,floats,floatfix,nobalancelastpage,twocolumn]{revtex4-2}

\usepackage{titlesec}
\usepackage{bm}
\usepackage{comment}
\usepackage{verbatim}

\usepackage{graphicx}
\usepackage{subfigure}
\usepackage{tabularx}

\usepackage[normalem]{ulem}

\usepackage{array}
\newcolumntype{L}[1]{>{\raggedright\let\newline\\\arraybackslash\hspace{0pt}}m{#1}}
\newcolumntype{C}[1]{>{\centering\let\newline\\\arraybackslash\hspace{0pt}}m{#1}}
\newcolumntype{R}[1]{>{\raggedleft\let\newline\\\arraybackslash\hspace{0pt}}m{#1}}

\usepackage{color}
\def\cred{\color{red}}

\def\cbl{\color{blue}}

\definecolor{darkgreen}{rgb}{0.02,0.45,0.0} 
\def\cgr{\color{darkgreen}}

\definecolor{violet}{rgb}{0.8,0.2,0.6}

\definecolor{light-gray}{gray}{0.5}

\usepackage[colorlinks,bookmarks=false,citecolor=blue,linkcolor=blue,urlcolor=blue]{hyperref}

\def\be{\begin{equation}}
\def\ee{\end{equation}}
\def\bea{\begin{eqnarray}}
\def\eea{\end{eqnarray}}

\def\vec{\mathbf}
\def\bs{\boldsymbol}
\def\mc{\mathcal}

\newcommand{\h}[1]{{#1}^{\dagger}} 

\newcommand{\cb}[1]{\bar{#1}}

\newcommand{\sgn}{{\rm sgn}}

\newcommand{\nairo}{Na$_2$IrO$_3$}
\newcommand{\liiro}{Li$_2$IrO$_3$}
\newcommand{\aliiro}{$\alpha$-Li$_2$IrO$_3$}
\newcommand{\bliiro}{$\beta$-Li$_2$IrO$_3$}
\newcommand{\gliiro}{$\gamma$-Li$_2$IrO$_3$}
\newcommand{\Hliiro}{H$_3$Li$_2$IrO$_6$}

\newcommand{\Agliiro}{Ag$_3$Li$_2$IrO$_6$}

\newcommand{\rucl}{$\alpha$-RuCl$_3$}

\setcitestyle{super}
\setcitestyle{numbers,square}

\allowdisplaybreaks

\usepackage{times}

\begin{document}
\date{\today}

\title{Beyond Kitaev physics in strong spin-orbit coupled magnets}

\author{Ioannis Rousochatzakis}
\affiliation{Department of Physics, Loughborough University, Loughborough LE11 3TU, United Kingdom}

\author{Natalia B. Perkins}
\affiliation{School of Physics and Astronomy, University of Minnesota, Minneapolis, Minnesota 55455, USA}
\affiliation{Technical University of Munich, Germany; Institute for Advanced Study, D-85748 Garching, Germany}

\author{Qiang Luo}
\affiliation{College of Physics, Nanjing University of Aeronautics and Astronautics, Nanjing, 211106, China}
\affiliation{Department of Physics, University of Toronto, Toronto, Ontario M5S 1A7, Canada}

\author{Hae-Young Kee}
\affiliation{Department of Physics, University of Toronto, Toronto, Ontario M5S 1A7, Canada}
\affiliation{Canadian Institute for Advanced Research, CIFAR Program in Quantum Materials, Toronto, Ontario M5G 1M1, Canada}

\begin{abstract}
We review the recent advances and current challenges in the field of strong spin-orbit coupled Kitaev materials, with a particular emphasis on the physics beyond the exactly-solvable Kitaev spin liquid point. To that end, we give a comprehensive overview of the most relevant exchange interactions in $d^5$ and $d^7$ iridates and similar compounds, an exposition of their microscopic origin, and a systematic attempt to map out the most interesting correlated regimes of the multi-dimensional parameter space, guided by powerful symmetry and duality transformations as well as by insights from  wide-ranging analytical and numerical studies. 
We also survey recent exciting results on quasi-1D models and discuss their relevance to higher-dimensional models.  
Finally, we highlight some of the key questions in the field as well as future directions.
\end{abstract}

\maketitle

\tableofcontents  

\pagebreak

\section{Introduction}\label{sec:intro}

Magnetic Mott insulators  that  combine  electronic  correlations with strong spin-orbit coupling (SOC) have recently attracted significant interest as a  prominent playground for quantum magnetism and, in particular, quantum spin liquids (QSLs)~\cite{Kitaev2006,Jackeli2009PRL,Chaloupka2010PRL,Chaloupka2010PRL,Krempa2014ARCMP,Rau2016ARCMP,Winter2016,Winter2017,Knolle2017ARCMP,Takagi2019NRP,Motome2019JPSJ,Takayama2021JPSJ,Trebst2022}. Due to the interplay between crystal field (CF) effects, strong electron-electron interactions, and SOC, entangled  spin  and  orbital  degrees  of  freedom are generically  described by the low-energy effective pseudo-spin models  with  bond-dependent  anisotropic-exchange  interactions~\cite{Khaliullin2005}.  

\begin{figure*}
\includegraphics[width=\textwidth]{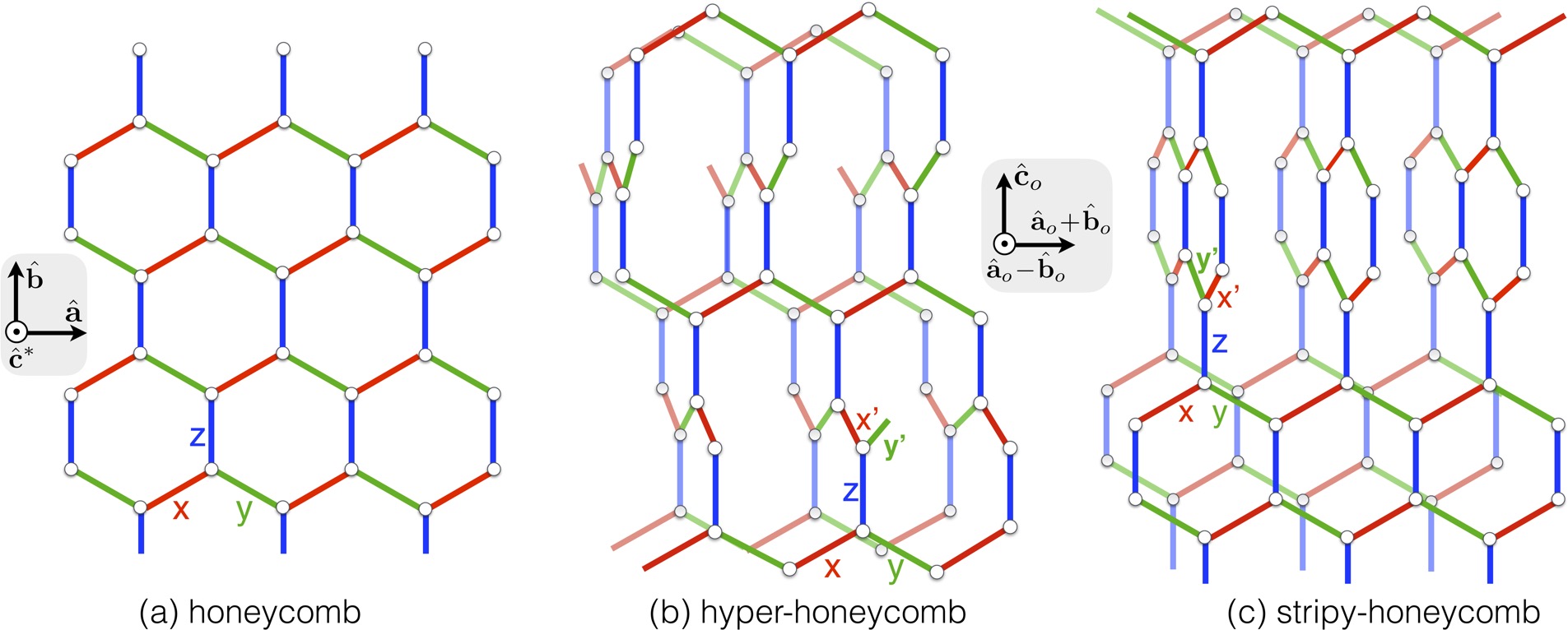}
\caption{The Kitaev model  is exactly solvable on a wide range of tricoordinated  lattices. Here are the examples of the harmonic honeycomb (H-$n$) lattices: (a) 2D honeycomb lattice (H-${\infty}$),  (b) 3D hyperhoneycomb lattice (H-$0$) , 3D stripy-honeycomb lattice (H-$1$). Here $n$ counts the number of rows along the $z$-axis before the orientation of the honeycomb plane switches between the two non-parallel chains of $x$ and $y$ bonds. The hyperhoneycomb lattice switches chains at every $z$-bond,  the stripy-honeycomb has one set of $z$-bonds making rungs of ladders before a bridge ($z$-bond) to the opposite ladder. The honeycomb lattice never switches ladders.}\label{fig:TriCoord}
\end{figure*}

In the last decade, a considerable theoretical and experimental effort has been devoted to a certain class of  two- (2D) and three-dimensional (3D) tricoordinated 4$d$ and 5$d$ materials that have been identified as being proximate to the Kitaev spin model, which is   famous for its exact solubility and the QSL ground state in both 2D and 3D tricoordinated lattices (e.g., see Fig.~\ref{fig:TriCoord}) ~\cite{Kitaev2006,Mandal2009PRB}. The realization of this model requires magnetic ions that are well described by pseudo-spin $J_{\text{eff}} = 1/2$ Kramers' doublets interacting via the nearest-neighbor (NN) Ising-like interactions along bond-dependent quantization axes. The compass form of this, so-called, Kitaev interaction stems from the highly entangled, spin-orbital nature of the Kramers' doublets and the particular geometry of the tricoordinated lattice structures composed of edge-sharing ligand octahedra~\cite{Chaloupka2010PRL,Rau2014,Katukuri2014NJP,Yamaji2014PRL,Sizyuk2014,HSKim2015PRB}. 

Most notably, a lot of effort has been devoted in the experimental investigation of the layered compounds A$_2$IrO$_3$ (A=Na, Li)~\cite{Singh2010PRB,Singh2012PRL,Liu2011PRB,Choi2012PRL,Ye2012PRB,Chun2015,Williams2016PRB} and $\alpha$-RuCl$_3$~\cite{Plumb2014PRB,Sears2015PRB,Majumder2015,Banerjee2016}, which are proximate to the honeycomb Kitaev QSL~\cite{Kitaev2006}, as well as the 3D harmonic-honeycomb iridates $\beta$-Li$_2$IrO$_3$ and $\gamma$-Li$_2$IrO$_3$ ~\cite{Biffin2014PRB,Biffin2014PRL,Modic2014NC,Takayama2015}, which are proximate to the Kitaev QSL on the hyper-honeycomb and stripy-honeycomb lattices, respectively~\cite{Hermanns2015,Obrien2016,Eschmann2020}. While there are numerous direct and indirect experimental indications (and first-principle calculations) that the Kitaev coupling $K$ is the dominant microscopic interaction in all these materials (hence the term `Kitaev materials')~ \cite{Trebst2022}, most of them are actually magnetically ordered at low enough temperatures ($k_B T_N\ll |K|$), which underlies the broader consensus that Kitaev QSLs are quite fragile against various types of perturbations. In turn, this exposes the realization that a fundamental understanding of the magnetism in these materials is far from complete without a careful exploration of  other exchange interactions that are allowed by their symmetry~\cite{Rau2014,Sizyuk2014,Rousochatzakis2015PRX,Winter2016,Winter2017}. 

This brings us to the primary scope of this review article, which is precisely to address the role of the most relevant anisotropic interactions beyond the Kitaev coupling $K$, and highlight some of the novel qualitative physics and exotic phases (including spin nematic and QSL ground states) that arise from their interplay.
To this end, we shall pay special attention to the role of the symmetric component of the off-diagonal exchange between nearest neighbours, commonly referred to as the $\Gamma$ coupling, which has emerged as another source of frustration that is essential for the complete understanding of Kitaev materials. Moreover, in most of these systems, $\Gamma$ is the second strongest interaction and its interplay with $K$ is often responsible for the highly anisotropic response to external magnetic fields and a variety of complex orders at intermediate field strengths~\cite{Modic2014NC,Janssen2016,Modic2017,Modic2018b,Sears2017,Gordon2019NC,Riedl2019,Ducatman2018,Rousochatzakis2018,Janssen2019,Li2020,Lee2020NC,Li2020b}.

Besides the $K$ and $\Gamma$ couplings we shall also discuss the impact of the NN Heisenberg exchange $J$, which is generically present in the Kitaev materials and plays a non-trivial role despite its relatively weak strength compared to $K$ and $\Gamma$. A more realistic model should also include the effect of the trigonal splitting of the $t_{2g}$ orbitals that modifies the pseudospin wavefunctions and brings a new interaction term $\Gamma'$ that is otherwise not symmetry-allowed in the absence of the trigonal distortion~\cite{Rau2014-arxiv}. We shall therefore limit ourselves to reviewing the physics of the NN Kitaev-Heisenberg-Gamma-Gamma$'$ ($J$-$K$-$\Gamma$-$\Gamma'$) model, which is broadly  considered as {\it the minimal model} for many Kitaev materials.

We will also briefly touch on the response of Kitaev materials under a magnetic field, as a more focused review of this topic can be found elsewhere~\cite{Janssen2019}.
Finally, we shall also discuss Kitaev materials with larger spins such as CrI$_3$~\cite{Lee20220PRL} and ZrCl$_3$~\cite{Yamada2018}, where the dominant Kitaev interaction among $S = 3/2$ spins was theoretically proposed. Simultaneously with significant theoretical developments in the study of the higher-spin Kitaev interaction~\cite{Baskaran2007PRL,Baskaran2008PRB,Rousochatzakis2018NC,Koga2018JPSJ,Oitmaa2018PRB}, there has been a growing interest in realizing such interactions in solid-state materials~\cite{Xu2020PRL,Stavropoulos2019PRL,Stavropoulos2019PRL}.

In this review we shall not discuss other types of perturbations that are present in real materials, such as further neighbour isotropic and anisotropic interactions ($J_2$, $J_3$, $K_2$, etc) \cite{Rousochatzakis2015PRX,Winter2017}, or the inequivalence between different bonds, i.e., we shall only consider models where the strengths of $J$, $K$, $\Gamma$ and $\Gamma'$ are the same on all types of NN bonds~\cite{Foyevtsova2013,Yamaji2014,Katukuri2014NJP,Katukuri2015,Winter2016,Winter2017}. Additionally, since most of Kitaev materials are centrosymmetric, NN Dzyaloshinskii–Moriya (DM) interactions are not present in at least one NN bond (the bond with the inversion center at its midpoint for $C_{2h}$ symmetry, and all NN bonds for $R\bar{3}$ symmetry), and we shall otherwise disregard the effect of the allowed DM terms on other bonds (e.g., the ones between second neighbors). %
A detailed study of the impact of Dzyaloshinskii–Moriya interactions on the Kitaev spin liquid (with novel phenomena such as turning the nonchiral Kitaev’s gapless QSL into a chiral one with equal Berry phases at the two Dirac points) has been reported in Ref. ~\cite{Ralko2020}.

The remaining part of this review is organized as follows.  In Sec.~\ref{sec:materials} we introduce the main lattice geometries and materials of interest. In Sec.~\ref{micro_theory}, we discuss the physical origin of bond-directional exchange anisotropies in spin-orbital materials and give an illustration via a step-by-step, microscopic derivation of these couplings in $d^5$ and $d^7$ systems. This leads us to the generic $J_{\text{eff}}\!=\!1/2$ nearest-neighbor (NN) $J$-$K$-$\Gamma$-$\Gamma'$ Hamiltonian, as a minimal model for the description of many Kitaev materials. We shall also briefly discuss higher-$S$ extensions of this model in Sec.~\ref{sec:generalS}. 
We then set out to review the various competing phases in this high-dimensional parameter space. 
To that end, we begin in Sec.~\ref{sec:sym_and_dual} with a discussion of the various (global and local) symmetries and duality transformations of the 2D Honeycomb model, which, in turn, are crucial for mapping out the highly-correlated regions (as well as a number of `hidden' isotropic points) in the parameter space. 
The exploration of the phase diagram of the planar model, which is presented in Sec.~\ref{sec:JKGammaPhaseDiagram}, proceeds via a separate discussion of some of the most relevant special limits and regions of the parameter space, before sensible connections are made for a qualitative understanding of the full picture. In this exploration we shall pay special attention to the interplay between the two known spin liquids of the phase diagram, the quantum spin liquid of the pure Kitaev model (and its higher-spin extensions in Sec.~\ref{sec:generalS2}) and the classical spin liquid of the pure $\Gamma$ model (Sec.~\ref{sec:Gpoints}). We shall also review the current state of affairs in the quantum $K$-$\Gamma$ model (Sec.~\ref{sec:KGmodel}), as well as the classical limit of this model, and the emergence of approximate, period-3 states with counter-rotating sublattices, which are observed experimentally in the three {\liiro} polymorphs. 
In Sec.~\ref{sec:3D} we provide a brief review of the magnetism of the 3D compounds {\bliiro} and {\gliiro}, which is driven, to a large extent, by the interplay between $K$ and $\Gamma$ interactions. 
We then review in Sec.~\ref{sec:1D} the recent developments on the physics of quasi-1D Kitaev-$\Gamma$ models, which have emerged as a theoretical platform to understand the physics of the corresponding 2D model. 
Finally, in Sec.~\ref{sec:Discussion} we give an overview of the outstanding questions that remain in the field, as well as our perspectives on potential future directions for research.

\section{Lattice geometries and materials of interest}\label{sec:materials}

We begin by highlighting some general structural and crystal symmetry aspects of the available Kitaev materials, in order to set the stage for the later discussion of the underlying microscopic processes in these materials.

Kitaev materials are the spin-orbit coupled Mott insulators that crystallize in the tricoordinated lattices of Fig.~\ref{fig:TriCoord},  namely the layered 2D honeycomb (panel a), the 3D hyper-honeycomb (b) and the 3D stripy-honeycomb (c). In these geometries, $J_{\rm{eff}}\!=\!1/2$ pseudospin degrees of freedom  interacting solely via Kitaev exchange form exactly-solvable QSL ground states with fractionalized excitations~\cite{Kitaev2006,Mandal2009PRB,Hermanns2015,Obrien2016}.

The layered materials include, most notably, the $d^5$ compounds {\nairo}~\cite{Singh2010PRB,Singh2012PRL,Liu2011PRB,Choi2012PRL,Ye2012PRB,Chun2015}, {\aliiro}~\cite{Williams2016PRB,Vale2019,Choi2019PRB}, {\rucl}~\cite{Plumb2014PRB,Sears2015PRB,Kubota2015,Johnson2015,Majumder2015,Banerjee2016}, {\Hliiro} \cite{Kitagawa2018spin,Pei2020,LeePRB2023}, {\Agliiro} \cite{Bahrami2019,Chakraborty2021,Bahrami2021} and Cu$_2$IrO$_3$ \cite{Choiprl2019,Kenney2019,Pal2021}, which typically have monoclinic ($C_{2/m}$ or $C_{2/c}$) or trigonal ($R$-$3m^\ast$ or $R$-$3$) symmetry~\cite{Takagi2019NRP}.
More recent proposals include the family of $d^5$-based ilmenites AIrO$_3$ with A=Mg, Zn, Cd~\cite{HaraguchiPRM2018,HaraguchiPRM2020}, as well as a family of $d^7$-based transition metal compounds~\cite{Liu2018PRB,Sano2018PRB,Liu2020PRL,Kim2021review}, such as the delafossites Na$_3$Co$_2$SbO$_6$ and Na$_2$Co$_2$TeO$_6$~\cite{Kim2021,Chen2021PRB,Samarakoon2021PRB,Lee2021PRB,VICIU2007,Yan2019PRM,Songvilay2020PRB,Janssen2023}, Li$_3$Co$_2$SbO$_6$~\cite{Vivanco2020PRB}, and also BaCo$_2$(AsO$_4$)$_2$ and BaCO$_2$(PO$_4$)$_2$~\cite{Nair2018PRB,Shi2018PRB,Ruidan2020SA}.

The 3D geometries of Fig.~\ref{fig:TriCoord}\,(b-c) are found in {\bliiro} and {\gliiro}, with orthorhombic $Fddd$ and $Cccm$ symmetry, respectively~\cite{Modic2014NC,Biffin2014PRB,Biffin2014PRL,Takayama2015,Ruiz2017,Modic2017,Modic2018b,Ruiz2019,Majumder2019,Majumder2020,Yang2022,Halloran2022,Tsirlin2022}. Together with {\aliiro}~\cite{Williams2016PRB}, they belong to the {\liiro} polymorph family~\cite{Tsirlin2022}, which comprises an infinite set of 3D lattices, the harmonic  honeycomb ($\mc{H}$-$n$) series~\cite{Kimchi2014b}. The $n$-th member of this series can be thought of as a sequence of $n$ rows of coplanar hexagonal plaquettes, followed by a bridge layer of $z$-bonds along  one of the crystallographic axes [the monoclinic {\bf b} axis in Fig.~\ref{fig:TriCoord}\,(a) or the orthorhombic ${\bf c}_o$ axis in Fig.~\ref{fig:TriCoord}\,(b-c)] and then by another $n$ rows of plaquettes in a new plane. The hyper-honeycomb and stripy-honeycomb correspond, respectively, to the $n\!=\!0$ and $n\!=\!1$ members of the series, whereas the 2D honeycomb is the $n\!=\!\infty$ member, see Fig.~\ref{fig:TriCoord}. One of the distinguishing features of the 3D lattices of Fig.~\ref{fig:TriCoord}\,(b-c) compared to the 2D lattice is that, while all $z$ bonds are aligned along the orthorhombic ${\bf c}_o$ axis, there are two different $xy$ chain directions, one along ${\bf a}_o$+${\bf b}_o$, and the other along the perpendicular direction ${\bf a}_o$-${\bf b}_o$. While the two chains are related to each other by crystal symmetries (in particular, the twofold rotations $C_{2{\bf a}}$ and $C_{2{\bf b}}$, see, e.g., discussion in Ref.~\cite{Ducatman2018}), the microscopic modeling on this structure necessitates that we treat them separately. So, unlike the honeycomb case which features three types of bonds,  labeled as $\gamma\!=\!x$, $y$ or $z$ [see Fig.~\ref{fig:TriCoord}\,(a)], the 3D lattices feature five types of bonds, labeled as $\gamma\!=\!x$, $y$, $x'$, $y'$,  and $z$ [see Fig.~\ref{fig:TriCoord}\,(b-c)].

\section{Origin of bond-directional interactions}\label{micro_theory}

\subsection{Historical perspective }
The  magnetic interactions in solids are generically anisotropic. Historically,  as early as  in 1937, Van Vleck showed that the SOC results in entangled  spin-orbit wave functions and anisotropic spin interactions~\cite{Vleck1937PR}.  However, in most systems the orbital degeneracy is lifted by Jahn-Teller distortions ~\cite{Jahn1937}, rendering SOC partially inactive and destroying the entangled nature of the spin-orbit wave function. This, in turn, often leads to dominant XY or Ising anisotropic spin models. 

It was then noted in early 2000 that another highly anisotropic bond-dependent interactions can appear in systems with partially filled t$_{\rm 2g}$ orbitals well separated from higher energy doublet of $e_g$ orbitals by octahedral crystal field ~\cite{Khaliullin2001PRB}. These bond-dependent interactions, i.e.,  when only particular components of the degrees of freedom interact on different bonds in the lattice, are also known as compass models~\cite{Nussinov2015}, among which  the Kugel-Khomskii model~\cite{Kugel1982} describing the interactions between orbital degrees of freedom in strongly correlated electron materials  is, perhaps, the most widely-studied in different contexts. Moreover,  in the presence of SOC, the spin ${\bf S}$ and orbital ${\bf L}$ degrees of freedom are no longer separated and instead  form  the total angular momenta ${\bf J}={\bf L}+{\bf S}$, dubbed pseudospins, interacting via entangled spin-orbital exchange interactions.

This physics is especially relevant to 4d and 5d electron compounds~\cite{Khaliullin2005}. Jackeli and Khaliullin have shown how the famous Kitaev interaction can be found in the edge-sharing octahedra honeycomb lattice,   opening a new road for QSLs~\cite{Jackeli2009PRL}. Shortly after, Rau {\it et al} reported another bond-dependent interaction denoted by $\Gamma$~\cite{Rau2014} not discussed in the earlier derivation of Jackeli and Khaliullin~\cite{Jackeli2009PRL}. Its presence was also reported by quantum chemistry~\cite{Katukuri2014NJP} and density functional theory calculations~\cite{Yamaji2014PRL}. Unlike the Heisenberg coupling, the $\Gamma$ interaction on tricoordinated lattices is highly frustrated, and has added significant challenges in understanding possible spin liquids via the interplay between Kitaev and $ \Gamma$ interactions. Since then, there have been intense studies on realization of Kitaev interaction and generalized bond-dependent spin models in Mott insulators with strong SOC. The particular form of the pseudospin interactions in these models is determined by
the 
filling factor $n$ of the corresponding $d$- or $f$-orbitals, which also  decides the shape and spin composition of the pseudospin wavefunctions. These interactions generally arise from the exchange processes that involve different orbital momenta, L$_x$, L$_y$, and L$_z$, which tie them to the bond direction, thus leading to the bond-dependent nature of the pseudospin interactions (spin for simplicity from now on).

\subsection{NN couplings in $d^5$ and $d^7$ systems}
To illustrate how the bond-directional interactions arise in the Kitaev materials we consider, as a platform, the broad family of $d^5$ and $d^7$ systems with pseudospin-$1/2$ degrees of freedom and  discuss the main steps in the derivation of their interactions. Since the cases of $d^2$ (quadrupole and octupole), $d^3$ ($S=3/2$), and $d^8$ ($S=1$)  were recently reviewed in Ref.~\cite{Takayama2021JPSJ}, we will only briefly discuss them in \ref{sec:generalS}.

\subsubsection{Local $J_{\rm eff}\!=\!1/2$ degrees of freedom}
A low-energy effective spin model can be derived when electron-electron interaction is stronger than hopping integrals between electrons (known as strong coupling expansion). To derive the effective spin model for multi-orbital Mott insulators such as $d^5$, one begins with a local Kanamori-Hubbard interaction, which has the following form at a given site $i$~\cite{Kanamori1963PTP,sugano1070multiplets,Rau2014}
\be\label{eq:atomic}
\!\!\mc{H}_{\text{loc},i}\!=\! \frac{U\!-\!3J_H}{2} (N_i\!-\!5)^2\!-\!2J_H {\bf S}_i^2\!-\!\frac{J_H}{2} {\bf L}_i^2\!+\!\lambda {\bf L}_i\cdot{\bf S}_i,
\ee
where $N_i$, ${\bf S}_i$, and ${\bf L}_i$ are the total number, spin, and (effective) orbital angular momentum operators at site $i$, $U$ is the Coulomb interaction, $J_H$ is Hund's coupling, and $\lambda$ is the SOC strength.

\begin{figure}[!t]
\includegraphics[width=0.35\textwidth]{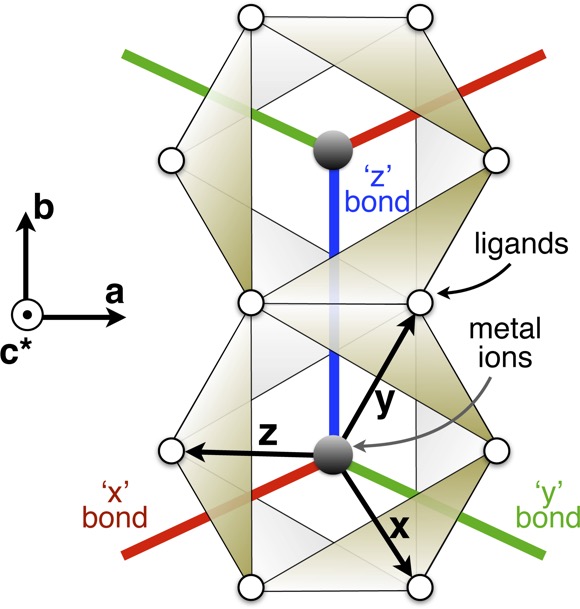}
\caption{Edge-sharing octahedral geometry of ligands (open circles) in layered honeycomb Kitaev materials, here illustrated for a `z'-bond of NN metal ions (grey spheres), as viewed along the cubic [111] axis. Also shown are the two most commonly used axes frames in the literature: i) the crystallographic $(a,b,c^\ast)$ frame of the monoclinic $C_{2/m}$ group, and ii) the `cubic' $(x,y,z)$ frame. The two frames are related by
 $\hat{{\bf a}}=\frac{\hat{{\bf x}}+\hat{{\bf y}}-2\hat{{\bf z}}}{\sqrt{6}}$, $\hat{{\bf b}}=\frac{\hat{{\bf y}}-\hat{{\bf x}}}{\sqrt{2}}$, and $\hat{{\bf c}}^\ast=\frac{\hat{{\bf x}}+\hat{{\bf y}}+\hat{{\bf z}}}{\sqrt{3}}$.
 }\label{fig:geometry}
 \end{figure}

For $4d^5$ and $5d^5$ systems, the single ion interactions give rise to a low-energy manifold with total spin $S\!=\!1/2$ and total angular momentum $L\!=\!1$. The strong SOC then gives rise to a Kramers doublet $J_{\text{eff}}=1/2$, whose member wavefunctions take the following form in the $|m_S, m_L\rangle$ basis (where $m_S$ and $m_L$ are the eigenvalues of $S_z$ and  $L_z$, respectively)~\cite{Jackeli2009PRL}: 
\bea
&&\underline{4d^5~\&~ 5d^5~\text{systems}:}\nonumber\\
\renewcommand{\arraystretch}{1.5}
&&\begin{array}{l}
|+\tilde{\frac{1}{2}}\rangle = \sqrt{\frac{2}{3}}|-\frac{1}{2},1\rangle-\sqrt{\frac{1}{3}}|\frac{1}{2},0\rangle,
\\
|-\tilde{\frac{1}{2}}\rangle = \sqrt{\frac{1}{3}}|-\frac{1}{2},0\rangle-\sqrt{\frac{2}{3}}|\frac{1}{2},-1\rangle.\\
\end{array}
\eea

For $3d^7$ systems, the large Hund's coupling gives rise to a high spin manifold with $S\!=\!3/2$ and total angular momentum $L\!=\!1$. The strong SOC then gives rise to a $J_{\rm eff}\!=\!\frac{1}{2}$ Kramer's doublet whose members take the following form in the 
$|m_S, m_L\rangle$ basis~\cite{Liu2018PRB,Sano2018PRB}
\bea
&&\!\!\!\underline{3d^7~\text{systems}:}\nonumber\\
\renewcommand{\arraystretch}{1.5}
&&\begin{array}{l}
|+\tilde{\frac{1}{2}}\rangle = \frac{1}{\sqrt{2}}|\frac{3}{2},-1\rangle-\frac{1}{\sqrt{3}}|\frac{1}{2},0\rangle+\frac{1}{\sqrt{6}}|-\frac{1}{2},1\rangle,
\\
|-\tilde{\frac{1}{2}}\rangle = \frac{1}{\sqrt{2}}|-\frac{3}{2},1\rangle-\frac{1}{\sqrt{3}}|-\frac{1}{2},0\rangle+\frac{1}{\sqrt{6}}|\frac{1}{2},-1\rangle.
\end{array}
\eea

In the following, we shall follow the usual convention in the literature and denote the pseudospin ${\bf J}_{\text{eff}}$ operators by the symbol ${\bf S}$.

\subsubsection{Hopping processes}
The effective hopping processes between two NN magnetic ions at sites $i$ and $j$ give rise to the kinetic terms for $t_{2g}$ and $e_g$ orbitals ($\{yz, xz, xy\}$ and $\{3z^2\!-\!r^2, x^2\!-\!y^2\}$, respectively): 
\be\label{eq:Hkin}
\mc{H}_{{\rm kin},ij}=  \sum\nolimits_{ab\sigma} t_{ij}^{ab} \h{d}_{ia\sigma} d_{jb\sigma},
\ee
where $\h{d}_{ia \sigma}$ are the creation and annihilation operators for the $a$-th orbital with spin $\sigma$ ($\uparrow$ or  $\downarrow$) at site $i$, and $t_{ij}^{ab}$ denotes the hopping amplitude from  $(ia\sigma)$ to $(jb\sigma)$.

For $4d^5$ and $5d^5$ systems with a single hole in a $t_{2g}$ orbital, one can disregard processes that involve $e_g$ orbitals. Furthermore, the symmetry of the local anion environment constrains the hopping terms between two magnetic sites. For example, for $t_{2g}$ orbitals surrounded by an octahedra cage with a slight trigonal distortion, there is a two-fold rotational symmetry about the $[1\cb{1}0]$ axis. This $C_{2{\bf b}}$ symmetry dictates that  $t^{zx,zx}\!=\!t^{yz,yz}\!\equiv\!t_1$,  $t^{zx,yz}\!=\!t^{yz,zx}\!\equiv\!t_2$ and $t^{yz,xy}\!=\!t^{zx,xy} \!\equiv\!t_4$. Defining further $t^{xy,xy}\!\equiv\!t_3$ and including time-reversal and inversion symmetries, leads to the following hopping matrix on a $z$-bond between sites $i$ and $j$~\cite{Rau2014}
\be\label{Eq:hoppingd5}
\underline{4d^5~\&~5d^5~\text{systems}:}~~~  {\bf T}_{ij,z} =   \left( \begin{tabular}{ccc}
      $t_1$ & $t_2$ & $t_4$ \\
      $t_2$ & $t_1$ & $t_4$ \\
      $t_4$ & $t_4$ & $t_3$ 
      \end{tabular}
    \right),
\ee
in the basis of the $t_{2g}$ orbitals $\{yz, zx, xy\}$. The corresponding matrices $T_{ij,x/y}$ for the bonds $\gamma\!=\!x$ and $y$ can be found by applying the $C_{3{\bf c}}$ rotation around the $[111]$ axis.
For the ideal octahedra without any trigonal distortion, there is an additional,  {\it local} $C_2$ symmetry~\cite{Rau2014-arxiv} around the axis perpendicular to  the bond  ($[001]$ axis for the  $z$-bond) and passing through its center which prevents any mixing between the $xy$ and the $zx$ and $yz$  orbitals, forcing $t_4\!=\!0$. Mott insulators with $R{\bar 3}$ space group do not feature such an additional $C_2$ symmetry around $[001]$, and can therefore have a nonzero $t_4$.

In $3d^7$ systems, there are three holes in the $t_{2g}$ and $e_{2g}$ orbitals, and one should therefore take into account, in principle, all possible $t_{2g}$-$t_{2g}$, $t_{2g}$-$e_g$ and $e_g$-$e_g$ hopping processes. The hopping matrix on a $z$-bond between sites $i$ and $j$ for the ideal honeycomb lattice takes the form
\be\label{eq:d-d}
\underline{3d^7~\text{systems}:}~~~    {\bf T}_{ij,z}=\left(
\begin{array}{ccccc}
    t_5 & 0 & 0 & 0 & 0  \\
    0 & {\tilde t}_4 & 0 & 0 & t_6\\
    0 & 0 & t_1 & t_2 & 0\\
    0 & 0 & t_2 & t_1 & 0\\
    0 & t_6 & 0 & 0 & t_3 
    \end{array}
\right),
\ee
in the basis $\{x^2\!-\!y^2, 3z^2\!-\!r^2, yz, xz, xy\}$. Here $t_1$, $t_3$, $\tilde{t}_4$ and $t_5$ denote hopping amplitudes between same-type orbitals ( $d_{yz/xz}$, $d_{xy}$, $d_{3z^2-r^2}$ and $d_{x^2-y^2}$, respectively); $t_2$ is the hopping between $d_{yz}$ and $d_{xz}$ (which includes both direct and indirect hoppings); $t_6$ is the hopping between $t_{2g}$ and $e_g$ manifolds (which also includes both direct and indirect hoppings). Processes on other type of bonds are related to the ones involving the $z$-bond by the $C_{3{\bf c}}$ symmetry, as above.

\begin{figure*}
 \includegraphics[width=\textwidth]{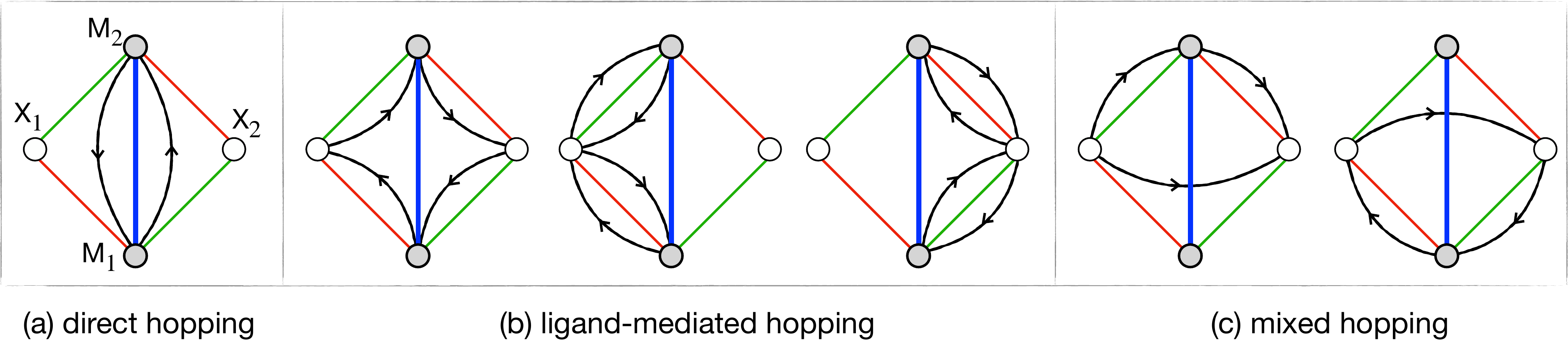}
 \caption{The three main types of superexchange processes contributing to the Hamiltonian (\ref{eq:JKGGpmodelij}) live on a (tilted) square plaquette, formed by two NN magnetic ions M$_1$ and M$_2$ (e.g., Ir$^{4+}$ or Ru$^{3+}$) and two ligand ions X$_1$ and X$_2$ (e.g., O$^{2-}$ or Cl$^{1-}$). (a) direct hopping. (b) ligand-mediated hopping. (c) mixed hopping.}\label{Fighopping}
 \end{figure*}

\subsubsection{Effective superexchange Hamiltonian}\label{sec:EffSuperHam}
A standard superexchange expansion, whereby one treats the kinetic terms $t^{ab}$ of (\ref{eq:Hkin}) as a small perturbation compared to the effect of SOC, crystal field splitting and correlations (in particular, $U\! \gg\! t^{ab}$ and $U\!> \!J_H\,,\lambda$), leads to an effective superexchange Hamiltonian 
\be\label{eq:JKGGpmodel}
\mc{H}=\sum\nolimits_\gamma\sum\nolimits_{\langle ij\rangle\in\gamma}\mc{H}_{ij,\gamma}+\mc{H}',
\ee
where the second term might include any longer-range couplings  (such as  $J_3$, $K_2$ or ${\bf D}_2$~\cite{Winter2016}) that we do not analyze here, and the first term incorporates the interactions between NN sites $i$ and $j$ on a bond of type $\gamma$:
\be\label{eq:JKGGpmodelij}
\renewcommand{\arraystretch}{1.5}
\begin{array}{l}
\mc{H}_{ij,\gamma}\!=\!J_\gamma \vec{S}_i\!\cdot\!\vec{S}_j\!+\!K_\gamma S^\gamma_i S^\gamma_j
\!+\!\Gamma_\gamma (S^{\alpha_\gamma}_iS^{\beta_\gamma}_j\!+\!S^{\beta_\gamma}_iS^{\alpha_\gamma}_j)\\
~~~~~~+\Gamma'_\gamma\big(S^{\alpha_\gamma}_i S^\gamma_j + S^\gamma_iS^{\alpha_\gamma}_j+S^{\beta_\gamma}_i S^\gamma_j + S^\gamma_iS^{\beta_\gamma}_j\big)\,,
\end{array}
\ee
where $(\alpha_\gamma,\beta_\gamma)\!=\!(y,z)$, $(z,x)$ and $(x,y)$ for $\gamma\!=\!x$ (and $x'$ for the 3D materials of Fig.~\ref{fig:TriCoord}\,(b-c)), $y$ (and $y'$), and $z$,  respectively~\cite{Rau2014,Rau2014-arxiv}, 

While detailed expressions for the dependence of the couplings $J$, $K$, $\Gamma$ and $\Gamma'$ on microscopic parameters  are provided  in Refs.~\cite{Rau2014,Sano2018PRB,Liu2018PRB,Liu2020PRL}, here, for illustration, we discuss virtual hopping processes that are responsible for the dominant contributions to these couplings in $d^5$ systems. As shown in   Fig.~\ref{Fighopping}, these processes are confined to a  plaquette consisting of two magnetic ions and two ligand ions, which we denote as M$_1$, M$_2$ and X$_1$, X$_2$, respectively. For iridium Kitaev materials, for example, the plaquette is formed by two iridium and two oxygen ions, while for $\alpha$-RuCl$_3$ it is formed by two ruthenium and two chlorine ions. For concreteness, we consider a $z$-bond with the hopping matrix elements (\ref{Eq:hoppingd5}) (the results for other types of bonds can be obtained in a similar way, or simply by symmetry, if present). There are three different types of paths on the M$_1$-X$_1$-M$_2$-X$_2$ plaquette:  direct hopping (panel a), ligand-mediated hopping (b), and mixed hopping (c).
The direct hopping  between $d$ orbitals (predominantly $d_{xy}$-$d_{xy}$ on the $z$-bond) gives a contribution to $J$ already in second-order of perturbation theory. The dominant contribution to $K$ arises in fourth-order, from the interference of the ligand-mediated hopping processes, whereas the off-diagonal exchanges $\Gamma$  and $\Gamma'$ appear in third order from the mixed direct/ligand-mediated hopping processes.

In general, the couplings depend on the bond type $\gamma$. We can reduce the number of independent coupling parameters by using crystal symmetries. Let us take for example the layered honeycomb materials. For the ideal geometry where the three types of bonds of Fig.~\ref{fig:geometry} are symmetry equivalent, the system has a $D_{3d}\!=\!C_{3v}\!\times\!I$ point group, generated by the following operations: i) inversion center $I$ on the middle of a $z$-bond; ii) a two-fold $C_{2{\bf b}}$ axis ($\pi$-rotation around the ${\bf b}$ axis going through a $z$-bond), which, in spin space, maps $(x,y,z)\!\to\!(-y,-x,-z)$. iii) a three-fold $C_{3{\bf c}^\ast}$ axis ($2\pi/3$-rotation around the ${\bf c}^\ast$ axis going through individual sites), which, in spin space, maps $(x,y,z)\!\to\!(y,z,x)$. For this ideal geometry, all bonds are equivalent and we end up with three independent couplings: the Heisenberg exchange $J$, the Kitaev exchange $K$ and the symmetric off-diagonal exchange $\Gamma$. The additional off-diagonal exchange $\Gamma'$ is introduced once the trigonal distortion is present, i.e, $t_4 \neq 0$. 

In {\nairo}, {\aliiro} and {\rucl}, the $C_{3{\bf c}^\ast}$ symmetry is actually absent and the three bonds are not all equivalent. In particular, the symmetry lowers to monoclinic $C_{2/m}$, with the twofold axis along ${\bf b}$ (see Fig.~\ref{fig:geometry}), so the $C_{2{\bf b}}$ rotation around the $z$ bond is still present and the $x$ and $y$ bonds remain equivalent to each other. Thus, the $C_{2/m}$ symmetry dictates that the exchange couplings on the $z$ bonds are different from those on $x/y$ bonds, although these differences can be small in some materials, see, e.g.,  discussion in Ref.~\cite{Winter2016PRB}.

Finally, it is worth noting that the signs of $\Gamma$ and $\Gamma'$ depend on the convention we use for the directions of the unit vectors $\hat{\bf x}$, $\hat{\bf y}$ and $\hat{\bf z}$ along the cubic axes $x$, $y$ and $z$, respectively. While this convention may seem innocuous, it can lead to quite different sign structures for the $\Gamma$ and $\Gamma'$ couplings across different bonds. Such different conventions have been used in the literature, e.g., for {\bliiro}, and care must be taken when comparisons are made. This issue is discussed in more detail in Appendix~\ref{app:GGpSignStructure}.

\subsubsection{Honeycomb Hamiltonian in the orthorhombic frame}\label{sec:abcframe}
Returning to the threefold-symmetric layered honeycomb case, it is worthwhile to point out that further insights can be gained by 
rewriting Eq.~(\ref{eq:JKGGpmodelij}) in the crystallographic $(a,b,c^\ast)$ frame, sometimes also referred to as the $(X,Y,Z)$ frame~\cite{Chaloupka2015PRB}:
\be\label{eq:JabJzABmodel}
\renewcommand{\arraystretch}{1.5}
\begin{array}{l}
\mc{H}_{ij,\gamma}\!=\!J_{ab}(S_{i}^{a}S_{j}^{a}+S_{i}^{b}S_{j}^{b})+J_{c}S_{i}^{c}S_{j}^{c}\\~~
+A\left[c_\gamma(S_{i}^{a}S_{j}^{a}-S_{i}^{b}S_{j}^{b})-s_\gamma(S_{i}^{a}S_{j}^{b}+S_{i}^{b}S_{j}^{a})\right]\\~~
-\sqrt{2}B\left[c_\gamma(S_{i}^{a}S_{j}^{c}+S_{i}^{c}S_{j}^{a})+s_\gamma(S_{i}^{b}S_{j}^{c}+S_{i}^{c}S_{j}^{b})\right],
\end{array}
\ee
where $c_\gamma\!\equiv\!\cos\phi_\gamma$, $s_\gamma\!\equiv\!\sin\phi_\gamma$, $\phi_{\gamma}\!=\!0$, $\frac{2\pi}{3}$, and $\frac{4\pi}{3}$ for $\gamma\!=\!z$-, $x$-, and $y$-bond, respectively, and 
\be\label{eq:JabJcABcouplings}
\begin{split}
A=&\frac{1}{3}K+\frac{2}{3}(\Gamma - \Gamma'), \;\; B=\frac{1}{3}K-\frac{1}{3} (\Gamma - \Gamma'),\\
J_{ab}&=J+B - \Gamma',\;\; J_{c}=J+A + 2 \Gamma'\,.
\end{split}
\ee
We note in passing that various notations for these constants have been used in the literature. For example, in Ref.~\cite{Chaloupka2015PRB}, $J_{ab}$ and $J_c$ are denoted by $J_{XY}$ and $J_Z$, respectively, and in Ref.~\cite{Onoda2011,Ross2011PRX}, $A$ and $B$ are denoted by $J_{\pm \pm}$ and $J_{z\pm}$, respectively, in the context of pyrochlores.

Equation~(\ref{eq:JabJzABmodel}) delivers the following insights. First, the bond-directionality in Kitaev materials enters only through the constants $A$ and $B$, since these are the only constants that contain the bond-dependent parameters $c_\gamma$ and $s_\gamma$~\cite{Chaloupka2015PRB}.
Second, in the absence of $A$ and $B$ (which translates into $K\!=\!0$ and $\Gamma\!=\!\Gamma'$), the Hamiltonian reduces to a $XXZ$ model (with no bond directionality).  
In this model, the in-plane vs out-of-plane anisotropy is determined by the combination $J_c\!-\!J_{ab}\!=\!\Gamma+2 \Gamma'$. In particular, a positive $\Gamma\!+\!2\Gamma'$ renders the $c^\ast$-axis (i.e, the $[111]$-axis) the hard axis. A magnetic field along this axis leads to an interesting competition, which has been explored in Ref.~\cite{Gordon2019NC}.
Finally, as advocated in Ref.~\cite{Maksimov2020PRR} for the case of {\rucl}, the use of the orthorhombic frame (in conjunction with constrains from experimental data and further duality transformations) can be useful for constraining the number of `free' (fitting) parameters of the model and for mapping to a simpler description.

\subsection{Typical values of microscopic couplings in $d^5$ systems}\label{sec:NumValues}

\begin{table*}[!t]
\setlength{\arrayrulewidth}{0.2mm}
\renewcommand{\arraystretch}{1.75}
\begin{tabular}{|C{5em}|C{3em}C{3em}|C{3em}C{3em}|C{3em}C{3em}|C{3em}C{3em}|C{3em}C{3em}|C{6em}|}
\hline
material& $K_z$ & $K_{x/y}$ & $\Gamma_z$ & $\Gamma_{x/y}$ & $\Gamma_z'$ & $\Gamma_{x/y}'$ & $J_z$ & $J_{x/y}$ & $J_{3,z}$ & $J_{3,x/y}$ & Refs.\\[0.5ex] 
\hline\hline
{\rucl} &-5.0 & -7.5 & 8.0 & 5.9 & -1.0 & -0.8 & -2.2 & -1.4 & 2.4 & 3.0 &\cite{Winter2016}\\
&\multicolumn{2}{c|}{$\in$ [-11,-3.8]} & \multicolumn{2}{c|}{$\in$~[3.9,5]} & \multicolumn{2}{c|}{$\in$~[2.2,3.1]} &  \multicolumn{2}{c|}{$\in$~[-4.1,-2.1]}& \multicolumn{2}{c|}{$\in$~[2.3,3.1]}&\cite{Maksimov2020PRR}\\
\hline 
{\nairo}&-17.9 & -16.2 & -0.1 & 2.1 & -1.8 & -2.3 & 1.6 & -0.1 & 6.8 & 6.7 &\cite{Winter2016}\\ 
&-20.5 & -15.2 & 0.5 & 1.2 & & & 5.0 & 1.5 &&&\cite{Katukuri2014NJP} \\
\hline
{\aliiro}&-4.2 & -13.0 & 11.6 & 6.6& -4.3 & -0.4 & -4.6 & -1.0 & 4.4 & 4.7 &\cite{Winter2016}\\ 
\hline 
{\bliiro}&\multicolumn{2}{c|}{-18} & -10 & $\pm$10 &&&\multicolumn{2}{c|}{0.4} &&&\cite{Rousochatzakis2018,Li2020,Ruiz2021}\\ 
&\multicolumn{2}{c|}{-24} & -9.3& $\pm$9.3 &&&\multicolumn{2}{c|}{0.4} &&&\cite{Halloran2022}\\
\hline 
\end{tabular}
\caption{Numerical values (in units of meV) for the most relevant microscopic parameters of four representative $d^5$ materials, as extracted from: i) exact diagonalizations of the Hubbard model on small clusters (first three materials~\cite{Winter2016}), ii) {\it ab initio} quantum chemistry calculations~\cite{Katukuri2014NJP}, or iii) fits to experiments based on minimal models of {\rucl}~\cite{maksimov2022PRB} and {\bliiro}~\cite{Rousochatzakis2018,Li2020,Ruiz2021,Halloran2022}. The values for the off-diagonal couplings $\Gamma$ and $\Gamma'$ correspond to the choice of the cubic frame shown in Fig.~\ref{fig:geometry} for the layered materials (first three rows) and to the frame of Eq.~(\ref{eq:bLi213frame}) for {\bliiro}. 
For the first three materials, the numbers include the longer-range Heisenberg exchange $J_3$ (across the diagonals of the hexagon plaquettes), which has been proposed to play a substantial role, at least in {\nairo} and {\aliiro}. 
Also, for these materials there are additional but much weaker interactions not shown here~\cite{Winter2016}.}\label{tab:NumValues}
\end{table*}

To understand the properties of the candidate Kitaev materials, it is desirable to estimate the exchange interactions as precisely as possible. 
This task is challenging for three main reasons: i) the relatively low crystal symmetry  renders different NN bonds non-equivalent (although this inequivalence is weak in many cases), ii) the strong spin-orbit coupling and the presence of multiple non-equivalent superexchange paths render the resulting spin Hamiltonian very complex~\cite{Sizyuk2014,Liu2018PRB,Maksimov2020PRR,Huimei2022PRB}, and iii) the fact that the dominant couplings $K$ and (often) $\Gamma$ are frustrated  evinces many of the subleading terms relevant, even if they are much weaker in strength.

The above task then requires bringing together insights from theory -- most notably, {\it ab initio}  quantum chemistry calculations (incorporating explicit many-body treatment of spin multiplets), exact diagonalizations of the Hubbard model on small clusters, perturbative superexchange expansions, and general symmetry analysis and duality mappings~\cite{Mazin2012PRL,Katerina2013,Yamaji2014PRL,Katukuri2014NJP,Satoshi2016,Kim2015,HSKim2016PRB,Winter2016,Winter2017,Hou2017PRB,Sugita2020,Eichstaedt2020} -- as well as important constraints from experiments~\cite{Singh2010PRB,Singh2012PRL,Liu2011PRB,Choi2012PRL,Ye2012PRB,Chun2015,Williams2016PRB,Vale2019,Choi2019PRB,Plumb2014PRB,Sears2015PRB,Kubota2015,Johnson2015,Majumder2015,Banerjee2016,Biffin2014PRB,Biffin2014PRL,Modic2014NC,Takayama2015,Ruiz2017,Ruiz2019,Majumder2019,Majumder2020,Yang2022,Halloran2022,Maksimov2020PRR}. As was pointed by Maksimov  and Chernyshev~\cite{Maksimov2020PRR}, these constrains still do not provide a unique set of model parameters because in most Kitaev materials the minimal models still require a high-dimensional parameter space (at least five-dimensional in {\rucl}). 

To give a rough idea of the typical energy scales and relative importance of the various microscopic couplings in Kitaev materials, we list in Table~\ref{tab:NumValues} some representative numerical values for the most relevant microscopic couplings in four representative $d^5$ materials, {\rucl}, {\nairo}, {\aliiro} and {\bliiro}. 
These numbers have been extracted from exact diagonalizations of the Hubbard model on small clusters~\cite{Winter2016}), or {\it ab initio} quantum chemistry calculations~\cite{Katukuri2014NJP}, or fits of minimal models to experiments~\cite{maksimov2022PRB,Rousochatzakis2018,Li2020,Ruiz2021,Halloran2022}. For a more complete overview of the numerical values extracted from various other methods see, e.g., Table~I of Ref.~\cite{maksimov2022PRB} and references therein.

The data of Table~\ref{tab:NumValues} provide the following insights. First, the Kitaev coupling is the dominant or one of the two dominant (in case of {\rucl}) interactions, as initially proposed by Jackeli and Khaliullin~\cite{Jackeli2009PRL}. The Kitaev coupling is also ferromagnetic. 

Second, except for {\nairo}, the off-diagonal coupling $\Gamma$ is quite substantial and is comparable in size or can even exceed $|K|$. 
The overall dominance of $K$ and $\Gamma$ stems from the fact that the superexchange via ligand $p$-orbitals often dominates the direct exchange contribution which leads to the dominant FM Kitaev coupling. As $\Gamma$ interaction arises from a combination of direct and ligand-mediated hopping, it is usually weaker that $K$, but is larger than other subdominant interactions allowed by symmetry (except for {\nairo}).

Third, the inequivalence between the $K$ and $\Gamma$ couplings on $z$ bonds and those on $x/y$ bonds in the layered materials with monoclinic symmetry seems to be relatively weak in {\rucl} and {\nairo}, but not in {\aliiro}.

Fourth, other interactions are generally much weaker than $K$, but can nevertheless play a decisive role in the low-temperature magnetic ordering because the $K$-$\Gamma$ model is frustrated, as we discuss in more detail below.
For example, the zigzag magnetic ordering in {\rucl}~\cite{Plumb2014PRB,Sears2015PRB,Kubota2015} and {\nairo}~\cite{Singh2010PRB,Singh2012PRL,Liu2011PRB,Choi2012PRL,Ye2012PRB,Chun2015} is selected by the competition of $K$  and $\Gamma$ interactions with smaller $\Gamma'$, $J$ and the longer-range Heisenberg coupling $J_3$~\cite{Winter2016PRB}, while the incommensurate (IC) modulation ~\cite{Biffin2014PRB} of the 3D hyperhoneycomb {\bliiro}, and many other  experimental observations in this compound~\cite{Ruiz2017,Ruiz2019,Majumder2019,Majumder2020,Yang2022,Halloran2022} can be well understood at the level of the NN $J$-$K$-$\Gamma$ model, with dominant ferromagnetic $K$, half-strong $\Gamma$ and a much weaker $J$~\cite{Ducatman2018,Rousochatzakis2018,Li2020,Li2020b}, see last row of Table~\ref{tab:NumValues}.

The situation for the $d^5$ ilmenites and $d^7$ cobaltates mentioned above is under active exploration. 
For the ilmenites AIrO$_3$ (A = Mg, Zn, Cd)~\cite{HaraguchiPRM2018,HaraguchiPRM2020}, whose electronic structure is similar to that of A$_2$IrO$_3$ (A=Na, Li), recent electronic band structures studies have shown that at least in A=Mg and Zn, the effective superexchange interactions are also dominated by a FM Kitaev and an AFM $\Gamma$ coupling~\cite{JangPRM2021,hao2022electronic}. 
Similarly, for the $d^7$ cobaltates, theoretical studies  suggest that, despite a weaker spin-orbit coupling (compared to their 4d and 5d counterparts), the bond-anisotropic $K$ and $\Gamma$ interactions can be realized in materials when the direct intra-orbital $t_{2g}$-$t_{2g}$ hopping ($t_3$) is negligible compared to indirect hoppings~\cite{Sano2018PRB, Liu2018PRB,Liu2020PRL}. While this is true for some cobaltates, it has recently been pointed out that $t_3$ is the largest hopping integral in  BaCo$_2$(XO$_4$)$_2$ with X = As, P leading to the dominant Heisenberg interaction~\cite{das2021PRB,winter2022arXiv,maksimov2022PRB,liu2022arXiv}. Further studies are currently underway to identify cobaltates with dominant Kitaev interaction.

\subsection{Higher-$S$ extensions}\label{sec:generalS}

To conclude this section, we will briefly discuss recent developments in the studies of  general spin-$S$ superexchange models. The latter have been only a theoretical interest until recently, as  an effective spin $S\!>\!1/2$ requires strong Hund's coupling, and the SOC required for the Kitaev interaction becomes inactive. For example, in $d^8$ and $d^3$ configurations, Hund's coupling gives $S\!=\!1$ and $S\!=\!3/2$, respectively. In these cases,  SOC has null effect, because of the quenched orbital angular momentum ($L\!=\!0$). While the atomic SOC, $\xi \sum_i {\bf l}_i \cdot {\bf s}_i$,  together with a small trigonal field  leads to a single ion anisotropy $({\bf S} \cdot {\hat c})^2$ where ${\hat c}$ is the trigonal field direction, it is natural to expect the absence of Kitaev interaction, as the Hund's coupling wins over the SOC at metal sites.

However, when anions are heavy, the bond-dependent interactions are generated by utilizing the SOC at anion sites~\cite{Xu2020PRL,Stavropoulos2019PRL}.  For example, a  strong coupling expansion for $d^8$ has revealed that the Kitaev interaction is twice larger than the Heisenberg interaction via indirect exchange paths~\cite{Stavropoulos2019PRL}. Furthermore,  the sign of the Kitaev interaction  is positive (AFM), and the $\Gamma$ interaction is absent up to the second order in the expansion. Combining both direct and indirect exchange contributions, the Kitaev coupling can be the dominant interaction in a $d^8$ Mott insulator. 

For $d^3$, a similar analysis has shown that  in this case the Kitaev interaction can  be generated via SOC at anion sites, but its strength depends crucially on the competition of the contributions from different  hopping paths~\cite{Stavropoulos2021PRR}. In particular, when the hopping  path involving both $t_{\rm 2g}$ and $e_{\rm g}$ is large,  it  might cancel the contribution from the path involving only $t_{\rm 2g}$ orbitals. Then, the Kitaev interaction is no longer dominant but is overshadowed by the Heisenberg interaction. Given its sensitivity to various hopping integrals, the strength of the Kitaev interaction in candidate materials with heavy anions such as CrI$_3$ is currently under debate~\cite{Lee20220PRL,Chen2018PRX,Chen2021PRX,Band2022PRB,Cen2023PRB} and yet to be fully determined. 

Finally, it is worth mentioning that bond-dependent interactions are not limited to dipole moments, but also occur in multipolar systems described by quadrupole and octupole moments. In $d^2$ Mott insulators, for example, such as Osmium double perovskites\cite{Chen2011PRB,Thompson2014,Marjerrison2016PRB}, a dipole moment is absent, because the total angular moment $J\!=\!2$ further splits into a non-Kramer doublet $E_g$ and a triplet $T_{2g}$ via the mixing of $t_{2g}$ and $e_g$ levels~\cite{Paramekanti2020PRB,Voleti2020PRB}.
The resulting effective Hamiltonian projected  into low-energy doublet states takes a form similar to the above $JK\Gamma\Gamma'$ model of the $d^5$ configuration. For example, a microscopic derivation for 90$^\circ$ bonding geometry leads to an effective pseudospin  $S\!=\!1/2$  description with Hamiltonian~\cite{Khaliullin2021PRR}
\be\label{eq:90}
\mc{H} = \sum_{\langle ij \rangle}
J_o\; S_i^yS_j^y + J_q ( S_i^xS_j^x + S_i^zS_j^z)+ J_\tau \; \tau_{i\gamma}\tau_{j\gamma}\,.
\ee 
Here the following notations are used: $\tau_\gamma \!=\!\cos\phi_\gamma S^z\!+\!\sin\phi_\gamma S^x$, $\phi_\gamma\!=\!(0, 2\pi/3, 4\pi/3)$ for the bond type $\gamma\!=\!z$, $x$, and $y$, respectively. Furthermore, for small $t_1$ and in the limit of $J_H \ll U$, we have
$J_o \!=\! - J_q \!=\!\frac{2}{3}\frac{t_2^2}{U}$ and $J_\tau\!=\!\frac{4}{9} \frac{t_3^2}{U}$ (we use the same notations for $t_1,t_2$ and $t_3$ as in Eq.~(\ref{Eq:hoppingd5}) and Ref.~\cite{Rau2014}). Including $t_1$ can however lead to a FM $J_o$, which is important for the octupolar ordering in Osmium double perovskites~\cite{Churchill2022PRB}. In bipartite lattices, such as the honeycomb, when $J_o=-J_q$, the Hamiltonian can be further transformed into the $JK\Gamma\Gamma'$ form with 
\bea\label{transformation}
K & =&  1, \; \;\;  \Gamma =  1- \frac23 (1-\delta),  \nonumber\\
J & = & \frac13 (1-\delta),  \;\; \; \Gamma' = - \frac23 (1-\delta), 
\eea
where $\delta \!=\! 2 J_o/J_\tau$. There is an exact degeneracy between quadrupole and octupolar orderings when $\delta=1$, which is equivalent to $K=\Gamma$ in the spin-1/2 $(J_{\rm eff} = 1/2)$ case where N\'eel and 120$^\circ$ orders are degenerate~\cite{Rau2014}. Under the magnetic field along [111]-axis, additional interaction between octupole and quadrupole occurs, which allows the Kitaev spin liquid out of non-Kramer doublet in the $d^2$ honeycomb Mott insulators~\cite{Rayyan2022arXiv}.

\section{The honeycomb  $JK\Gamma\Gamma'$ model (I):\\ Symmetries \& dualities}\label{sec:sym_and_dual}

Having illustrated the origin of the various microscopic interactions in Kitaev materials, we shall now turn to the analysis of the generic $JK\Gamma\Gamma'$ model of Eq.~(\ref{eq:JKGGpmodel}), paying special attention to the 2D honeycomb geometry with pseudospin $S\!=\!1/2$ degrees of freedom. 
In some special cases, we will also discuss the semiclassical (large-$S$) limit as a means to understand the magnetic phases and the origin of frustration in some regions of the phase diagram. 

Before we embark on reviewing the phase diagram, it is worthwhile to first map out some of the special, highly-correlated regions in the parameter space, based on general symmetry considerations. We shall begin with the two highly-frustrated points of major interest, the pure Kitaev and $\Gamma$ points, and focus on the corresponding local symmetries that are responsible for driving spin liquidity. We shall then move onto identifying other special points with hidden symmetries, or whole regions that are related by duality transformations.

\subsection{Local symmetries}\label{sec:localsymmetries}
Quite generally, spin liquidity arises from the presence of local symmetries that are either explicitly present in the Hamiltonian, or emerge at low energies in the form of energetic constraints. 
Indeed, according to Elitzur's theorem~\cite{Elitzur1975}, such symmetries cannot be broken spontaneously, and this automatically excludes conventional magnetic long-range ordering, and opens the door for fluctuating, spin-liquid-like phases. 
Local symmetries or constraints are, in addition, responsible for two of the most characteristic signatures of spin liquids: i) the spectral downshift which is responsible for the persistence of large entropy down to very low temperatures~\cite{RamirezBook,Rousochatzakis2019PRB}, and ii) the existence of long-lived fractionalized excitations, which arises from the fact that conventional {\it single}-particle excitations are generally incompatible with local symmetries.

Such local symmetries exist in the pure Kitaev model at both the classical and quantum level, and in the pure $\Gamma$ model at the classical level.

\subsubsection{Pure Kitaev model}\label{sec:localsymmetriesK}

The Kitaev point ($J=\Gamma=\Gamma'=0$) hosts the Kitaev QSL phases for $S=1/2$~\cite{Kitaev2006} and similar QSL phases for higher spin~\cite{Baskaran2008PRB,Rousochatzakis2018NC,Koga2018JPSJ,Minakawa2019,Stavropoulos2019PRL,Lee2020PRR,Khait2021PRR,Jin2022NC,Bradley2022}. For the $S=1/2$ honeycomb model, the local symmetries underlying the QSL have been identified in the seminal work of Kitaev~\cite{Kitaev2006}, and correspond to local transformations that reside on individual hexagon plaquettes.  
Using the following site-labelling for any given hexagon plaquette,
\be\label{eqfig:hex}
\includegraphics[width=1.0in]{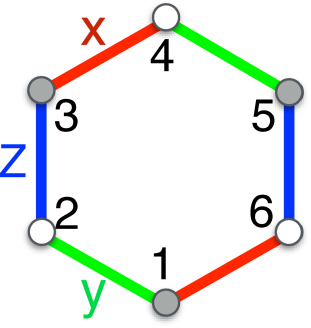}
\ee 
these transformations take the form~\cite{Kitaev2006}:
\be\label{eq:WhS1o2}
W_h = 2^6~S_1^z S_2^x S_3^y S_4^z S_5^x S_6^y\,,
\ee
and $[W_h, \mc{H}]=0$ for every separate hexagon $h$. Such local symmetries have also been identified for various tricoordinated  3D lattices~\cite{Mandal2009PRB,Obrien2016}, as well as in 1D versions of the model ~\cite{Feng2007PRL,WU2012}, which we will discuss in Sec.\ref{sec:1D}. 

Interestingly, higher-spin $S$ extensions of the Kitaev model also feature an extensive number of local symmetries. Indeed, as shown in Ref.~\cite{Baskaran2008PRB}, for general $S$, the plaquette operators take the form 
\be\label{eq:WhhighS}
W_h = - \exp[i \pi (S_1^z+S_2^y+S_3^x+S_4^z+S_5^y+S_6^x)]\,,
\ee
which reduces to Eq.~(\ref{eq:WhS1o2}) for $S=1/2$. As discussed briefly in Sec.~\ref{sec:localsymmetries}, Elitzur's theorem prohibits magnetic long-range ordering and opens the door to QSL-like fluctuating states for any $S$~\cite{Rousochatzakis2018NC,Koga2018JPSJ,Minakawa2019}. 
It was also shown that these operators  are fermions in the half integer spin model, but are bosons in the integer spin model  \cite{HanMa2023}.

\subsubsection{Pure $\Gamma$ model} \label{sec:Gammaclassicalsymmetry}
At the classical level, the $\Gamma$ point ($J=K=\Gamma'=0$) marks a strongly correlated regime (both in 2D and 3D three-coordinated Kitaev lattices) with infinite number of classical ground states and a distinctive pattern of anisotropic spin-spin correlations (see Sec.~\ref{sec:Gpoints} below). 
The reason is that, for classical spins, the pure $\Gamma$ model has an extensive number of local symmetry transformations, of three types, that affect the six spins of an individual hexagon plaquette $h$~\cite{Rousochatzakis2017PRL}.  
Using the site labeling of (\ref{eqfig:hex}), these transformations amount to the following three types of transformations 
\be
\Theta(h)\!\cdot\!\mc{R}_x(h),~~
\Theta(h)\!\cdot\!\mc{R}_y(h),~~
\Theta(h)\!\cdot\!\mc{R}_z(h),~~
\ee
where $\mc{R}_{x,y,z}(h)$ are the operations 
\bea\label{eq:Rxyz}
\renewcommand{\arraystretch}{1.5}
\begin{array}{l}
\mc{R}_x(h) \!=\! 
\mathsf{C}_{2x}(1)
\mathsf{C}_{2z}(2)
\mathsf{C}_{2z}(3)
\mathsf{C}_{2y}(4)
\mathsf{C}_{2y}(5)
\mathsf{C}_{2x}(6),
\\
\mc{R}_y(h) \!=\! 
\mathsf{C}_{2y}(1)
\mathsf{C}_{2y}(2)
\mathsf{C}_{2x}(3)
\mathsf{C}_{2x}(4)
\mathsf{C}_{2z}(5)
\mathsf{C}_{2z}(6),
\\
\mc{R}_z(h) \!=\! 
\mathsf{C}_{2z}(1)
\mathsf{C}_{2x}(2)
\mathsf{C}_{2y}(3)
\mathsf{C}_{2z}(4)
\mathsf{C}_{2x}(5)
\mathsf{C}_{2y}(6),
\end{array}~~~~~
\eea 
and the symbol $\Theta(h)$ denotes the time-reversal operation, applied {\it only} to the six spins of the given hexagon $h$, without affecting the remaining spins of the lattice.
Since $\mc{R}_{x,y,z}(h)$ correspond to successive $\pi$-rotations which reverse two Cartesian components of the hexagon spins, the combinations $\Theta(h)\!\cdot\!\mc{R}_{x,y,z}(h)$ reverse a single Cartesian component. For example, 
\be\label{eq:ThetaRGamma}
\begin{array}{c}
 \Theta(h)\!\cdot\!\mc{R}_{z}(h)\\
\text{(classical spins)}
\end{array}:~~
\renewcommand{\arraystretch}{1.5}
\begin{array}{c |  rrr}
\text{label}~j& S_j^x & S_j^y & S_j^z \\
\hline
1 & \widetilde{S}_1^x& \widetilde{S}_1^y&-\widetilde{S}_1^z\\
2 & -\widetilde{S}_2^x& \widetilde{S}_2^y&\widetilde{S}_2^z\\
3 & \widetilde{S}_3^x& -\widetilde{S}_3^y&\widetilde{S}_3^z\\
4 & \widetilde{S}_4^x& \widetilde{S}_4^y&-\widetilde{S}_4^z\\
5 & -\widetilde{S}_5^x& \widetilde{S}_5^y&\widetilde{S}_5^z\\
6 & \widetilde{S}_6^x& -\widetilde{S}_6^y&\widetilde{S}_6^z\\
\hline
\end{array}
\ee
Importantly, the situation in the quantum case is qualitatively different, because, for extended systems, the time-reversal operator $\Theta$ is a global transformation, and thus it cannot affect only a part of the system (here the six spins of an individual hexagon plaquette $h$). In other words, the operation $\Theta(h)$ [and hence $\Theta(h)\!\cdot\!\mc{R}_{x,y,z}(h)$ as well] simply {\it does not exist} for quantum spins.

\subsection{Hidden SU(2) symmetry points}\label{subsec:hiddenSU2}

Besides the two special points discussed above, the parameter space hosts a number of hidden, continuous and discrete symmetries, as well as a number of dualities.

We first discuss the hidden SU(2) symmetry points~\cite{Chaloupka2010PRL,Chaloupka2013PRL,Chaloupka2015PRB}, at which the Hamiltonian can be re-written as an isotropic Heisenberg Hamiltonian,
\be
\mc{H}=\sum\nolimits_{\langle ij \rangle} \widetilde{J}~ \widetilde{\bf S}_i\cdot\widetilde{\bf S}_j,
\ee
under a generally site-dependent transformation ${\bf S}_i\!\mapsto\!\widetilde{\bf S}_i$. These special points can be associated with 2-, 4- or 6-sublattice transformations as detailed below.

\subsubsection{Hidden SU(2) point in the $K\Gamma$ plane:\\ Six-sublattice mapping $\mc{T}_6$}\label{Sec:HiddenSU2KG}
Consider the six-sublattice decomposition and the associated transformation for the six spins of every third hexagon (here the ones shaded in blue): 
\be\label{eqfig:T6transfhex}
\includegraphics[width=2.5in]{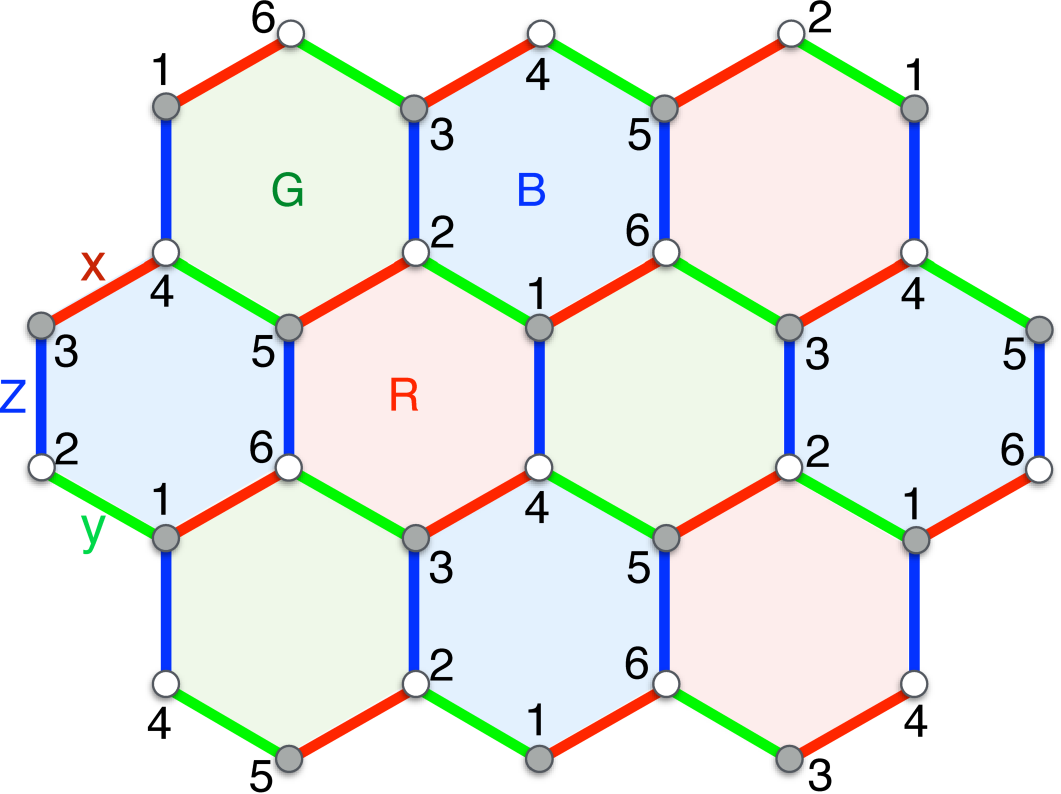}
\ee 
\be
\begin{array}{c}
\text{6-sublattice}\\
\text{transf.}~\mc{T}_6
\end{array}:~~
\renewcommand{\arraystretch}{1.5}
\begin{array}{c |  rrr}
\text{label}~j& S_j^x & S_j^y & S_j^z \\
\hline
1 & \widetilde{S}_1^x&  \widetilde{S}_1^y& \widetilde{S}_1^z \\
2 &- \widetilde{S}_2^z&- \widetilde{S}_2^y& - \widetilde{S}_2^x\\
3 & \widetilde{S}_3^y& \widetilde{S}_3^z& \widetilde{S}_3^x \\
4 &-\widetilde{S}_4^y&- \widetilde{S}_4^x&- \widetilde{S}_4^z \\
5 & \widetilde{S}_5^z& \widetilde{S}_5^x& \widetilde{S}_5^y\\
6 &-\widetilde{S}_6^x&-\widetilde{S}_6^z&-\widetilde{S}_6^y\\
\hline
\end{array}
\label{eq:T6transfhex}
\ee
Note that this 6-sublattice decomposition corresponds to a trimerization of the lattice (see shaded hexagons). Therefore, there are, in total, three different transformations of this kind.

Now, one can show that the special point $J\!=\!\Gamma'\!=\!0$ and $K\!=\!\Gamma$ is a hidden SU(2) point with 
\be\label{eq:hiddenSU2fromT6}
\widetilde{J}=-K\,.
\ee
To see this, consider, e.g., the interactions on the $y$-bond (1,2) in (\ref{eqfig:T6transfhex}). Under $\mc{T}_6$, these map as follows
\be
K S_1^y S_2^y \!+\! \Gamma (S_1^x S_2^z \!+\!S_1^z S_2^x ) \mapsto 
-K \widetilde{S}_1^y \widetilde{S}_2^y \!-\! \Gamma (\widetilde{S}_1^x \widetilde{S}_2^x \!+\!\widetilde{S}_1^z \widetilde{S}_2^z )\,,\nonumber
\ee
which further maps to $-K \widetilde{\bf S}_1\!\cdot\!\widetilde{\bf S}_2$ for $K\!=\!\Gamma$. The same holds for the other bonds.

\subsubsection{Hidden SU(2) point in the $JK$ plane:\\ Four-sublattice mapping $\mc{T}_4$}\label{sec:HiddenSU2JK}
Consider the following four-sublattice decomposition
\be\label{eqfig:T4transf}
\includegraphics[width=2.0in]{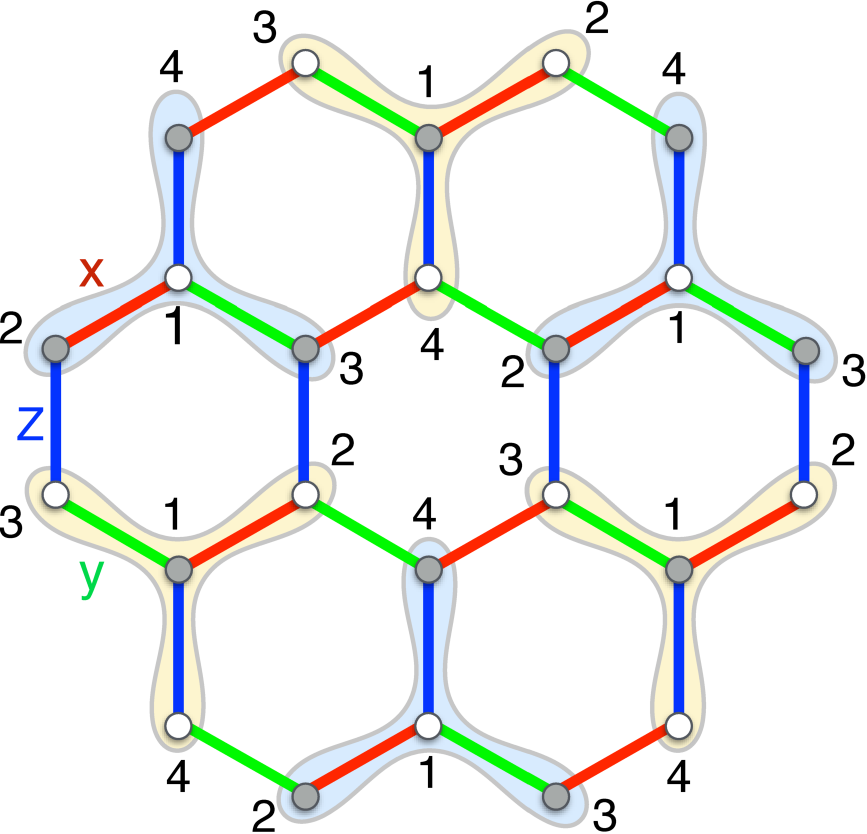}
\ee 
and the associated transformation
\be\label{eq:T4transf}
\begin{array}{c}
\text{4-sublattice}\\
\text{transf.}~\mc{T}_4
\end{array}:~~~
\renewcommand{\arraystretch}{1.5}
\begin{array}{c |  rrr}
\text{label}~j& S_j^x & S_j^y & S_j^z \\
\hline
1 & \widetilde{S}_1^x& \widetilde{S}_1^y&\widetilde{S}_1^z\\
2 & \widetilde{S}_1^x& -\widetilde{S}_1^y&-\widetilde{S}_1^z\\
3 & -\widetilde{S}_1^x& \widetilde{S}_1^y&-\widetilde{S}_1^z\\
4 & -\widetilde{S}_1^x& -\widetilde{S}_1^y&\widetilde{S}_1^z\\
\hline
\end{array}
\ee
For $\Gamma\!=\!\Gamma'\!=\!0$, this is a duality that maps $K$ and $J$ to
\be\label{eq:JKduality}
\widetilde{K}=K+2J\,,~~
\widetilde{J}=-J\,,
\ee
from which it follows that the special point
\be
\Gamma=\Gamma'=0\,,~~
K=-2J
\ee 
is a hidden $SU(2)$ point with 
\be\label{eq:hiddenSU2fromT4}
\widetilde{J}=-J\,.
\ee

\subsubsection{Hidden SU(2) point in the $J\Gamma\Gamma'$ space:\\ Two-sublattice mapping $\mc{T}_2$}
The first type of hidden SU(2) points are along the special line 
$K=0$, $\Gamma=\Gamma'=-2J$, 
for which 
\be\label{eq:hiddenSU2fromT2}
\widetilde{J}=-3J\,.
\ee
To see this, we note that along this line, the couplings $A$ and $B$ of Eq.~(\ref{eq:JabJcABcouplings}) vanish identically, while $J_{ab}=-J_c=3J$, and therefore Eq.~(\ref{eq:JabJzABmodel}) reduces to 
\be 
\mc{H}_{ij,\gamma} = 3J (S_i^aS_j^a+S_i^bS_j^b-S_i^cS_j^c)\,,
\ee
(the same for all $\gamma$), which is the XXZ model mentioned in Sec.~\ref{sec:abcframe}.  This model can, in turn, be mapped to an isotropic Heisenberg form with $\widetilde{J}\!=\!-3J$, by performing a $C_{2{\bf c}^\ast}$ rotation (in spin space alone) on one of the two sublattices of the honeycomb model.

\subsubsection{Hidden SU(2) points from transformations $\mc{T}_1\mc{T}_4$ \& $\mc{T}_2\mc{T}_6$}\label{subsec:hidden}
As shown by Chaloupka and Khaliullin~\cite{Chaloupka2015PRB}, the point
\be\label{eq:HiddenSU2pointfromT1T4}
J=K/6=-\Gamma/8=\Gamma'/4 
\ee
is a hidden SU(2) point with $\widetilde{J}=-J$, corresponding to the transformation $\mc{T}_1\mc{T}_4$, where $\mc{T}_1$ is a global $\pi$-rotation around the ${\bf c^\ast}$-axis and $\mc{T}_4$ is given in Eq.~(\ref{eq:T4transf}). 
According to a recent study by Kr\"uger {\it et al}~\cite{Janssen2023}, the proximity of the $d^7$ compound Ni$_2$Co$_2$TeO$_6$ to precisely this hidden SU(2) point is responsible for the stabilization of a triple-${\bf Q}$ magnetic order with finite scalar spin chirality.

Similarly, the point
\be\label{eq:HiddenSU2pointfromT2T6} 
J=-2K/3=-2\Gamma=\Gamma'
\ee 
is another hidden SU(2) point with $\widetilde{J}=K$, corresponding to the combined transformation $\mc{T}_2\mc{T}_6$, where $\mc{T}_2$ is a $\pi$-rotation around the ${\bf c^\ast}$-axis applied to one sublattice only, and $\mc{T}_6$ is given by Eq.~(\ref{eq:T6transfhex}).

\subsection{Hidden discrete symmetries}\label{sec:Hiddendiscrete}
As pointed out in Ref.~\cite{Rousochatzakis2017PRL} (see also Ref.~\cite{Samarakoon2018PRB}), the entire $K\Gamma$ line hosts 
a hidden $C_{2v}$ symmetry group, composed of the following three, global, $\pi$-rotations
\be\label{eq:RxyzFull}
\mc{R}_x\!=\!\prod_{h\in B}\!\mc{R}_x(h),~
\mc{R}_y\!=\!\prod_{h\in B}\!\mc{R}_y(h),~
\mc{R}_z\!=\!\prod_{h\in B}\!\mc{R}_z(h),
\ee
where the six-body operations $\mc{R}_{x,y,z}(h)$ were defined in Eq.~(\ref{eq:Rxyz}) above and the products over $h$ are over the `blue' shaded hexagons of (\ref{eqfig:T6transfhex}).
Note that $\mc{R}_z$ can also be thought of as the product of the plaquette operators $W_h$ of Kitaev~\cite{Kitaev2006} over the `blue' shaded hexagons of (\ref{eqfig:T6transfhex}): 
\be
\mc{R}_z = \prod_{h \in B} (-\sigma_1^z \sigma_2^x \sigma_3^y \sigma_4^z \sigma_5^x \sigma_6^y)_h = \prod_{h\in B} (-W_h) \,.
\ee
Similarly, $\mc{R}_x$ and $\mc{R}_y$ are the corresponding products of Kitaev's plaquettes defined on the hexagons labeled by `G' and `R', respectively, in (\ref{eqfig:T6transfhex}), that is
\be
\mc{R}_x = \prod_{h\in G} (-W_h),~~
\mc{R}_y = \prod_{h\in R} (-W_h).
\ee

Note further that in the frame corresponding to the six-sublattice transformation $\mc{T}_6$ discussed above, the operations $\mc{R}_x$, $\mc{R}_y$ and $\mc{R}_z$ correspond to two-fold rotations around the (local) axes $\widetilde{x}$, $\widetilde{y}$ and $\widetilde{z}$, respectively:
\be\label{Ralpha}
\mc{R}_\alpha = \prod_{i=1}^N \sigma_i^{\widetilde{\alpha}} = (-1)^{N/6} \prod_{i=1}^N  \mathsf{C}_{2\widetilde{\alpha}}(i),~~\widetilde{\alpha}=\widetilde{x},\widetilde{y},\widetilde{z},
\ee
where $N$ is the total number of sites. Hence, this symmetry is a $C_{2v}$ group in the rotated frame.

\subsection{Dualities}
Dualities are transformations that preserve the form of the Hamiltonian but alter the values of the parameters, and are very common in spin-orbital models~\cite{Khaliullin2005PTPS,Chaloupka2010PRL,Chaloupka2015PRB,Rau2014,Kimchi2014a,Yang2020}. 
The isolated hidden SU(2) points of Eqs.~(\ref{eq:hiddenSU2fromT2}), (\ref{eq:hiddenSU2fromT4}),  (\ref{eq:hiddenSU2fromT6}), (\ref{eq:HiddenSU2pointfromT1T4}) and (\ref{eq:HiddenSU2pointfromT2T6}) -- resulting, respectively, from the transformations $\mc{T}_2$, $\mc{T}_4$, $\mc{T}_6$, $\mc{T}_1\mc{T}_4$ and $\mc{T}_2\mc{T}_6$ -- provide examples of such dualities at special, isolated points.
The more general duality of Eq.~(\ref{eq:JKduality}) associated with $\mc{T}_4$ maps an extended region of parameter space in the $JK$ plane to another extended region in the same plane.
Below we discuss a few more relevant examples.

\subsubsection{Duality that maps $K\mapsto-K$}\label{sec:dualityKtomK}
For $J=\Gamma=\Gamma'=0$, the ferromagnetic and antiferromagnet Kitaev points are related to each other by the four-sublattice duality transformation~\cite{Chaloupka2010PRL,Rousochatzakis2015PRX} 
\be\label{eq:T4ptransf}
\begin{array}{c}
\text{4-sublattice}\\
\text{transf.}~\mc{T}_4'
\end{array}:~~~
\renewcommand{\arraystretch}{1.5}
\begin{array}{c |  rrr}
\text{label}~j& S_j^x & S_j^y & S_j^z \\
\hline
1 & \widetilde{S}_1^x& \widetilde{S}_1^y&\widetilde{S}_1^z\\
2 & -\widetilde{S}_2^x& \widetilde{S}_2^y&-\widetilde{S}_2^z\\
3 & -\widetilde{S}_3^x& -\widetilde{S}_3^y&\widetilde{S}_3^z\\
4 & \widetilde{S}_4^x& -\widetilde{S}_4^y&-\widetilde{S}_4^z\\
\hline
\end{array}
\ee
associated with the four-sublattice decomposition shown in (\ref{eqfig:T4transf}) above. This operation simply maps $K\!\to\!-K$. This duality disappears for nonzero $J$, $\Gamma$ or $\Gamma'$.

\subsubsection{$\mc{T}_1$ duality}
As shown by Chaloupka and Khaliullin ~\cite{Chaloupka2015PRB}, the general $(J,K,\Gamma,\Gamma')$ point of the parameter space maps to the point $(\widetilde{J},\widetilde{K},\widetilde{\Gamma},\widetilde{\Gamma'})$ under the $\pi$-rotation around the ${\bf c}$-axis (denoted by $\mc{T}_1$ in \cite{Chaloupka2015PRB}), with 
\be
\left(
\begin{array}{c}
\widetilde{J}\\
\widetilde{K}\\
\widetilde{\Gamma}\\
\widetilde{\Gamma'}
\end{array}
\right)=
\left(
\begin{array}{rrrr}
1 & 4/9 & -4/9 & 4/9\\
0 & -1/3 & 4/3 & -4/3\\
0 & 4/9 & 5/9 & 4/9\\
0 & -2/9 & 2/9 & 7/9
\end{array}
\right)
\left(
\begin{array}{c}
J\\
K\\
\Gamma\\
\Gamma'
\end{array}
\right)
\ee
This tells us, e.g., that the Kitaev point $J=\Gamma=\Gamma'=0$ maps to the point
\be
\widetilde{J}=4K/9,~
\widetilde{K}=-K/3,~
\widetilde{\Gamma}=4K/9,~
\widetilde{\Gamma'}=-2K/9,
\ee
and, conversely, that the special point
\be
J=\Gamma=-2\Gamma'=-4K/3
\ee
is a hidden Kitaev point with coupling $\widetilde{K}=-3K$.

Likewise, the special point
\be
J=-K/3=-4\Gamma/5=-2\Gamma'
\ee
is a hidden $\Gamma$ point with coupling $\widetilde{\Gamma'}=9\Gamma/5$.

\subsubsection{Classical duality}\label{sec:ClassicalDuality}
Finally, there is a duality present only for classical spins, and amounts to a sublattice spin inversion $\left(\vec{S}_{\vec{r},1},\vec{S}_{\vec{r},2}\right)\!\to\!\left(\vec{S}_{\vec{r},1},-\vec{S}_{\vec{r},2}\right)$. This transformation maps 
\be
(J,K,\Gamma,\Gamma')\!\to\!(-J,-K,-\Gamma,-\Gamma')~.
\ee
So the classical phase diagram for a given set of couplings can be obtained from that of the opposite couplings. 
This is not true for quantum spins however, because the above transformation requires acting with the complex conjugation on half of the system, which is not possible for quantum spins~\cite{Rousochatzakis2015PRX}.

\section{The honeycomb  $JK\Gamma\Gamma'$ model (II):\\ ~~~Exploration of the phase diagram}\label{sec:JKGammaPhaseDiagram}

The extended $J$-$K$-$\Gamma$-$\Gamma'$ Hamiltonian is too complex to solve analytically, and numerical studies have been playing significant role in offering some physical insights into its phases and providing experimental predictions for  Kitaev materials in which all these interactions might be present simultaneously. The zero and finite magnetic field phase diagram of the $J$-$K$-$\Gamma$-$\Gamma'$ model and its response to various perturbations have been studied extensively using a wide range of techniques, including  classical energy minimization~\cite{Rau2014,Rau2014-arxiv,Sizyuk2014,Lee2016,Plakida2016}, classical Monte Carlo simulations combined with the spin wave analysis~\cite{Price2012PRL,Price2013PRB,Sizyuk2014,Janssen2016,Chern2017,Li2020,Zhang2020}, exact diagonalizations and cluster mean-field theories~\cite{Chaloupka2010PRL,Chaloupka2013PRL,Gotfryd2017PRB,Catuneanu2018npj}, density-matrix renormalization  group~\cite{Gohlke2017PRL,Gohlke2018PRBa,Gohlke2018PRBb,Gordon2019NC,Gohlke2020PRR}, tensor-network methods~\cite{Osorio2014,Czarnik2019,Lee2020NC}, functional renormalization group studies~\cite{Buessen2021PRB} slave-particle mean-field  theories~\cite{Burnell2011PRB,Schaffer2012PRB,Johannes2018PRB}, variational methods~\cite{Shang-Shun2021}, variational Monte Carlo~\cite{Normand2019} and machine learning~\cite{Rau2021PRR,Liu2021PRR}.
All these methods have their particular strengths and weaknesses, and are complementary to each other and to analytical theories. However, there are limitations too and therefore, not surprisingly, these methods often disagree in predicting properties of such a complicated system. 

In the following, we set out to explore the various competing phases of the $J$-$K$-$\Gamma$-$\Gamma'$ model. Given the high-dimensional parameter space of the model, we shall organize the discussion around special limits and regions of the parameter space, paying special attention to the highly-correlated regions of the model (in particular the ones around the pure Kitaev and the pure $\Gamma$ point), and building on insights from the symmetries and dualities discussed in the previous section.

\subsection{The Kitaev points}\label{sec:Kpoints}  

\subsubsection{Quantum $S=1/2$ limit}
We start with the celebrated Kitaev model ($J$=$\Gamma$=$\Gamma'$=0)~\cite{Kitaev2006}, which is an exactly solvable yet realistic spin model with a QSL ground state. Neighboring $S=1/2$ spins of the honeycomb lattice are coupled via different spin components along the three bonds connected to any given site: 
\be
\mc{H}_K =\sum\nolimits_\gamma
\sum\nolimits_{\langle ij \rangle \in \gamma} K_\gamma S_i^\gamma  S_j^\gamma, 
\ee
where $K_\gamma$ are the coupling constants for the three types of bonds $\gamma\!=\!x$, $y$, and $z$. Note that, due to the duality transformation of Sec.~\ref{sec:dualityKtomK}, the  Kitaev  model is symmetric  with respect to the overall sign of $K_\gamma$.

The above model falls under the general family of compass models~\cite{Nussinov2015}, originally discussed in the content of Kugel-Khomskii Hamiltonians in systems with orbital degrees of freedom \cite{Kugel1982}. The special aspect of the honeycomb lattice model (and its 3D extensions) is the presence of {\it local} symmetries $W_h$ discussed in Sec.~\ref{sec:localsymmetriesK}, which lead to fluctuating QSL-like states consistent with Elitzur's theorem~\cite{Elitzur1975}.
For $S=1/2$, the nature of the QSL states has first been revealed by Alexey Kitaev in his original work~\cite{Kitaev2006}.
In particular, the model has two different types of QSL phases, gapped and gapless QSLs, depending on the relative strength of the $K_{\gamma}$ couplings.
These phases can be understood in terms of emergent fermionic degrees of freedom appearing from the spin fractionalizion into  two types of elementary excitations -- Majorana fermions and magnetic fluxes \cite{Kitaev2006}. In both phases, Majorana fermions (introduced in various  spin representations~\cite{Kitaev2006,Chen2008,Chen2012,Fu2018}) move in the background of emergent magnetic fluxes.
Thermal fractionalization manifests itself in successive entropy releases at two well-separated energy scales $T_L\simeq 0.15 K$ and $T_H\simeq K$. At low temperatures ($T < T_L$), the fluxes are mostly frozen and only  Majorana fermions around the  ${\rm K}$ Dirac points in the BZ (with vanishing density of states)  are thermally excited~\cite{Nasu2015PRB}. At intermediate temperatures ($T_L < T < T_H$), thermal energy goes into both fluxes and Majorana fermions, but their fractionalized behavior remains readily observable. Finally, at high temperatures ($T > T_H$), fluxes and Majorana fermions recombine into spins, and the system crosses over to a conventional paramagnetic regime.

The response of the Kitaev magnet to an external magnetic field is nontrivial and depends on the overall sign of $K_\gamma$, as the above duality ceases to exist in the presence of the field~\cite{Kitaev2006,Zhu2018PRB,Gohlke2018PRBb,Hickey2019NC,Liang2018PRB,Berke2020PRB}. For the symmetric model where all $K_\gamma$ are equal, and for weak enough fields, the gapless Kitaev spin liquid turns into a gapped QSL and hosts non-Abelian anyonic excitations~\cite{Kitaev2006}. However, they behave very differently when are subjected to a stronger magnetic field. In particular, for  the  antiferromagnetic  Kitaev interaction ($K>0$), a variety of numerical techniques show that a distinct gapless QSL phases appear in between the non-Abelian topological phase at low fields and the high-field polarized phase~\cite{Zhu2018PRB,Gohlke2018PRBb,Hickey2019NC,Liang2018PRB,Berke2020PRB}.

\subsubsection{Classical limit: The `K recipe'}\label{sec:KclassGSs}
While the quantum spin-1/2 model is exactly solvable, it is instructive to discuss the purely classical limit of this model and the structure of the corresponding ground state manifold. The resulting insights will be useful for the discussion of various classical phases  in Secs.~\ref{sec:KGmodel} and \ref{sec:JKGmodel}, and will also form the basis for the semiclassical analysis of the large-$S$ Kitaev model in Sec.~\ref{sec:generalS2}.

Classically, the Kitaev model is characterized by a huge ground state degeneracy, which was first analysed in the seminal work of Baskaran, Sen and Shankar (BSS)~\cite{Baskaran2008PRB}. The authors identified an infinite number of so-called `Cartesian' states, which map to the dimer coverings of the honeycomb lattice, modulo a factor of two for the orientation of the two spins per dimer. They further showed that the Cartesian states are connected to each other by continuous valleys of other ground states, leading to a huge ground state degeneracy. Shortly after, Chandra, Ramola and Dhar~\cite{Chandra2010} showed that the manifold actually consists of infinitely more ground states and possesses an emergent gauge structure that leads to power-law correlations.

The general structure of the classical ground states can be seen in the parametrization proposed in Ref.~\cite{Rousochatzakis2018NC}, which in addition reveals the topological terms arising from the leading quantum fluctuations in an explicit way. 
In this parametrization, one writes the Cartesian components of each spin as $\vec{S}_i\!=\![x_i,y_i,z_i]$ with $x_i^2\!+\!y_i^2\!+\!z_i^2\!=\!S^2$, and then, for every pair of NN sites ${\bf S}_i$ and ${\bf S}_j$, imposes the constraint that $x_i\!=\!\kappa x_j$ or $y_i\!=\!\kappa y_j$ or $z_i\!=\!\kappa z_j$, if the two sites share, respectively, an `x' or `y' or `z' type of bond, where $\kappa=-\sgn(K)$. For $K<0$, the resulting structure is shown below for the 4-site building block of the lattice  
\be\label{eqfig:KmodelClassGSs}
\includegraphics[width=1.75in]{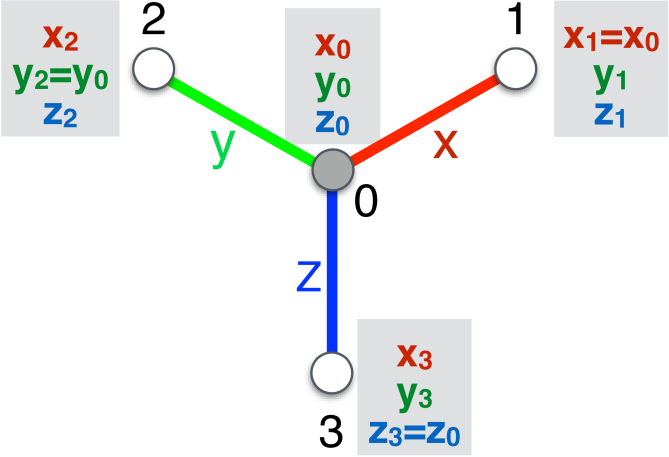}
\ee
The energy contribution from the three bonds emanating from each site $i$ add up to $-|K| (x_i^2\!+\!y_i^2\!+\!z_i^2)\!=\!-|K|S^2$. And since each bond is shared by two sites, these configurations saturate the lower bound $E_{\text{min}}/N\!=\!-|K| S^2/2$~\cite{Baskaran2008PRB}, and are therefore ground states.

The Cartesian states of BSS form the subset of ground states with only one non-vanishing Cartesian component. For the above cluster, a state of such a type is, e.g.,
\be\label{eqfig:KmodelCartesianGSs}
\includegraphics[width=1.75in]{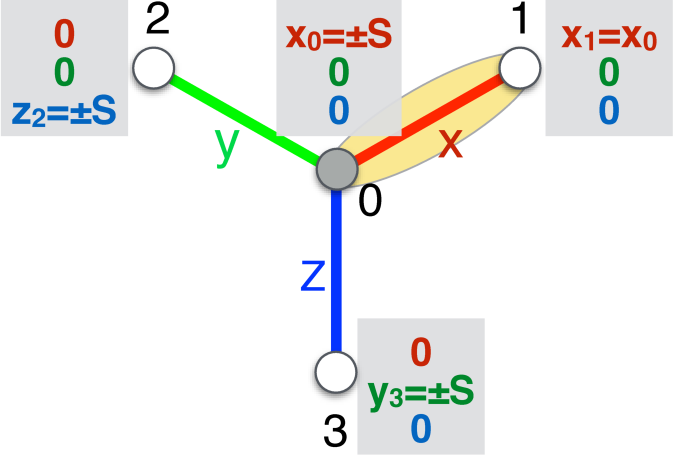}
\ee
where the yellow oval (or dimer) denotes the pair of spins that share the same non-vanishing Cartesian component, here the `x' component of ${\bf S}_0$ and ${\bf S}_1$. The spins ${\bf S}_2$ and ${\bf S}_3$ form similar dimers (not shown) with their neighbouring sites, here along the `z' and `y' type of bond, respectively. Hence, the Cartesian states of BSS map to dimer coverings of the lattice, modulo the sign of the nonvanishing spin components on each dimer. The latter introduces an Ising degree of freedom residing on each dimer.

\subsubsection{Semiclassical approach to the spin-$S$ Kitaev honeycomb model: `QSL out of disorder'}\label{sec:generalS2}

It is worth staying a little longer with the large-$S$ Kitaev model, to discuss the nature of the spin liquid phases [which arise from the presence of local symmetries of Eq.~(\ref{eq:WhhighS})], and their crossover to the more well-known spin-1/2 counterpart. 
To that end, we shall first briefly review the main insights that derive from semiclassical analysis~\cite{Baskaran2008PRB,Rousochatzakis2018NC}.

{\it Selection of Cartesian states.} 
The first effect of the semiclassical corrections is the lifting of the classical degeneracy and the selection of the submanifold of the above BSS Cartesian states, via the order by disorder mechanism. 
Referring back to the general parametrization of (\ref{eqfig:KmodelClassGSs}), the leading, short-wavelength corrections from second-order real space perturbation theory gives an anisotropy term~\cite{Rousochatzakis2018NC} 
\be
E_{\text{ani}}=-\frac{|K|S}{16} \sum\nolimits_i (\tilde{x}_i^4+\tilde{y}_i^4+\tilde{z}_i^4),
\ee
similar to the ones in Eqs.~(\ref{eq:EaniG}) and (\ref{eq:KGObDdE2}), where $\tilde{x}_i\!=\!x_i/S$, $\tilde{y}_i\!=\!y_i/S$, and $\tilde{z}_i\!=\!z_i/S$. 
This term selects the cubic directions, which corresponds to the $(1.662)^N$ different Cartesian states of BSS~\cite{Baskaran2008PRB}.

\begin{figure}[!t]
{\includegraphics[width=0.75\linewidth]{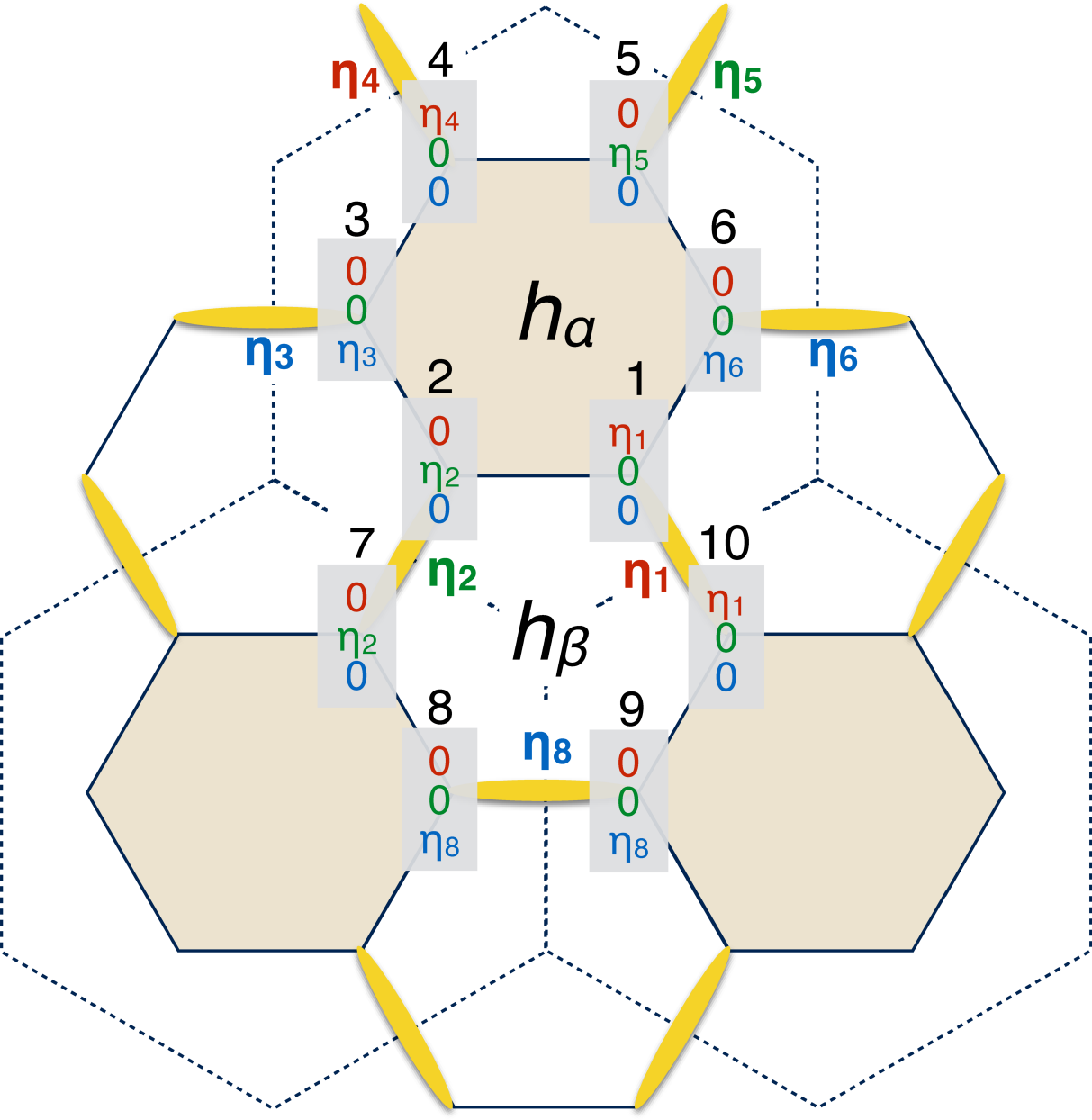}}
\caption{The star dimer pattern of Cartesian classical ground states selected from quadratic spin wave fluctuations~\cite{Baskaran2008PRB} for $S\!>\!3/2$ and $K\!<\!0$. The Ising-like $\eta$ variables reside on the  midpoints of the NN bonds of a honeycomb superlattice (dashed lines) and their dynamics is governed by a Toric code Hamiltonian. Figure adapted from Ref.~\cite{Rousochatzakis2018NC}. 
}\label{fig:KmodelstarPattern}
\end{figure}

{\it Dimer freezing.} As shown by BSS~\cite{Baskaran2008PRB}, the linear spin-wave Hamiltonian around any of the Cartesian states leads to non-interacting magnons propagating along loops that do not contain dimers, and the minimum zero-point energy arises by maximizing the number of the shortest such `empty' loops, which are the elementary hexagonal plaquettes. This leads to the selection of the `star' dimer pattern of Fig.~\ref{fig:KmodelstarPattern}, in which the shortest empty loops are the the shaded hexagons. In this pattern, the only dynamical degrees of freedom remaining are the Ising-like variables $\eta\!=\!\pm1$, which specify the direction of the two spins forming the dimer. 
Higher-order corrections give rise to significant quantitative changes but do not alter this picture qualitatively~\cite{Rousochatzakis2018NC}.  

The star dimer pattern breaks the translational symmetry as there are three ways to place this pattern in the lattice. Furthermore, each dimer has two configurations, so, at first sight, the number of selected spin states is $3\times 2^{N/2}$. BSS showed, however, that the minimum zero-point energy is associated with spin-wave modes that have antiperiodic boundary conditions around the empty hexagons, which reduces the number of states to $3\times2^{N/3}$~\cite{Baskaran2008PRB}. However, this is not the full story yet. As it turns out, the boundary condition on the spin-wave modes endows the selected manifold with a topological magnetic flux term~\cite{Rousochatzakis2018NC}.

{\it Effective model of $\eta$ variables.}  The $\eta$ variables reside on the midpoints of the bonds of a hexagonal superlattice, see dashed lines in Fig.~\ref{fig:KmodelstarPattern}.  To leading order, the effective model that describes the low-energy dynamics of the $\eta$ variables takes the form of a Toric code~\cite{Rousochatzakis2018NC} 
\be
\mc{H}_{\text{eff}}(\{\eta\}) = J_e \sum_v \eta_{v_1}^x\eta_{v_2}^x\eta_{v_3}^x+J_m\sum_p \eta_{p_1}^z\cdots \eta_{p_6}^z\,.
\ee
Here $v$ and $p$ label, respectively, the vertices and plaquettes of the honeycomb superlattice, the indices $v_1-v_3$ are the three bond variables emanating form $v$, and the indices $p_1-p_6$ are the six bond variables around $p$. The first term of this Hamiltonian comes from the quantum-mechanical tunneling processes that connect one configuration of $\eta$ variables to another, and is known as the electric charge term of the Toric code 
\cite{Kitaev2003,Kitaev2006,Savary2016}. The second term is the magnetic flux term of the Toric code and comes from the topological terms mentioned above, related to the special anti-periodic boundary conditions of the magnons.

The above Toric code Hamiltonian is an example of a $Z_2$ gauge theory, which hosts quantum spin liquid ground states with topological degeneracy ($2^{2g-1}$, where $g$ is the genus of the system), and fractionalized excitations \cite{Kitaev2003,Kitaev2006,Savary2016}. 
So, the spin-$S$ version of the Kitaev model is  a  topological $Z_2$ QSL even in the semiclassical  limit~\cite{Rousochatzakis2018NC}. This $Z_2$ gauge structure descends from the gauge structure of the original spin-$S$ Kitaev model, identified by BSS~\cite{Baskaran2008PRB}.

{\it Breakdown of semiclassical picture.}  
We now turn our discussion to what can go wrong with the above semiclassical picture as we lower $S$. The dimer freezing in the star pattern of Fig.~\ref{fig:KmodelstarPattern} stems from the zero-point energy of spin waves. However, this analysis disregards the quantum tunneling between different dimer patterns. A shown in Ref.~\cite{Rousochatzakis2018NC}, the leading such process is the one that shifts dimers around elementary hexagon plaquettes, and  the relevant tunneling amplitude $t_d$  can be estimated by
\be\label{eq:td}
|t_d| / |K|= 3 S^2 2^{-6S} / (1-2^{-12S}).
\ee
At large $S$, $t_d$ is extremely small, and the dimer pattern of Fig.~\ref{fig:KmodelstarPattern} is stable. However, $t_d$ becomes relevant below $S\!\sim\!3/2$. 
It follows that in order to understand the physics of the $S\!=\!3/2$ and $S\!=\!1$ cases, we need to allow both the position of the dimers and the orientation of the $\eta$ variables that reside on top of these dimers to resonate. 
Such a `decorated quantum dimer' picture remains to be confirmed by numerical studies, and is a very active line of research.

{\it Numerical and other studies.}
For the spin-1 Kitaev model, an iDMRG study has shown that the system is a gapless QSL, and that the position and number of gapless modes are quite different from the corresponding spin-1/2 case~\cite{Dong2020PRB}. By contrast, a tensor network analysis suggest that the ground state of the spin-1 Kitaev model is a gapped quantum spin liquid with $Z_2$ gauge structure and Abelian quasiparticles~\cite{Lee2020PRR}.

The spin-3/2 Kitaev honeycomb model is unique among the spin-$S$ Kitaev models due to a massive ground state quasi-degeneracy that has hampered numerical and analytical studies~\cite{Jin2022NC,Natori2023}. In a recent work~\cite{Jin2022NC}, the authors have used an SO(6) Majorana parton mean-field theory  and showed that the anomalous features of S-3/2 Kitaev QSL can be understood in terms of
an emergent low-energy Majorana flat band. Away from the isotropic limit, when  the $S\!=\!3/2$ KSL is supplemented with single-ion anistoropy term, it
generally displays a quadrupolar order with gapped or gapless Majorana excitations, features that
were quantitatively confirmed by DMRG simulations \cite{Jin2022NC}.

\subsection{The $\Gamma$ points}\label{sec:Gpoints} 
 The off-diagonal exchange coupling $\Gamma$ has emerged in recent years as another way to drive Kitaev materials to a strongly correlated behaviour, which is in fact universal across the 2D and 3D geometries of Fig.~\ref{fig:TriCoord}~\cite{Rousochatzakis2017PRL}. 
The pure, threefold symmetric $\Gamma$ model on the honeycomb lattice is described by the Hamiltonian
\be 
\mc{H}_\Gamma=\sum\nolimits_\gamma \sum\nolimits_{\langle ij \rangle\in\gamma}
\Gamma_\gamma (S_i^{\alpha_\gamma}S_j^{\beta_\gamma}+S_i^{\beta_\gamma} S_j^{\alpha_\gamma}),
\ee 
where, as previously,  $(\alpha_\gamma,\beta_\gamma)=(y,z)$, $(z,x)$ and $(x,y)$ for $\gamma=x$, $y$ and $z$, respectively, and $\Gamma_\gamma\!=\!\Gamma$ has a uniform sign across all bonds, using the cubic axes frame of Fig.~\ref{fig:TriCoord}\,(a).

Since the quantum  behavior of this model is less understood, it is instructive to first discuss the classical  limit of the model and then move on to discuss the leading effect of semiclassical corrections, before we review the numerical results from the quantum spin-1/2 limit.

\subsubsection{Classical limit: The `$\Gamma$ recipe'}\label{sec:GammaClass}
Just like the classical Kitaev model~\cite{Baskaran2007PRL,Baskaran2008PRB,Chandra2010,Rousochatzakis2018NC}, the classical $\Gamma$-model has a classical spin liquid ground state, characterized by an infinite number of classical ground states~\cite{Rousochatzakis2017PRL,Samarakoon2018PRB,Chern2019PRL}. 
The general structure of these states can be seen in the parametrization proposed in Ref.~\cite{Rousochatzakis2017PRL}. 
One again writes the Cartesian components of each spin as $\vec{S}_i\!=\![x_i,y_i,z_i]$ with $x_i^2\!+\!y_i^2\!+\!z_i^2\!=\!S^2$, and then: i) for every pair of NN sites ${\bf S}_i$ and ${\bf S}_j$, one imposes the constraint that $(y_j,z_j)=\zeta (z_i,y_i)$ or $(x_j,z_j)=\zeta(z_i,x_i)$ or $(x_j,y_j)=\zeta(y_i,x_i)$, if the two sites share, respectively, an `x' or `y' or `z' type of bond, and where $\zeta=-\sgn(\Gamma)$. Since this amounts to swapping two out of three components (and multiplying by $\zeta$), the third components must have the same magnitude, in order to satisfy the spin length constraint on both sites.
This `$\Gamma$ recipe' is illustrated below for the 4-site building block of the lattice for $\Gamma<0$ ($\zeta\!=\!1$): 
\be\label{eqfig:GmodelClassGSs}
\includegraphics[width=1.75in]{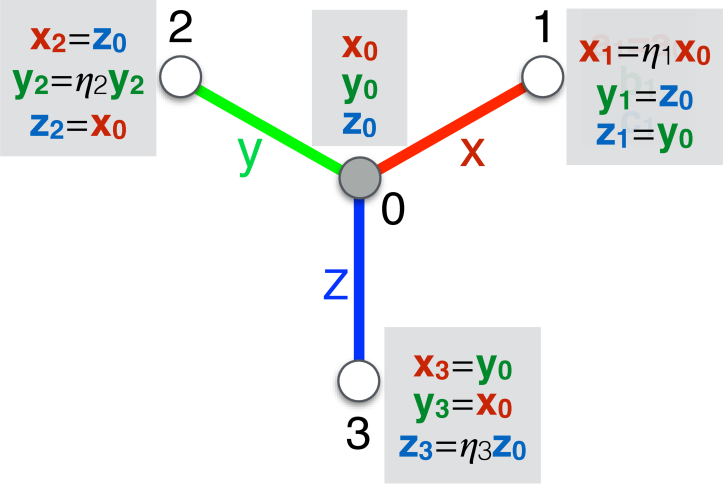}
\ee
where $\eta_{1-3}\!=\!\pm1$ are Ising degrees of freedom.
The energy contribution from the three bonds emanating from each site $i$ add up to $-2|\Gamma| (x_i^2\!+\!y_i^2\!+\!z_i^2)\!=\!-2|\Gamma|S^2$. And since each bond is shared by two sites, these configurations saturate the lower bound $E_{\text{min}}/N\!=\!-|\Gamma| S^2$~\cite{Rousochatzakis2017PRL}, and are therefore ground states.

To get the global lattice structure of the ground states, we take a reference site, say $i\!=\!0$, and re-write $(x_0,y_0,z_0)\!=\!({\cred\eta_1a}, {\cgr \eta_2b}, {\cbl \eta_3c})$, where $({\cred a},{\cgr b},{\cbl c}) \!\equiv\!(|x_0|,|y_0|,|z_0|)$, and then proceed to apply the `$\Gamma$ recipe' to the first neighbours of $i=0$, and then to its second neighbours, etc, until we cover the whole lattice. This recipe leads to the following structure, 
\be\label{eqfig:GmodelClassGSs2}
\includegraphics[width=2.75in]{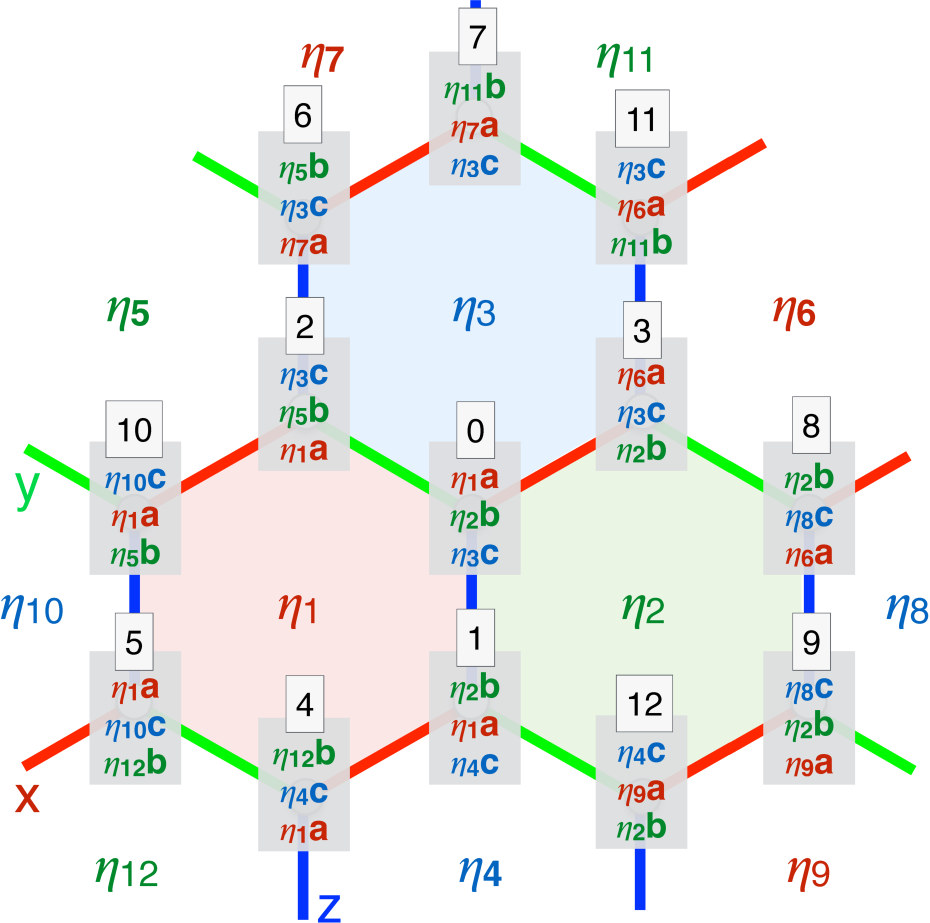}
\ee
which can be described in terms of the three positive numbers $({\cred a},{\cgr b},{\cbl c}) $, which satisfy ${\cred a}^2+{\cgr b}^2+{\cbl c}^2\!=\!S^2$,  plus a set of $N/2$ degrees of freedom, $\eta_i\!=\!\pm1$, 
residing on the hexagons of the lattice. 
Each such Ising degree of freedom appears only on the Cartesian components of the spins of the given hexagon, and nowhere else. For example, the variable $\eta_1$ of (\ref{eqfig:GmodelClassGSs2}) appears only on the $z$, $x$, $y$, $z$, $x$, $y$ components of the sites labelled by 4, 5, 10, 2, 0, and 1, respectively. 
Conversely, the three Ising variables that carry the sign of the three Cartesian components of each site, `migrate' on the three hexagons sharing the spin. 

Another aspect of the $\eta$ variables is that they split into three types that occupy the vertices of three inter-penetrating triangular sublattices, denoted by `{\cred R}' (shaded in red), `{\cgr G}' (green) and `{\cbl B}' (blue) in (\ref{eqfig:T6transfhex}). This structure is reflected directly in the classical value of Kitaev's fluxes $W_h$, since (for general $\zeta$)
\be\label{eq:GstatesWh}
W_{h\in R}\!=\!\zeta \tilde{{\cred a}}^6,~
W_{h\in G}\!=\!\zeta \tilde{{\cgr b}}^6,~
W_{h\in B}\!=\!\zeta \tilde{{\cbl c}}^6,~
\ee
where
$\tilde{{\cred a}}={\cred a}/S$, $\tilde{{\cgr b}}={\cgr b}/S$ and $\tilde{{\cbl c}}={\cbl c}/S$.

According to the above, the classical ground state manifold has a discrete degeneracy of $2^{N/2}$ different states, on top of the degeneracy associated with the choice of $({\cred a},{\cgr b},{\cbl c}) $. 
The local zero-energy modes responsible for this degeneracy correspond to flipping one particular component for each of the six spins of a hexagon. Taking again the ${\cred \eta_1}$ hexagon of (\ref{eqfig:GmodelClassGSs2}) as an example, the zero mode amounts to flipping the signs of $S_4^z$, $S_5^x$, $S_{10}^y$, $S_2^z$, $S_0^x$ and $S_1^y$. 
These zero modes have also been confirmed numerically by Landau-Lifshitz dynamics simulations~\cite{Samarakoon2018PRB}.

Importantly, the above operations of flipping individual $\eta$ variables are precisely the local symmetries $\Theta(h)\!\cdot\!\mc{R}(h)_{x,y,z}$ of  Sec.~\ref{sec:Gammaclassicalsymmetry}. Therefore, the discrete degeneracy associated with the $\eta$'s is not accidental but symmetry related (this is different for quantum spins, see below).

\subsubsection{Effect of thermal fluctuations:\\ Classical spin liquid \& finite-$T$ transition}\label{sec:Gpoints-TObD}

According to Elitzur's theorem~\cite{Elitzur1975}, the local symmetries $\Theta(h)\cdot\mc{R}(h)_{x,y,z}$ cannot break spontaneously at any finite temperature. In other words, the $\eta$ variables are expected to fluctuate and average to zero at all $T$.
At low enough temperatures ($T\lesssim\Gamma$), the  system is then expected to crossover into a `classical spin liquid' regime, with short-range spin-spin correlations, between spins belonging to the same hexagon only, and with a distinctive anisotropic pattern, reflecting the above `locking' of the Cartesian components $z$, $x$, $y$, $z$, $x$, $y$ of the six spins around the hexagon.

As it turns out, the physics of the classical $\Gamma$ model becomes more interesting at even lower temperatures: 
Unlike the degeneracy associated with the $\eta$ variables, the degeneracy associated with the choice of $({\cred a},{\cgr b},{\cbl c}) $ is accidental,  and, as such, it can be lifted by thermal order-by-disorder. Indeed, as shown in Ref.~\cite{Chern2019PRL}, the classical spin liquid regime includes a thermal phase transition at $T_c\simeq 0.04\,\Gamma$, associated with the spontaneous trimerization of the lattice, which, in the above parametrization, amounts to selecting the so-called `cubic' (or `Cartesian') states with $({\cred a},{\cgr b},{\cbl c}) \!=\!(S,0,0)$, or $(0,S,0)$ or $(0,0,S)$. According to Eq.~(\ref{eq:GstatesWh}), such a trimerization gives $W_{h}\!=\!\zeta$ on 1/3 of the hexagons ({\cred R} or {\cgr G} or {\cbl B}, respectively) and $W_h\!=\!0$ on the remaining 2/3 of the hexagons. 
So, the low-$T$ phase shows long-range order in the fluxes $W_h$ (with the $\eta$ variables still fluctuating), and has thus been termed `hidden plaquette order'~\cite{Chern2019PRL}.

It is worth noting that the above transition breaks a {\it discrete global} symmetry (the translations) and is therefore not in conflict with Elitzur's theorem. 
Such a transition is analogous to the thermal transitions proposed by Chandra, Coleman, and Larkin~\cite{CCL1990} in two-dimensional systems where a spontaneous symmetry breaking of a {\it continuous global}  symmetry are prohibited  due to Hohenberg and Mermin-Wagner theorem~\cite{Hohenberg1967,MerminWagner1966}, see also Ref.~\cite{Cedric2003}.

\subsubsection{Effects of quantum fluctuations from semiclassics:\\
magnetic order vs algebraic liquid}\label{sec:Gpoints-QObD}

We now turn to the quantum spin-1/2 case, and discuss a semiclassical approach~\cite{Rousochatzakis2017PRL} which addresses the role of quantum fluctuations within the classical ground state manifold. As such, this approach is variational in nature.

We begin by recalling the important fact that the local symmetries $\Theta(h)\!\cdot\!\mc{R}(h)_{x,y,z}$ do not exist for quantum spins (see  Sec.~\ref{sec:Gammaclassicalsymmetry}). This means that the $2^{N/2}$ degeneracy associated with the $\eta$ variables is actually accidental, and can be 
lifted too, like the one associated with the choice of $({\cred a},{\cgr b},{\cbl c}) $. Hence, magnetic long-range order is possible~\cite{Rousochatzakis2017PRL}, in stark contrast to the classical version of the model discussed above.  

The lifting of the classical degeneracy can be captured by an effective model, that gives the quantum energy corrections in terms of $({\cred a},{\cgr b},{\cbl c})$ and the $\eta$ variables~\cite{Rousochatzakis2017PRL}.  
The general form of this effective model can be guessed using the symmetries of the quantum model, which include the $D_{3d}$ group discussed in Sec.~\ref{sec:EffSuperHam}, the time-reversal operation, as well as the discrete operations $\mc{R}_{x,y,z}$ of Eq.~(\ref{eq:RxyzFull}). The $D_{3d}$ group includes the threefold rotation  around sites, in combined spin-orbit space. This maps $({\cred a},{\cgr b},{\cbl c})\to({\cgr b},{\cbl c},{\cred a})$ in spin space, and one $\eta$ sublattice (colour) to another in real space. The symmetry group also includes reflections that change the signs of two of $({\cred a},{\cgr b},{\cbl c})$. 
On the other hand, the operations $\mc{R}_{x,y,z}$ effectively flip the signs of all $\eta$'s of a given type.
These symmetries necessitate the following:

i) The leading symmetry-allowed anisotropy term that lifts the degeneracy associated with $({\cred a},{\cgr b},{\cbl c})$ must be of the type
\be\label{eq:EaniG}
E_{\text{ani}}=A^{(4)}_{\text{ani}}~({\cred a}^4+{\cgr b}^4+{\cbl c}^4)\,,
\ee
since the combination ${\cred a}^2+{\cgr b}^2+{\cbl c}^2=S^2$ is a constant. 

ii) Interactions that contain an odd number of $\eta$'s of the same type are excluded from the effective model. So, the leading interactions are bilinear terms of the Ising type $\eta_i\eta_j$, where both $\eta_i$ and $\eta_j$ belong to the same type. And, by virtue of the threefold rotation symmetry, we expect three Ising models, one for each $\eta$ sublattice. 
The leading effective terms that lift the accidental degeneracy associated with the $\eta$ variables must then take the form of three decoupled triangular Ising models
\be\label{eq:deltaEeta}
\delta E(\{\eta\})\!=\!
J_{\cred R}\!\!\!\sum_{\langle ij\rangle\in {\cred R}}\!\!{\cred \eta_i\eta_j}
+J_{\cgr G}\!\!\!\sum_{\langle ij\rangle\in {\cgr G}}\!\!{\cgr \eta_i\eta_j}
+J_{\cbl B}\!\!\!\sum_{\langle ij\rangle\in {\cbl B}}\!\!{\cbl \eta_i\eta_j}.
\ee
The leading corrections from the so-called real space perturbation theory (a type of short-wavelength expansion that captures spin-wave corrections at a local level~\cite{Lindgard1988,Long1989,Heinila1993,Chernyshev2014}) confirms this general picture, and gives, to leading order~\cite{Rousochatzakis2017PRL}:
\be
\renewcommand{\arraystretch}{1.5}
\begin{array}{c}
A^{(4)}_{\text{ani}}/N = -|\Gamma|S/32 < 0 \\
J_{\cred R}=\Gamma S \tilde{\cred a}^2/8,~
J_{\cgr G}=\Gamma S \tilde{\cgr b}^2/8,~
J_{\cbl B}=\Gamma S \tilde{\cbl c}^2/8\,.
\end{array}
\ee
These expressions give the following insights:

i) The fourth-order cubic anisotropy $E_{\text{ani}}$ is minimized when $\{{\cred\tilde{a}}, {\cgr\tilde{b}}, {\cbl\tilde{c}}\}\!=\!\{1,0,0\}$, $\{0,1,0\}$ or $\{0,0,1\}$. Then, 2/3 of the $\eta$'s become idle, and only the behavior of the remaining 1/3 stay dynamical. This amounts to a spontaneous trimerization of the lattice, with broken translations and threefold rotations around sites (but not around hexagon centers).

ii) For $\Gamma\!<\!0$, we get a FM Ising model on one of the three $\eta$ sublattices, below a scale set by $A_{\text{ani}}^{(4)}$. In terms of spins, this is a three-sublattice non-coplanar state, with spins pointing along the cubic axes, and a nonvanishing total moment along one of the $\langle111\rangle$ axes. 

iii) For $\Gamma\!>\!0$, we end up with an AFM Ising model on the triangular lattice, which is the prototype of classical spin liquids~\cite{Wannier1950,Houtappel1950}. The system then remains highly frustrated even well below the scale set by $A_{\text{ani}}^{(4)}$ and $J_{R,G,B}$, and shows algrebraic spin liquid behaviour down to very low $T$.

Higher order corrections analyzed in Ref.~\cite{Rousochatzakis2017PRL}, include sixth order anisotropies such as ${\cred a}^2{\cgr b}^2{\cbl c}^2$, a coupling between different $\eta$ sublattices, of the type $({\cbl \eta_3\eta_4})({\cred \eta_1\eta_9})$, as well as kinetic (off-diagonal) terms that map one $\{\eta_i\}$ configuration to another. The proper description of such tunneling terms requires promoting the $\eta_i$ variables  to spin-1/2 operators $\eta_i^z$, and, to leading order, the tunneling terms take the form of transverse couplings of the type $J_x{\cred \eta_1^x\eta_9^x}$ and $J_y{\cred \eta_1^y\eta_9^y}$, which together with the dominant terms of Eq.~(\ref{eq:deltaEeta}) lead to an XYZ model.
While the detailed role of these tunneling events is out of the scope of this review~\cite{quantum2023}, the key point is that the tunneling amplitudes $J_{x,y}$ are extremely small, because flipping two $\eta$ variables amounts to flipping a Cartesian component for {\it twelve} spins. 

Altogether then, the variational semiclassical approach predicts: i) the emergence of $\eta$ degrees of freedom that live on the hexagon plaquettes, ii) a spontaneous trimerization of the lattice, where one of the three triangular superlattices of $\eta$'s is selected, iii) for negative $\Gamma$, these $\eta$'s order magnetically below an energy scale set by $\Gamma$, whereas iv) for positive $\Gamma$, the $\eta$ variables show a classical spin liquid behaviour with algebraic correlations down to extremely low temperatures $k_BT\ll\Gamma$.

\subsubsection{Quantum $S=1/2$ model: Thermodynamics and dynamical spin correlations} 

While the exact nature of the ground state of the pure, spin-1/2 AFM $\Gamma$-model is not yet fully understood (a review of the main relevant numerical results will be provided in Sec.~\ref{sec:KGquantum} below, as part of the $K$-$\Gamma$ phase diagram), here we review some of the most interesting thermodynamic and dynamical spin properties of the spin-1/2 $\Gamma$ model.

As shown in Ref.~\cite{Catuneanu2018npj} (see also discussion in \cite{Samarakoon2018PRB}), the specific heat of the spin-1/2 $\Gamma$ model shows a two-peak structure, with the two temperature scales given by the peaks being $T_L\simeq 0.03 \,\Gamma$ and $T_H\simeq 0.4\,\Gamma$. 
This two-peak structure is similar to what has been found in the  Kitaev model~\cite{Nasu2015PRB} (and in its proximity~\cite{Yamaji2016}), although the balance of the entropy released at the two peaks, and its overall $T$ dependence differs from that of the Kitaev model.

Results for the dynamical spin structure factors of the quantum model~\cite{Samarakoon2018PRB} show a characteristic classical to quantum crossover as we lower the temperature. Specifically, at high temperatures the correlations show distinct signatures of the zero mode structure of the degenerate manifold of classical ground states of Sec.~\ref{sec:GammaClass},which persist even down to $T_L<T<T_H$. 
This structure gradually crosses over to the quantum regime for $T<T_L$, in which the dynamical spin correlations show features that are not present in the classical model.

A separate ED study~\cite{Rousochatzakis2020KITP,quantum2023} shows evidence that the ground state short-range spin-spin correlation pattern of the spin-1/2 $\Gamma$ model is in fact consistent with the distinctive anisotropic correlations of the $\eta$ variables of  Sec.~\ref{sec:GammaClass}. These compound degrees of freedom then seem to survive in the quantum ground state (see further discussion in Sec.~\ref{sec:KGquantum}).

\subsection{The $K$-$\Gamma$ model}\label{sec:KGmodel}
Next and, perhaps, the most interesting regions of the parameter space are the $K$-$\Gamma$ lines, which connect the two types of strongly correlated regimes discussed so far, the Kitaev quantum spin liquid and the $\Gamma$ classical spin liquid.
As mentioned above, the $K$-$\Gamma$ model seems to play a central role in several compounds~\cite{Rousochatzakis2017PRL,Samarakoon2018PRB,Chern2019PRL,Gordon2019NC,Gohlke2020PRR,Wachtel2019PRB,Chern2020PRR,Yamada2020PRB,Yang2020PRL,Liu2021PRR,Luo2021PRB,Sorensen2021PRX,Buessen2021PRB}, and although it is deceptively simple, it is not yet fully understood.

It is instructive to discuss the classical limit first, before we present what is known for the quantum spin-1/2 version of the model. In doing this, we wish to highlight one of the key ramifications of the interplay between $K$ and $\Gamma$ interactions: {\it The emergence, in the classical limit, of `period-3 chain' orders, or long-wavelength modulations thereof, with two counter-rotating spin sublattices}. It is noteworthy that this characteristic type of ordering can manifest irrespective of the relative signs of $K$ and $\Gamma$, although their stability can be proven analytically only when the two couplings have the same sign, as we discuss below. Moreover, these orders appear to be universal in all 2D and 3D geometries (see Sec.~\ref{sec:3D}), and have in fact been observed in the three available {\liiro} polymorphs~\cite{Biffin2014PRB,Biffin2014PRL,Williams2016PRB,Tsirlin2022}.

Classically, the points $(K,\Gamma)$ and $(-K,-\Gamma)$ map to each other under the sublattice rotation of Sec.~\ref{sec:ClassicalDuality}. This leads to two, qualitatively different regions in parameter space, the ones where $K$ and $\Gamma$ have the same sign, and the ones where they have opposite signs. The ground states of the former regions are known exactly, whereas the ones of the latter are not fully understood~\cite{Rau2014,Rousochatzakis2020KITP,Rayyan2021,Liu2021PRR,li2022tangle,Chen2023NJP,classical2023}.

\subsubsection{Semiclassical phase diagram (I): $K$, $\Gamma$ of the same sign. \\
The $K\Gamma$ recipe}\label{sec:KGclassI}  

When $K$ and $\Gamma$ are negative, we get a family of six-sublattice ground states, which map to a collinear N\'eel AFM under the duality $\mc{T}_6$ of (\ref{eq:T6transfhex}). This dual N\'eel phase includes the hidden SU(2) point $K\!=\!\Gamma$ (with $K\!<\!0$) of Sec.~\ref{Sec:HiddenSU2KG}. 
When $K$ and $\Gamma$ are positive, we get a dual FM phase, by virtue of the sublattice rotation of Sec.~\ref{sec:ClassicalDuality}.
The two dual phases feature a number of interesting aspects, most notably the period-3 structure mentioned above,  with counter-rotating sublattices.

{\it The `$K\Gamma$ recipe'.}  We will now analyze these features and reveal their physical origin, by combining the `$K$ recipe' of  (\ref{eqfig:KmodelClassGSs}) and the `$\Gamma$ recipe' of (\ref{eqfig:GmodelClassGSs}), which are,  compatible when $K$ and $\Gamma$ have the same sign. 
To proceed, one writes again the Cartesian components of each spin as $\vec{S}_i\!=\![x_i,y_i,z_i]$ with $x_i^2\!+\!y_i^2\!+\!z_i^2\!=\!S^2$, and then, for every pair of NN sites ${\bf S}_i$ and ${\bf S}_j$, imposes the constraint that $[x_j,y_j,z_j]=\kappa [x_i,z_i,y_i]$ or $\kappa[z_i,y_i,x_i]$ or $\kappa[y_i,x_i,z_i]$, if the two sites share, respectively, an `x' or `y' or `z' type of bond, and where $\kappa\!=\!-\sgn(K)$. For $\kappa\!=\!1$, for example, this gives, for the 4-site building block of the lattice:
\be\label{fig:KnegGnegmodelClassGSs}
\includegraphics[width=1.75in]{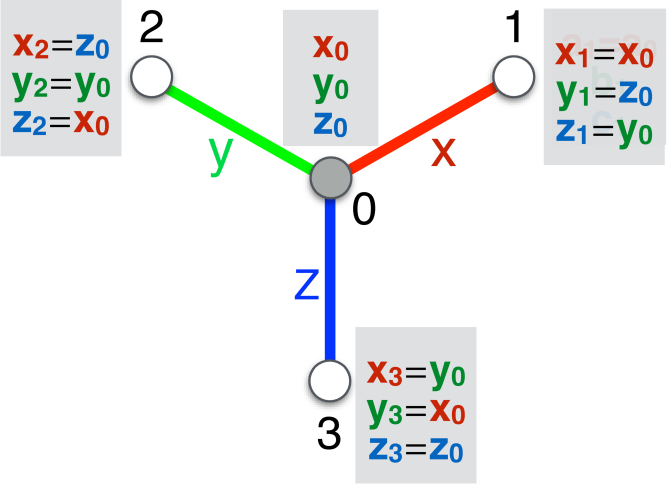}
\ee
The energy contributions from the three bonds emanating from each site $i$ add up to $-(|K|+2|\Gamma|) (x_i^2\!+\!y_i^2\!+\!z_i^2)\!=\!-(|K|+2|\Gamma|)S^2$. Since each bond is shared by two sites, these configurations saturate the energy lower bound $E_{\text{min}}/N\!=\!-(|K|/2+|\Gamma|) S^2$, and are, therefore, ground states.

{\it The dual AFM phase.} When $K$ and $\Gamma$ are both negative, the ground states take the general form
\be\label{eqfig:KGClassGSsSameSign}
\includegraphics[width=0.95\columnwidth]{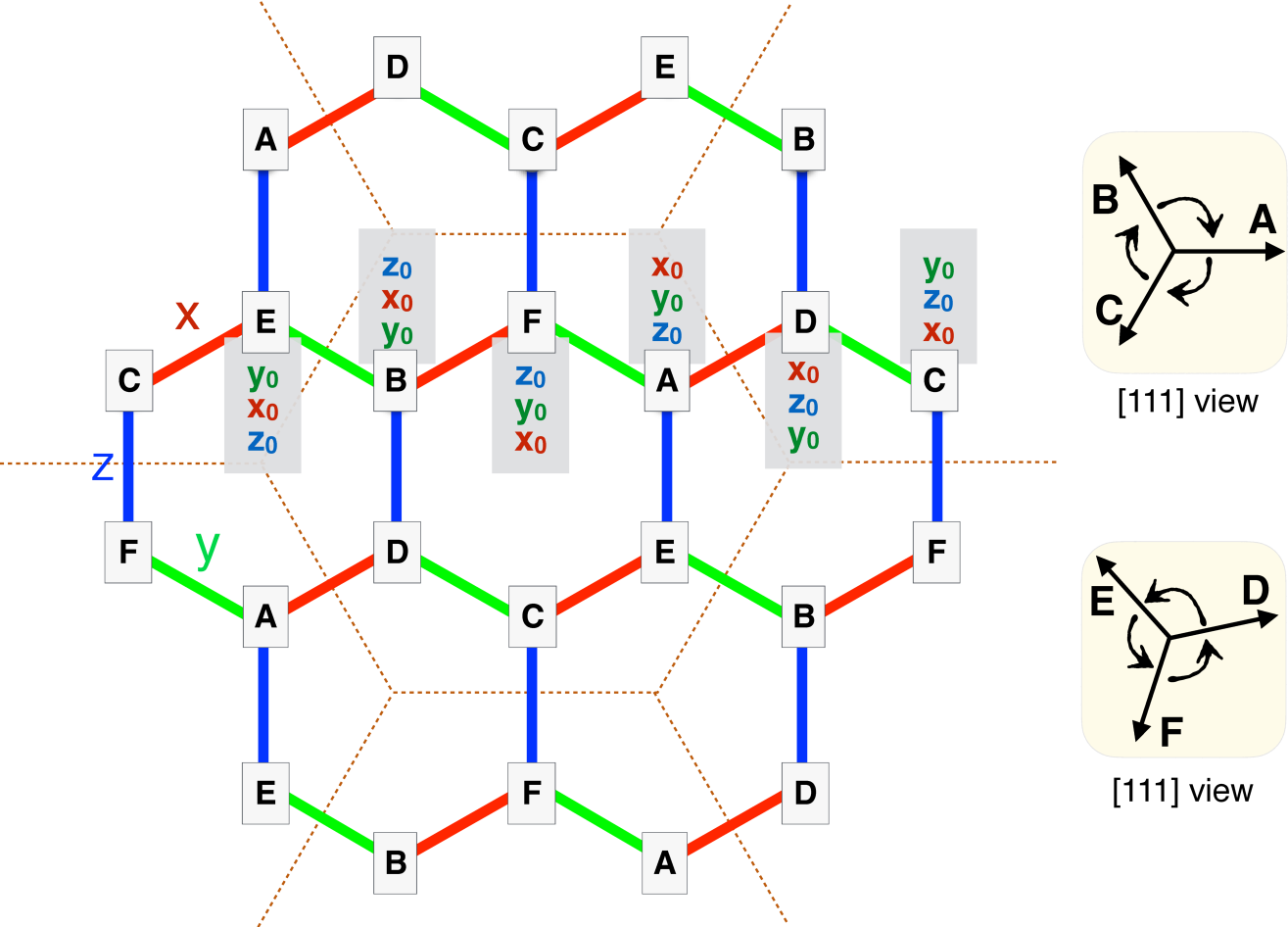}
\ee
and have the following features, for general $(x_0,y_0,z_0)$: 

i) the states break translational symmetry, with the underlying superlattice shown by dashed lines in (\ref{eqfig:KGClassGSsSameSign}).

ii) the states form a two-parameter manifold, characterized by the Cartesian components $[x_0,y_0,z_0]$ of the reference site. 

iii) there are six spin-sublattices, denoted by A, B, C, D, E and F in (\ref{eqfig:KGClassGSsSameSign}), which point along the directions 
\be\label{eq:ABCDEF}
\renewcommand{\arraystretch}{1.5}
\begin{array}{lll}
{\bf A}\!=\![x_0,y_0,z_0], 
&{\bf B}\!=\![z_0,x_0,y_0], 
&{\bf C}\!=\![y_0,z_0,x_0],
\\
{\bf D}\!=\![x_0,z_0,y_0], 
& {\bf E}\!=\![y_0,x_0,z_0], 
&{\bf F}\!=\![z_0,y_0,x_0]\,.
\end{array}
\ee

iv) The directions ${\bf A}$, ${\bf B}$ and ${\bf C}$ 
are related to each other by three-fold rotations around the [111] axis, and similarly for the directions ${\bf D}$, ${\bf E}$ and ${\bf F}$. 
As we travel along a specific chain of the lattice, spins on the first sublattice of the honeycomb rotate  from ${\bf B}\to{\bf A}\to{\bf C}\to{\bf B}\to\cdots$, which amounts to successive {\it clockwise} 120$^\circ$ rotations around [111] in spin space. By contrast, the spins on the second sublattice rotate from  ${\bf D}\to{\bf E}\to{\bf F}\to{\bf D}\to\cdots$, which amounts to successive {\it counter-clockwise} 120$^\circ$ rotations around [111],  see right panels in (\ref{eqfig:KGClassGSsSameSign}). So, for general $[x_0,y_0,z_0]$, the classical ground states feature a period-3 modulation with two counter-rotating sublattices, as mentioned above. This is one of the key ramifications of the interplay of $K$ and $\Gamma$, and here it  arises simply from the requirement to saturate the energy contributions from both couplings along all bonds (the `$K\Gamma$ recipe'). For the honeycomb lattice, this requirement can be fulfilled when $K$ and $\Gamma$ have the same sign. This result has broader generality, as we shall see in Secs.~\ref{sec:3D} and \ref{sec:1D}.

vi) In the local frames defined by the six-sublattice transformation $\mc{T}_6$ of (\ref{eq:T6transfhex}), the above states map to the (much simpler) collinear N\'eel states, with moments along $\widetilde{S}_i=\pm [x_0,y_0,z_0]$. So, for $K$ and $\Gamma$ both negative, we end up with a dual N\'eel AFM, which includes the hidden SU(2) point $K\!=\!\Gamma$ (with $\widetilde{J}\!=\!-K$), discussed in Sec.~\ref{Sec:HiddenSU2KG}. The two-parameter degeneracy associated with the direction of $[x_0,y_0,z_0]$ is symmetry related only at $K\!=\!\Gamma$, but it is accidental everywhere else inside this dual phase.

{\it The dual FM phase.} 
A similar analysis can be carried out for positive $K$ and $\Gamma$, in which case the classical ground states map to a fully polarized state in the frame defined by $\mc{T}_6$.

{\it Semiclassical corrections.} In the two dual phases discussed above, the choice of the direction of the reference spin $[x_0,y_0,z_0]$ is completely arbitrary. This degeneracy is accidental everywhere except at the two special hidden SU(2) points, where it is symmetry related.
Away from these points, quantum corrections are expected to lift the accidental degeneracy and select a discrete set of `easy axes', that are related to each other by the true, discrete symmetry of the model. 
As it turns out, the leading corrections from real space perturbation theory take the form of a fourth-order anisotropy~\cite{Rousochatzakis2020KITP} 
\be\label{eq:KGObDdE2}
\delta E^{(2)}/N = -\frac{(\Gamma-K)^2S}{32 |\Gamma+2K|}(\tilde{x}_0^4+\tilde{y}_0^4+\tilde{z}_0^4)
\ee
where $\tilde{x}_0\!=\!x_0/S$, $\tilde{y}_0\!=\!y_0/S$ and $\tilde{z}_0\!=\!z_0/S$. This anisotropy selects the cubic directions $[\tilde{x}_0,\tilde{y}_0,\tilde{z}_0]=\pm[1,0,0]$, $\pm[0,1,0]$, or $\pm[0,0,1]$. And, as expected, this anisotropy vanishes by symmetry at the hidden SU(2) points $K\!=\!\Gamma$, where all directions are equivalent.

\begin{figure*}[t]
\centering
\includegraphics[width=0.97\textwidth]{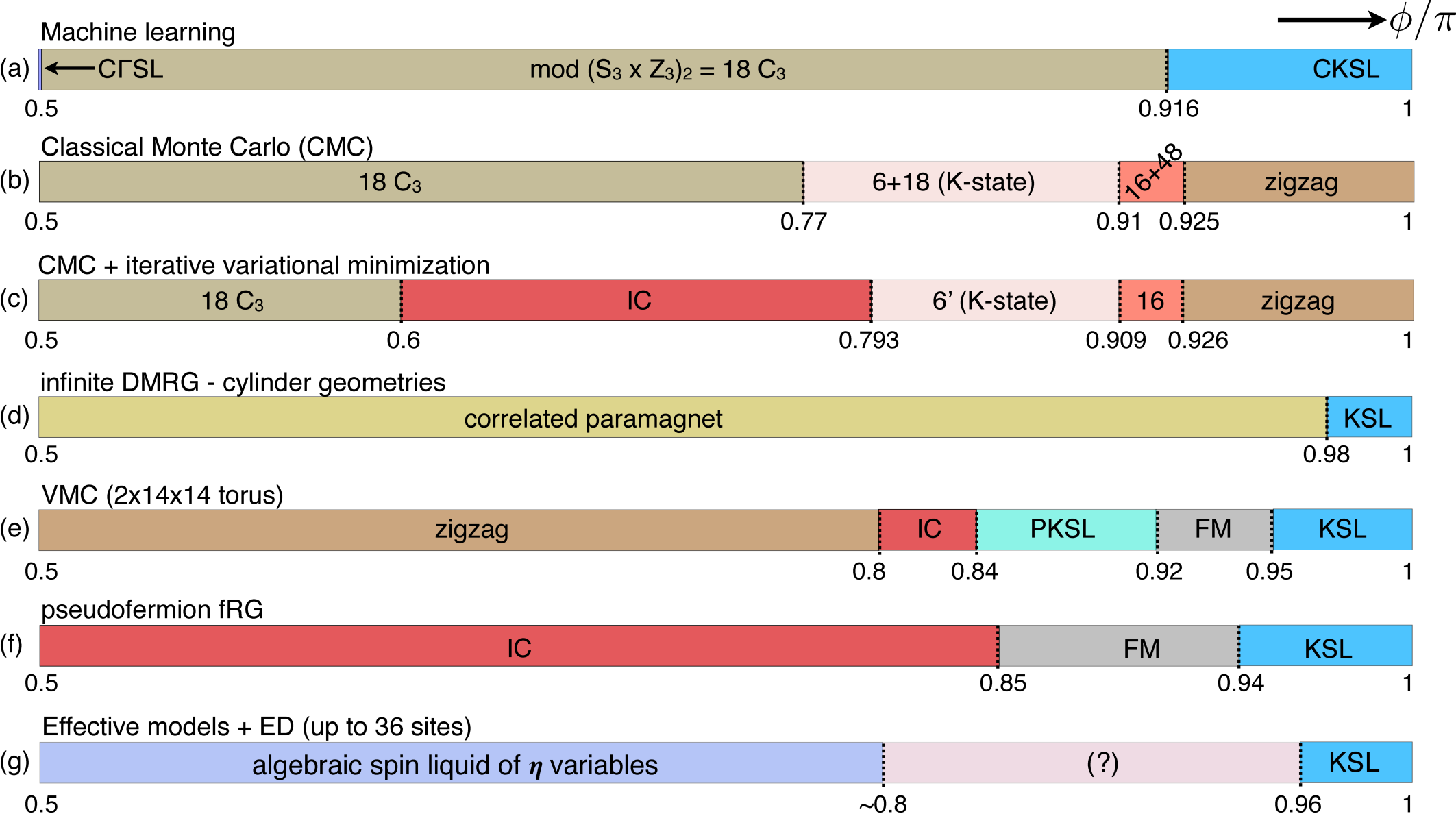}
\caption{Classical (a-c) and quantum spin-1/2 (d-g) phase diagrams of the $K$-$\Gamma$ model inside the frustrated region $\phi\in[\pi/2,\pi]$, where $K\!=\!\cos\phi$ and $\Gamma\!=\!\sin \phi$, as obtained by: (a) machine learning~\cite{Liu2021PRR}, (b) Classical Monte Carlo simulations~\cite{Rayyan2021,Chen2023NJP}, (c) hybrid Classical Monte Carlo and iterative  variational minimizations~\cite{classical2023}, (d) infinite DMRG~\cite{Gohlke2018PRBa,Gohlke2020PRR}, (e) Variational Monte Carlo (VMC)~\cite{Normand2019}, (f) pseudofermion functional renormalization group (fRG)~\cite{Buessen2021PRB}, and (g) combined effective models and exact diagonalizations ~\cite{Rousochatzakis2020KITP,quantum2023}.}\label{fig:KGphasediagram}
\end{figure*}

\subsubsection{Semiclassical phase diagram (II): $K$, $\Gamma$ of opposite signs}\label{sec:KGclassII}

The $K\Gamma$ recipe does not work when the two couplings are of opposite sign. While a slight variation of this recipe can still be applied on individual chains  (see Secs.~\ref{sec:3D} and \ref{sec:3D}), the bonds connecting the chains remain frustrated, which gives rise to a rich interplay of various complex phases. Figure~\ref{fig:KGphasediagram}\,(a-c) shows the classical phase diagrams for positive $\Gamma$ and negative $K$, as proposed from various classical energy minimization approaches, including machine learning~\cite{Liu2021PRR} (Fig. \ref{fig:KGphasediagram} (a)), Monte Carlo simulations~\cite{Rayyan2021,Chen2023NJP} (Fig. \ref{fig:KGphasediagram} (b)), hybrid Monte Carlo and iterative variational minimizations~\cite{classical2023} (Fig. \ref{fig:KGphasediagram} (c)). Let us summarise the main features:

i) In the region around the AFM $\Gamma$ point, there is some consensus that a small negative $K$ stabilizes a threefold-symmetric order with 18 spin sublattices, whose stability region is varied across the different studies, see Fig.~\ref{fig:KGphasediagram}\,(a-c). This state is dubbed 18C$_3$ order in Refs.~\cite{Rayyan2021,classical2023}, modulated S$_3\times$Z$_3$ phase in Ref.~\cite{Liu2021PRR}, and triple-meron crystal in Ref.~\cite{Chen2023NJP}. Applying the symmetries $R_{x,y,z}$ of Eq.~(\ref{eq:RxyzFull})  to the 18C$_3$ state delivers other ground states with 54 sublattices.

ii) In the region around the FM Kitaev point, a small positive $\Gamma$ seems to stabilize a zigzag (ZZ) order with a 4-site unit cell that breaks the threefold symmetry. This state is one of the classical ground states of the FM Kitaev point of Sec.~\ref{sec:KclassGSs}. Applying the symmetries $\mc{R}_{x,y,z}$ of Eq.~(\ref{eq:RxyzFull}) to the ZZ state delivers other ground states with 12 sublattices, because these operations trimerize the lattice. 

iii) The number and exact nature of the phases that appear in between the 18C$_3$ and the ZZ rgions are debated.
Nevertheless, it seems that this intermediate region is occupied by phases with large magnetic unit cells or long-wavelength, IC modulations thereof~\cite{Chern2020PRR,Rousochatzakis2020KITP,Rayyan2021,Liu2021PRR,Liu2021PRR,li2022tangle,Chen2023NJP,classical2023}. While such orders with large (or infinite)  unit cells are challenging to detect in finite-size simulations, the associated multi-peaked spin structure factors~\cite{classical2023} offer distinctive fingerprints for their experimental detection by, e.g., neutron diffraction, NMR or $\mu{SR}$.

iv) One of the proposed intermediate phases is the so-called `$K$-state' of  Fig.~\ref{fig:KGphasediagram}\,(b-c), discussed in Refs.~\cite{Rousochatzakis2020KITP,Rayyan2021,li2022tangle, Chen2023NJP,classical2023}, which is analogous to the $K$-state proposed in Ref.~\cite{Ducatman2018} for {\bliiro} (see Sec.~\ref{sec:3D}). 
This state has a six-site magnetic unit cell and shows a characteristic `period-3 chain' structure with counter-rotating spin sublattices, that has the following form 
\be\label{eqfig:KGClassGSsOppSign}
\includegraphics[width=0.75\columnwidth]{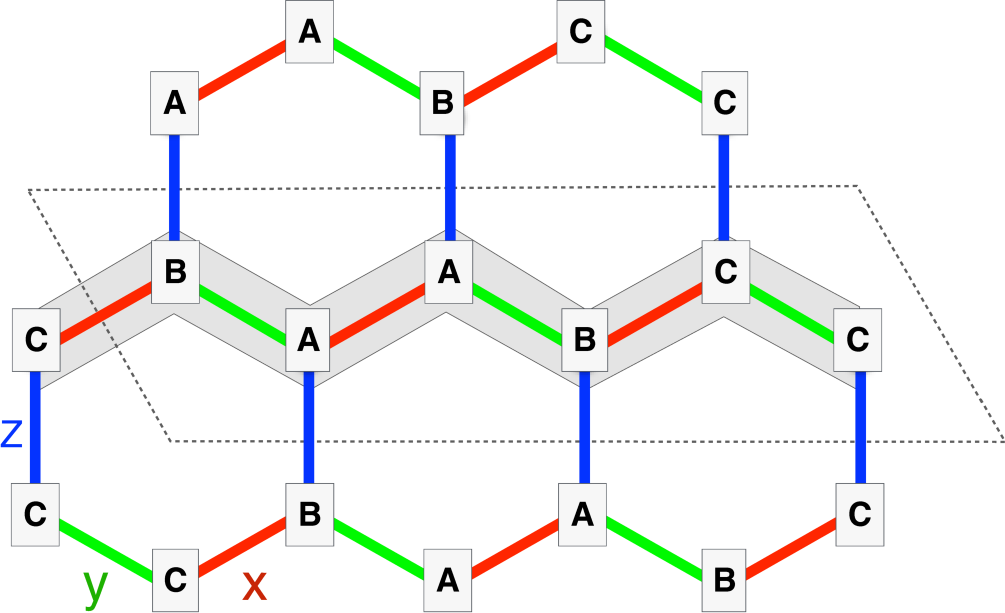}
\ee
This differs from (\ref{eqfig:KGClassGSsSameSign}) in that we now have three (instead of six) spin sublattices, A, B and C, whose directions 
\be\label{eq:ABC}
\!\!\!{\bf A}\!=\![x_A,y_A,z_A],  {\bf B}\!=\![x_B,y_B,z_B], {\bf C}\!=\![x_C,y_C,z_C],
\ee
are not given by Eq.~(\ref{eq:ABCDEF}) and are not related to each other by threefold rotations around [111], but can be found numerically by minimization. 
Finally, we note that applying the symmetries $\mc{R}_{x,y,z}$ of Eq.~(\ref{eq:RxyzFull}) to the $K$-state delivers other ground states with 18 sublattices.

v) Another proposed intermediate phase is the state denoted by `IC' in Fig.~\ref{fig:KGphasediagram}\,(c), which appears to be a long-wavelength modulation of the `$K$-state'~\cite{li2022tangle,classical2023}.

\subsubsection{Quantum $S=1/2$ limit: Numerical results}\label{sec:KGquantum}

As discussed above, the non-trivial part of the phase space of the $K$-$\Gamma$ model occurs when $\Gamma$ and $K$ have opposite signs. Here we will focus on the regime with FM Kitaev and AFM $\Gamma$ couplings, which is relevant to candidate materials.
Several numerical attempts have been made on the spin-1/2 $K$-$\Gamma$ model, and the phase diagrams obtained using different techniques are shown in Fig.~\ref{fig:KGphasediagram} (d)-(g), with the parametrization
\be
K=\cos\phi,~~
\Gamma=\sin\phi
\ee
and $\phi\in[\pi/2,\pi]$. 
A series of two infinite density matrix renormalization group studies~\cite{Gohlke2018PRBa,Gohlke2020PRR} have given evidence for highly correlated paramagnetic states, irrespective of the ratio $\Gamma/K$ for negative $K$, see panel (d) in Fig.~\ref{fig:KGphasediagram}. Namely,  they observed 
that the entanglement entropy remains very high in this entire region.
Specifically, the work of Ref.~\cite{Gohlke2018PRBa} suggests that there are two distinct spin liquid phases with a sharp
transition between them at around $\phi\simeq 0.98 \pi$, which is seen as a small
discontinuity both in the entanglement
entropy and in the spin structure factor. The state below $0.98\pi$ might be a nematic paramagnet with a spontaneous broken 3-fold rotation symmetry~\cite{Gohlke2020PRR}.
A similar picture for an extended spin liquid phase has been suggested by earlier exact diagonalization (ED) results on a 24-site cluster, in the presence of a weak bond anisotropy~\cite{Catuneanu2018npj}. 
Finally, another DMRG study, on a modulated $J$-$\Gamma$ model~\cite{luo2021npjqm}, gives evidence that the $\Gamma$ point features a gapless quantum spin liquid ground state. 

The picture for an extended spin liquid phase is at variance with other approaches, such as variational Monte Carlo~\cite{Normand2019} and pseudofermion functional renormalization group (pf-FRG)~\cite{Buessen2021PRB} calculations, see panels (e-g) in Fig.~\ref{fig:KGphasediagram}. 
Specifically, the VMC study of Ref.~\cite{Normand2019} delivers four ordered states outside the stability region of the Kitaev spin liquid ($\pi/2\le\phi\le0.95\pi$). These include the zigzag phase for $\pi/2\le\phi\le0.8\pi$, an incommensurate (IC) spiral phase for $0.84\pi\le\phi\le0.84\pi$, a proximate Kitaev spin liquid phase (PKSL) for $0.84\pi\le\phi\le0.92\pi$, and finally a FM phase for $0.92\pi\le\phi\le0.95\pi$.

The pf-FRG study of Ref.~\cite{Buessen2021PRB} suggests that, away from the Kitaev spin liquid phase ($\pi/2\le\phi\le0.94\pi$), the system features two ordered phases, a FM phase for $0.85 \pi\leq\phi\leq 0.94 \pi$ and an IC order for $0.5 \pi\leq\phi\leq 0.85 \pi$, whose ordering wavevector evolves continuously with $\phi$, along the line that connects the BZ center to the midpoint of the BZ edge ($M$-point). 
In the more extended phase diagram that includes the region of positive $K$, the pf-FRG results show a variety of other magnetic phases, including IC and vortex-like phases. 

Finally, a study that combines effective strong coupling expansions and ED up to 36 sites~\cite{Rousochatzakis2020KITP,quantum2023} gives evidence that the classical spin liquid of Sec.~\ref{sec:Gpoints-QObD} (see also \cite{Rousochatzakis2017PRL}) survives in a wide region surrounding the pure AFM $\Gamma$ point, from $\phi\simeq 0.41\pi$ up to approximately  $\phi\sim0.8\pi$. The results also show that this phase is followed by at least one more phase (of hitherto unknown origin) before we reach the Kitaev spin liquid, see panel (g) in Fig.~\ref{fig:KGphasediagram}. 

While different methods come with their own advantages and disadvantages, and further studies are definitely needed to reach a broad consensus, the multitude of competing phases depicted in Fig.~\ref{fig:KGphasediagram} illustrates the strong frustration that arises from the interplay of FM Kitaev and AFM $\Gamma$ interactions in the region where the two couplings have different signs. The fact that the available layered Kitaev materials seem to be in the vicinity of precisely this coupling regime (with dominant FM $K$ and subdominant AFM $\Gamma$) underlies the strong theoretical interest to this open problem over the last years.

\subsection{The Kitaev-Heisenberg model}\label{sec:JKmodel}  
 
Historically, the Kitaev-Heisenberg (KH) model was the first extension of the Kitaev model, in which one includes the isotropic NN Heisenberg interaction $J$, that arises from the direct overlap of the $d$-orbitals. The KH model was already proposed in the seminal paper of Jackeli and Khaliullin~\cite{Jackeli2009PRL}, and has been pivotal in demonstrating that the Kitaev QSL can survive in an extended region of parameter space (instead of a single point), which immediately attracted a lot of attention ~\cite{Chaloupka2010PRL,Chaloupka2013PRL,Jiang2011PRB,Reuther2011PRB,Price2012PRL,Price2013PRB,Shinjo2015PRB,Chern2017,Gohlke2017PRL,Gotfryd2017PRB,Czarnik2019,Janssen2019,Zhang2020}, thus giving birth to the field of Kitaev materials~\cite{Takagi2019NRP,Janssen2019,Trebst2022,Tsirlin2022}.

One of the special features of the KH model is the duality transformation $\mc{T}_4$ of (\ref{eq:JKduality}). As discussed in  Sec.~\ref{sec:HiddenSU2JK}, this duality maps the general parameter point $(K,J)$ to $(\widetilde{K},\widetilde{J})=(K+2J,-J)$. On one hand, this effectively allows to study only half of the phase diagram. At the same time, it reveals the two hidden SU(2) points with $K=-2J$. Altogether then, the KH model has four SU(2) symmetry points. Using  the parametrization 
\be
K=\sin\phi,~~J=\cos\phi
\ee
with $\phi\in[0,2\pi)$, these SU(2) symmetry points reside at: $\phi\!=\!0$, $\pi$, $-\tan^{-1}(2)=-63.43^\circ$ and $-\tan^{-1}(2)+\pi=116.56^\circ$. The presence of the four SU(2) points governs much of the phase diagram, as we discuss below.

\subsubsection{{Classical \& quantum spin-1/2 phase diagram}} 
\begin{figure}[!t]
\includegraphics[width=0.95\linewidth]{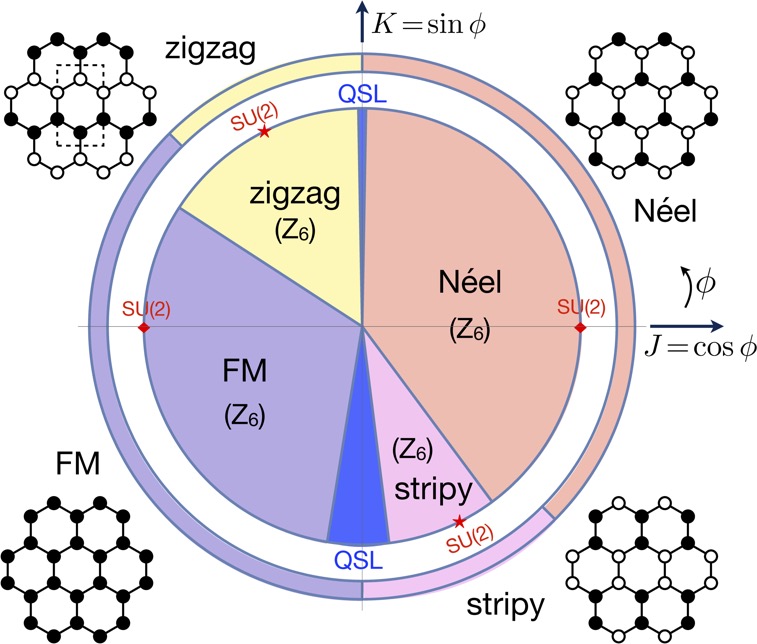}
\caption{Classical (outer ring) and quantum spin-1/2 phase diagram (inner disc, adapted from Ref.~\cite{Chaloupka2013PRL}) of the $J$-$K$ model, parametrized as $J\!=\!\cos\phi$  and $K\!=\!\sin\phi$. The star symbols indicate the two hidden SU(2) points discussed in Sec.~\ref{sec:HiddenSU2JK}.}\label{fig:JKPhaseDiag2013}
\end{figure}

Figure~\ref{fig:JKPhaseDiag2013} shows the classical (outer ring) and quantum spin-1/2 phase diagram (as adapted from Ref.~\cite{Chaloupka2013PRL} (where the authors use a slightly different parametrization for $K$ and $J$).
The main features can be summarized as follows: 

i) Besides the special Kitaev points which feature an infinite number of classical ground states, the classical phase diagram comprises four collinear magnetic orders: The N\'eel AFM, a zigzag AFM (the dual of the N\'eel), the FM, and the stripy AFM (the dual of the FM). Each of these phases surrounds one of the four SU(2) points of the model. Away from these points, all phases are gapped.  More specifically, the {\it linear} spin-wave spectrum has a quasi-Goldstone mode at the ordering wavevectors, despite the absence of continuous symmetry, and the {\it actual} discrete symmetry and the nonzero gap are recovered by anharmonic spin wave fluctuations~\cite{Chaloupka2010PRL,Chaloupka2013PRL}.


ii) The direction of the moments in any of the four collinear states does not affect the classical energy, so the system has an $\mc{S}^2$ ground state degeneracy. Away from the SU(2) symmetry points, this degeneracy is accidental and
is therefore lifted by quantum (and thermal) fluctuations.  
Indeed, as shown in several works (including semiclassical spin-wave expansions, real space perturbation theory and variational minimization over spin coherent states, Hubbard-Stratonovich approach~\cite{Chaloupka2010PRL,Chaloupka2013PRL,Sizyuk2016,Peter2017,Sela2014PRB,Chaloupka2016,Gotfryd2017PRB})
the magnetic moments in each of the four magnetic phases point along one of the six cubic directions $\pm\hat{{\bf x}}$, $\pm\hat{{\bf y}}$ or $\pm\hat{{\bf z}}$, except at the four SU(2) points where all directions are equally likely. The selection of the cubic axes can be captured by a fourth-order anisotropy similar to that of Eq.~(\ref{eq:KGObDdE2}), see, e.g., Ref.~\cite{Chaloupka2016}.  Similar analysis was later performed to the  extended models including $\Gamma$, further neighbor interactions or external field \cite{Winter2017,Maksimov2020PRR,Sizyuk2016,Smit2020,Janssen2016,Chern2017,Maksimov2022}.

iii) In the quantum limit, we have, in addition, the two Kitaev QSL phases, which extend in a narrow neighborhood of the pure Kitaev points. The stability region of the FM Kitaev QSL is much more extended than that of the AF Kitaev point. Specifically, the AF Kitaev QSL is stable for $\phi \in (89^\circ,91^\circ)$, whereas the FM Kitaev QSL is stable for $\phi \in (-99.23^\circ,-82.89^\circ)$.  The physical reason for the asymmetric stability of the Kitaev spin liquid phases around the FM and AFM Kitaev limits can be understood from the fact that the Heisenberg interaction drives the condensation of  a bound state of the fractionalized excitations, which signals the transition to the magnetically ordered state \cite{Shang-Shun2021}. Remarkably, this bound state appears as a sharp mode in the dynamical spin structure factor, while its condensation patterns at the appropriate phase transitions provide a simple explanation for the
magnetically ordered phases surrounding each Kitaev spin liquid phase.

iv) Quantum corrections shift the boundaries between the magnetic orders: the boundary between the FM and the zigzag phase shifts from $\phi\!=\!3\pi/4$ to $\phi\!\simeq\!146.98^\circ$, and the boundary between the stripy and the N\'eel phase shifts from $\phi\!=\!-\pi/4$ to $\phi\!\simeq\!-53.45^\circ$.

v) Finally, while the phase transitions between the magnetic phases appear to be discontinuous, numerical studies based on density matrix renormalization group~\cite{Shinjo2015PRB} and exact diagonalizations~ \cite{Gotfryd2017PRB} suggest that the transition between the QSL and the stripy state is continuous~\cite{Chaloupka2013PRL} or weakly first-order.

\begin{figure*}[!t]
\includegraphics[width=\linewidth]{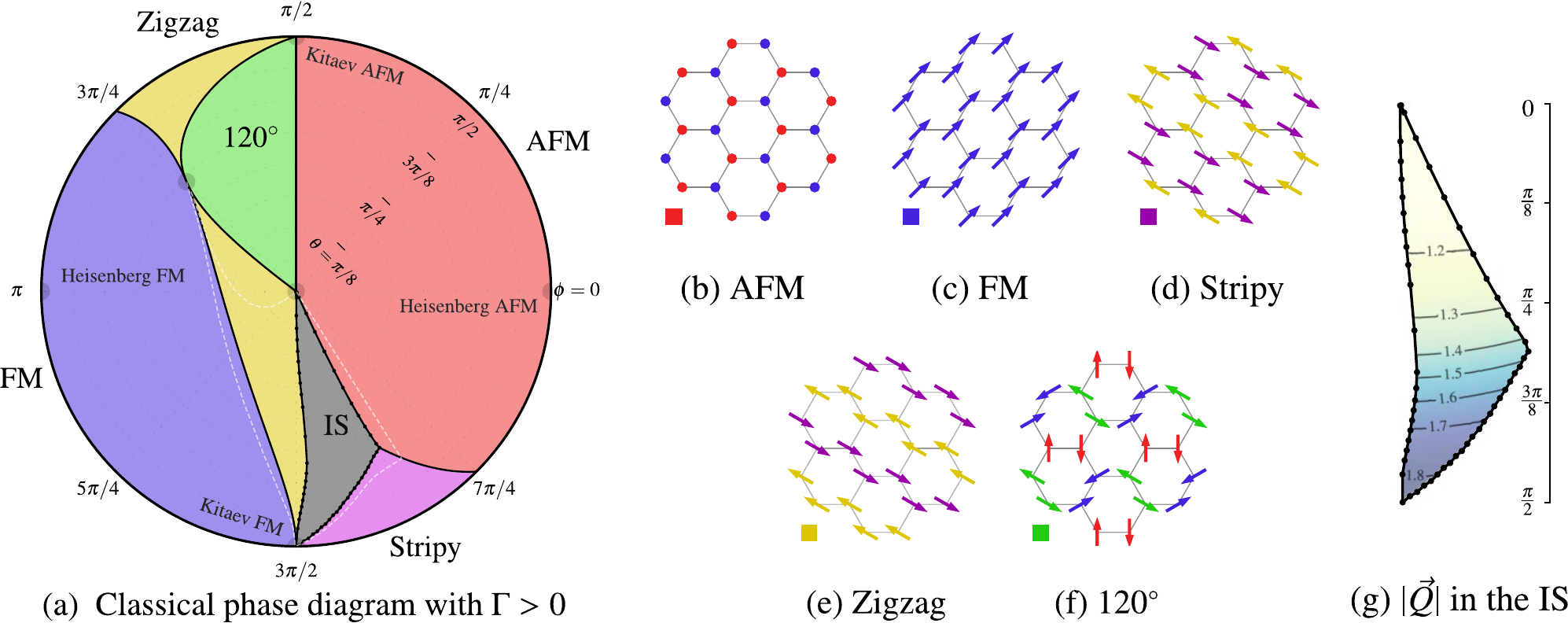}
\caption{(a) Classical phase diagram of the $J$-$K$-$\Gamma$ model for $\Gamma>0$. The couplings are parametrized in terms of two angles $\theta\in[0,\pi/2]$ and $\phi\in[0,2\pi)$, with $J=\sin\theta\cos\phi$, $K=\sin\theta\sin\phi$ and $\Gamma=\cos\theta$. 
Solid lines correspond to the boundaries from the Luttinger-Tisza (LT) method, while the region indicated by the white dashed line is where the LT method fails, and a single-${\bf Q}$ analysis is carried out instead. (b-f) Pictorial representation of the main ground states. (g) Magnitude of the ordering wave vector ${\bf Q}$ in the IS region. Adapted from Ref.~\cite{Rau2014}.  
}\label{fig:JRauJKGPhaseDiagram}
\end{figure*}

\subsubsection{{Finite-temperature transitions}} 
The frustration in the classical KH model manifests itself in very non-trivial finite temperature properties~\cite{Price2012PRL,Price2013PRB}, which are similar to those of the six-state
clock model~\cite{Jose1977}. For all values of parameters, except for the pure Kitaev limit and points of hidden symmetries which we discussed above in Sec.~\ref{subsec:hiddenSU2}, the KH model undergoes two continuous phase transitions as a function of temperature;separating three phases: a low-$T$ ordered phase, an intermediate critical phase, and a high-$T$ disordered phase. 

The presence of long range magnetic order in two dimension at finite temperatures requires a discreteness of the order parameter, which means that the direction of the order parameter must also be selected.  As we discussed above, higher order anharmonic modes of spin fluctuations indeed select collinear magnetic states at $T=0$. At finite temperatures,  magnetically ordered states are  stabilized entropically by an order by disorder mechanism where thermal fluctuations of classical spins  select collinear magnetic states in which magnetic moments point along one of the cubic directions,  in a similar fashion  to  quantum   spin fluctuations~\cite{Price2012PRL,Price2013PRB,Sizyuk2015}.

Regardless of the specific kind of magnetic order, the low-$T$ ordered phase is separated from the high-$T$ paramagnetic phase by the intermediate critical phase.  The critical phase has an emergent, continuous $U(1)$ symmetry,  which is reminiscent of the intermediate phase of the six-state clock model~\cite{Jose1977}, and is similar to the low-$T$ phase of the $XY$ model, a well-known Berezinskii-Kosterlitz-Thouless phase of critical points with floating exponents and algebraic correlations. To see this analogy to the six-state clock model, one can map the order parameter describing the any of the magnetically ordered phase of the KH model onto a 2D complex order parameter $m_{N(S)}=\sum_{i=1}^6 |m_{i,N(S)}| e^{ \imath\theta_i}$, where the phase $\theta$ is such that the minimal-energy states of the order parameter, which point along the cubic axes, will be labeled by the values $\theta_i=\pi n_i/3$, $n_i=0,..5$~\cite{Price2012PRL,Price2013PRB}.
In general, the long-range low-temperature magnetic order can be destroyed in one of several ways, i.e., by a discontinuous first-order transition, by two continuous phase transitions with an intermediate,  or by two BKT phase transitions with an intermediate critical phase~\cite{Jose1977,Isakov2003,ChernGW2012}. Which way is actually realized depends on on the relative strength of the spin stiffness in different directions. If the stiffness of thermal fluctuations along the circle is softer than the stiffness of fluctuations in the direction transverse to the circle, the long-range order may be destroyed by the last scenario. In this case, the low-T transition is between the magnetically ordered phase and the BKT  critical phase, and the high-T transition is  between the BKT phase  and a disordered phase where fluctuations are primarily two-dimensional. The crossover to the three-dimensional paramagnet occurs at even higher temperatures~\cite{Price2012PRL,Price2013PRB}. While the ordered phase is destroyed by the appearance of the topological excitations in the form of discrete vortices whose existence is directly related to the emergence of a continuous symmetry in the critical phase, the BKT phase is destroyed by their proliferation~\cite{Jose1977}.

\subsection{The $JK\Gamma$ model}\label{sec:JKGmodel}

Let us now put together the above insights from the various effects of $K$, $\Gamma$ and $J$, and map out the phase diagram of the $JK\Gamma$ model. We shall begin with the semiclassical limit, which features a rich interplay between a variety of classical magnetic orders.  
Figure~\ref{fig:JRauJKGPhaseDiagram}\,(a) shows the early classical phase diagram proposed by Rau, Lee and Kee~\cite{Rau2014}. The diagram shown covers the regions of parameter space with positive $\Gamma$. The corresponding space with $\Gamma\!<\!0$ can be obtained by the sublattice duality of Sec.~\ref{sec:ClassicalDuality}, which maps $(J,K,\Gamma)\!\mapsto\!(-J,-K,-\Gamma)$. 
Note that this diagram is shown using the parametrization 
\be\label{eq:JKGparametrization1}
J=\sin\theta \cos\phi,~
K=\sin\theta \sin\phi,~
\Gamma=\cos\theta
\ee
with $\theta\in[0,\pi/2]$, $\phi\in[0,2\pi)$. Compared to the parametrization of the KH model of Fig.~\ref{fig:KGphasediagram}, there is a second angle ($\theta$) which quantifies the strength of the $\Gamma$ coupling, and varies from $\theta\!=\!0$ (center of the disc) to $\theta\!=\!\pi/2$ (outer perimeter of the disc). In this parametrization, the center of the disc corresponds to the pure $\Gamma$ model, the outer perimeter of the disc corresponds to the KH model, and the vertical line $\phi\!=\!\pm\pi/2$ corresponds to the $K\Gamma$ model with $\Gamma\!>\!0$.

The main features of the classical phase diagram are:

i) Around the outer perimeter of the disc, we recover the four collinear orders of the KH model of Fig.~\ref{fig:KGphasediagram}. These phases occupy a large portion of the phase diagram, especially the FM and the N\'eel AFM phase. 

ii) The introduction of the $\Gamma$ coupling affects the direction of the moments in these collinear phases already at the mean field level (see Ref.~\cite{Chaloupka2015PRB,Chaloupka2016}), where it is further demonstrated that the moment direction can be used as a sensitive probe of the model parameters in real materials.

iii) Around the vertical half-line $\phi\!=\!\pi/2$, we get the so-called 120$^\circ$ state of Fig.~\ref{fig:JRauJKGPhaseDiagram}\,(f), for positive $K$ and negative $J$, with spins lying on the $(111)$ plane. 
This state is in fact a member of the classical ground state manifold of Sec.~\ref{sec:KGclassI}, for positive $K$ and $\Gamma$. To see why this phase emerges for negative $J$, we use the six-spin-sublattices ansatz
of Eq.~(\ref{eq:ABCDEF}) corresponding to $K$ and $\Gamma$ positive, namely 
\be\label{eq:ABCDEFposKG}
\renewcommand{\arraystretch}{1.5}
\begin{array}{lll}
\!{\bf A}\!=\![x_0,y_0,z_0], 
\!&\!{\bf B}\!=\![z_0,x_0,y_0], 
\!&\!{\bf C}\!=\![y_0,z_0,x_0],
\\
\!{\bf D}\!=\!-[x_0,z_0,y_0], 
\!&\!{\bf E}\!=\!-[y_0,x_0,z_0], 
\!&\!{\bf F}\!=\!-[z_0,y_0,x_0],
\end{array}
\ee
and express the Heisenberg energy in terms of $x_0$, $y_0$ and $z_0$: 
\be
E_{J}\!=\!
\frac{J}{6}
({\bf A}+{\bf B}+{\bf C})\cdot({\bf D}
+{\bf E}+{\bf F})
\!=\!-\frac{J}{6} (x_0+y_0+z_0)^2.
\ee
This tells us that a negative $J$ necessitates that $x_0+y_0+z_0=0$, i.e., the spins will be on the (111) plane, and form a `period-3 chain' state with two sublattices that are counter-rotating by successive 120$^\circ$, compare Eq.~(\ref{eqfig:KGClassGSsSameSign}) and Fig.~\ref{fig:JRauJKGPhaseDiagram}\,(f).

iv) By contrast, a positive $J$ maximizes $(x_0+y_0+z_0)^2$, and this leads to the N\'eel state with spins along the $[111]$ axis.
Note that this mean-field result  competes with the effect of quantum fluctuations close to the circumference of the disc, where spins prefer to point along the cubic axes. So, one expects a rotation of the moments from $[111]$ to one of the cubic axes as we reduce the strength of $\Gamma$.

iv) There is an extended region, delineated by a white dashed line in Fig.~\ref{fig:JRauJKGPhaseDiagram}\,(a), and centered around the vertical half-line $\phi\!=\!-\pi/2$,  where the classical minimization by various methods shows incommensurate ground state(s). This incommensurate phase actually emerges from the phases of  Fig.~\ref{fig:KGphasediagram}\,(a-c), and it also remains coplanar despite its pitch vector characterising it varies throughout the phase~\cite{Rau2014}. 

{\it Quantum limit.} 
A complete mapping of the region delineated  by a white dashed line in Fig.~\ref{fig:JRauJKGPhaseDiagram}\,(a) remains a matter of active investigation. As explained in Sec.~\ref{sec:KGclassII}, this is the region where $K$ and $\Gamma$ compete the most, and it is where we expect quantum fluctuations to have the strongest effect. 
It is therefore not surprising that this region of the phase diagram has attracted a lot of attention in the literature, also  given  the relevance to {\rucl} and {\aliiro}, see Table~\ref{tab:NumValues}. The possible presence of an additional QSL phase somewhere in this region -- besides the QSL phases surrounding the two Kitaev points -- is an exciting possibility that has been evidenced in some numerics (see discussion in Sec.~\ref{sec:KGquantum}), but a more conclusive picture warrants further investigations.

\subsection{Other special limits}\label{sec:Otherlimits} 
Even in the absence of Kitaev interaction, the interplay of Heisenberg and off-diagonal $\Gamma$ and $\Gamma'$ interactions leads to some non-trivial physics.
The huge ground-state degeneracy of the pure AFM $\Gamma$ model is a source of exotic states of matter,
which could be selected due to the interplay of exchange frustration and competing interactions.
The $\Gamma'$ interaction serves as a representative perturbation to stabilize novel phases \cite{Rau2014-arxiv}. 
In the following, we will consider a few interesting  limits. 

i) The first example is the anisotropic Heisenberg-$\Gamma$ model where the $z$-bond strength is different from the other bonds~\cite{Suzuki2021PRB}. The model thus interpolates between isolated bonds at one limit and a set of decoupled chains at the other limit, with the isotropic exchanges in between. It has a very rich phase diagram indicating the existence of ten phases: three ferromagnetically ordered, three dimerized, two spiral, one antiferromagnetically ordered, and one stripy phases. In the spin-chain limit, there is also a Luttinger liquid in addition to two magnetically ordered states. In the isolated dimer limit, there are three distinct dimerized phases among which a triplet dimer phase can sustain up to the isotropic interaction limit with dominant $\Gamma$ interaction.

ii) The second example is the anisotropic Heisenberg-$\Gamma'$ model in which the anisotropy is exerted solely on the Heisenberg term~\cite{Zhao2022PRB}. It is found that the ground state is the AFM phase with the spin aligning with the [111] direction near the isotropic limit, which is replaced by the dimerized phase as the anisotropy increases above certain critical strength. In an out-of-plane magnetic field, the model shows a field-induced AFM-type order occupying a large parameter region, before entering into the polarized phase. Nature of the intermediate phase can be understood from the picture of magnon condensation~ \cite{Zhao2022PRB}.

iii) The third example is the bond-modulated Heisenberg-$\Gamma$ model, another interesting example which displays several phase transitions~\cite{luo2021npjqm}. In this model,  there is a  competition between the  fixed $\Gamma$ term and   a staggered Heisenberg interaction, in which there is a sign difference between the coupling on the $z$-bond and on the other bonds. The phase  diagram of this model contains the zigzag order for positive Heisenberg interaction and stripy order  for the negative one. There is also an intermediate phase sandwiching between the two. Though exhibiting magnetic order at the classical level, quantum fluctuations suppress such ordering since it acquires a large energy according to the spin-wave result. In the quantum case, it turns out to be disordered and is separated from its two neighbors by first-order transitions~\cite{luo2021npjqm}. Though indirectly, the numerical results presented in Ref.~\cite{luo2021npjqm} emphasize the quantum nature of the $\Gamma$ model.

iv) The fourth example is the   $\Gamma$-$\Gamma'$ model on the honeycomb lattice~\cite{Luo2022PRR}. This model is particularly interesting in the vicinity of the AFM $\Gamma$ interaction. There the spin liquid phase of the classical $\Gamma$ model is separated by the first order phase transitions from the 120$^{\circ}$ ordered  phase for $\Gamma'<0$ and the AFM phase for $\Gamma'>0$. In the quantum limit, a chiral-spin ordering with a spontaneous time-reversal symmetry breaking  is stabilized in a wide parameter space proximate to the pure $\Gamma$ model. The chiral-spin ordering is characterized by a nonzero scalar spin chirality that is defined as
\be\label{EQ:ChiIJK}
\hat{\chi}^{\triangle}_{ijk} = \hat{\mathbf{S}}_i\cdot(\hat{\mathbf{S}}_j\times\hat{\mathbf{S}}_k),
\ee
where sites ($i, j, k$) are three neighboring sites that belong to the same sublattice.  
The intriguing question is  whether the chiral-spin ordering is a QSL or not. The ED calculation and the large-scale DMRG calculations~\cite{Luo2022PRR} show 
 the vanishing magnetization in a large enough
system size indicates that the CS phase is a magnetically
disordered state, in striking contrast to a classical chiral-spin ordering 
that possesses a long-range magnetic order. 
Also, dynamic structure factor calculations find a broad continuous feature in the low-frequency region, which is likely the evidence of the QSL phase~\cite{Luo2022PRR}. However, the nature of this phase is still unclear.



\begin{figure}[!b]
{\includegraphics[width=\linewidth]{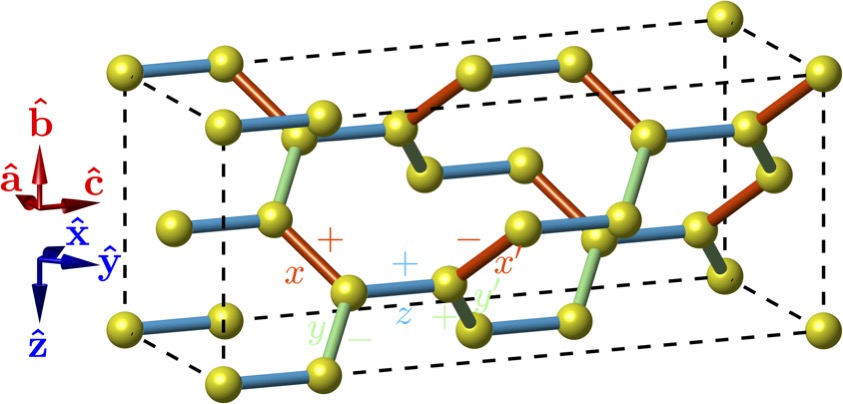}}
\caption{Orthorhombic unit cell of the hyperhoneycomb lattice. The five NN bonds of the $JK\Gamma$ model are marked in red for bonds of type $\gamma\in\{x,x'\}$, green for $\gamma\in\{y,y'\}$, and blue for $\gamma=z$. The $\pm$ signs on the bonds $\gamma$ indicate the sign $\sigma_\gamma$ in $\Gamma_\gamma\!=\!\sigma_\gamma\Gamma$.}
\label{fig:hyperhoneycomblattice}
\end{figure}


\section{Interplay between $K$ and $\Gamma$ in 3D compounds}\label{sec:3D}
The rich interplay between $K$ and $\Gamma$ is not specific to 2D, but manifests also in the 3D geometries of Fig.~\ref{fig:TriCoord}\,(b-c), which are realized, respectively, in {\bliiro} and {\gliiro}, the only 3D tricoordinated Kitaev materials studied extensively so far~\cite{Modic2014NC,Biffin2014PRB,Biffin2014PRL,Modic2017,Modic2018b,Tsirlin2022}. As mentioned earlier,  together with {\aliiro}~\cite{Williams2016PRB,Hermann2018PRB,Choi2019PRB,Vale2019,Shen2022}, these compounds are special members of the {\liiro} polymorph family of crystal structures~\cite{Kimchi2014b}. The Kitaev model can be solved exactly on all these geometries~\cite{Mandal2009PRB,Hermanns2015,Obrien2016}, despite the fact that Lieb's theorem~\cite{Lieb1994} does not constrain the ground state flux sectors of every one of them (in this case, the ground state flux sector have been found numerically).

\subsection{Universality across the {\liiro} polymorphs}\label{sec:Li213universality}
Contrary to {\nairo} and {\rucl}, which display collinear zigzag order at low temperatures~\cite{Ye2012PRB,Sears2015PRB,Choi2012PRL}, the magnetic states of the {\liiro} polymorphs involve non-collinear incommensurate (IC) spin spirals~\cite{Williams2016PRB,Biffin2014PRL,Biffin2014PRB,Modic2014NC}, characterised by two counter-rotating spin sublattices. Locally, these states resemble the `period-3 chain' structure discussed above, which is an indication for sizeable off-diagonal $\Gamma$ interactions.

The origin of this universality across the three {\liiro} polymorphs has been the subject of several works~\cite{Kimchi2014b,Lee2015,Lee2016,Ducatman2018,KrugerPRR2021}, see also Ref.~\cite{Tsirlin2022}. One key ingredient seems to be the common local geometry of the bonds, which suggests that the minimal microscopic description is similar in these compounds. Indeed, the orders observed in {\bliiro} and {\gliiro} have been rationalized~\cite{Lee2015,Lee2016,Ducatman2018} within the framework of the minimal NN $J$-$K$-$\Gamma$ model, with large FM Kitaev interaction, sizeabe $\Gamma$ interactions (with $|\Gamma|\!<\!|K|$) and much weaker AFM Heisenberg exchange $J\!\ll|\Gamma|$.

Another key ingredient behind this universality seems to be the physics along individual $xy$ chains and how the corresponding modulated orders on these chains tile in the whole 3D netowork~\cite{Kimchi2014b,Ducatman2018}. The work of Ref.~\cite{KrugerPRR2021} has revealed a more general mapping that connects most of the ordered states of the $JK\Gamma$ phase diagram (and not just the ones relevant to {\bliiro}) to their 2D counterparts on the honeycomb lattice, with the classical energetics being identical in 2D and 3D.

\subsection{The paradigm of {\bliiro}}\label{sec:bLi213paradigm}
In the following we focus on the paradigm of {\bliiro}~\cite{Biffin2014PRB,Takayama2015,Ruiz2017,Ruiz2021,Halloran2022,Yang2022PRB}, which stands out among the three polymorphs for its quantitative level of description~\cite{Ducatman2018,Rousochatzakis2018,Li2020,Li2020b,Yang2022PRB}.

{\it Model.} Here, the $JK\Gamma$ model takes the general form of Eqs.~(\ref{eq:JKGGpmodel}) and (\ref{eq:JKGGpmodelij}), where, as mentioned above, the bond index $\gamma$ runs over five types of bonds, $x$, $y$, $x'$, $y'$ and $z$ (see also Fig.~\ref{fig:hyperhoneycomblattice}), with
\be\label{eq:3d-JKG}
K_\gamma=K,~~~
\Gamma_\gamma=\sigma_\gamma \Gamma,~~~
\Gamma'_\gamma=0,
\ee
and $\sigma_\gamma\!=\!\pm1$ incorporates the sign structure of the $\Gamma$ couplings, as discussed in App.~\ref{app:GGpSignStructure}. For the cubic axes frame of Eq.~(\ref{eq:bLi213frame}) taken in Ref.~\cite{Ducatman2018}, or the equivalent frame of Refs.~\cite{Lee2015,Lee2016}, $ \sigma_{z/x/y'}\!=\!1$ and $\sigma_{x'/y}=-1$, see  Fig.~\ref{fig:hyperhoneycomblattice}.

\begin{figure}[!t]
\includegraphics[width=0.85\linewidth]{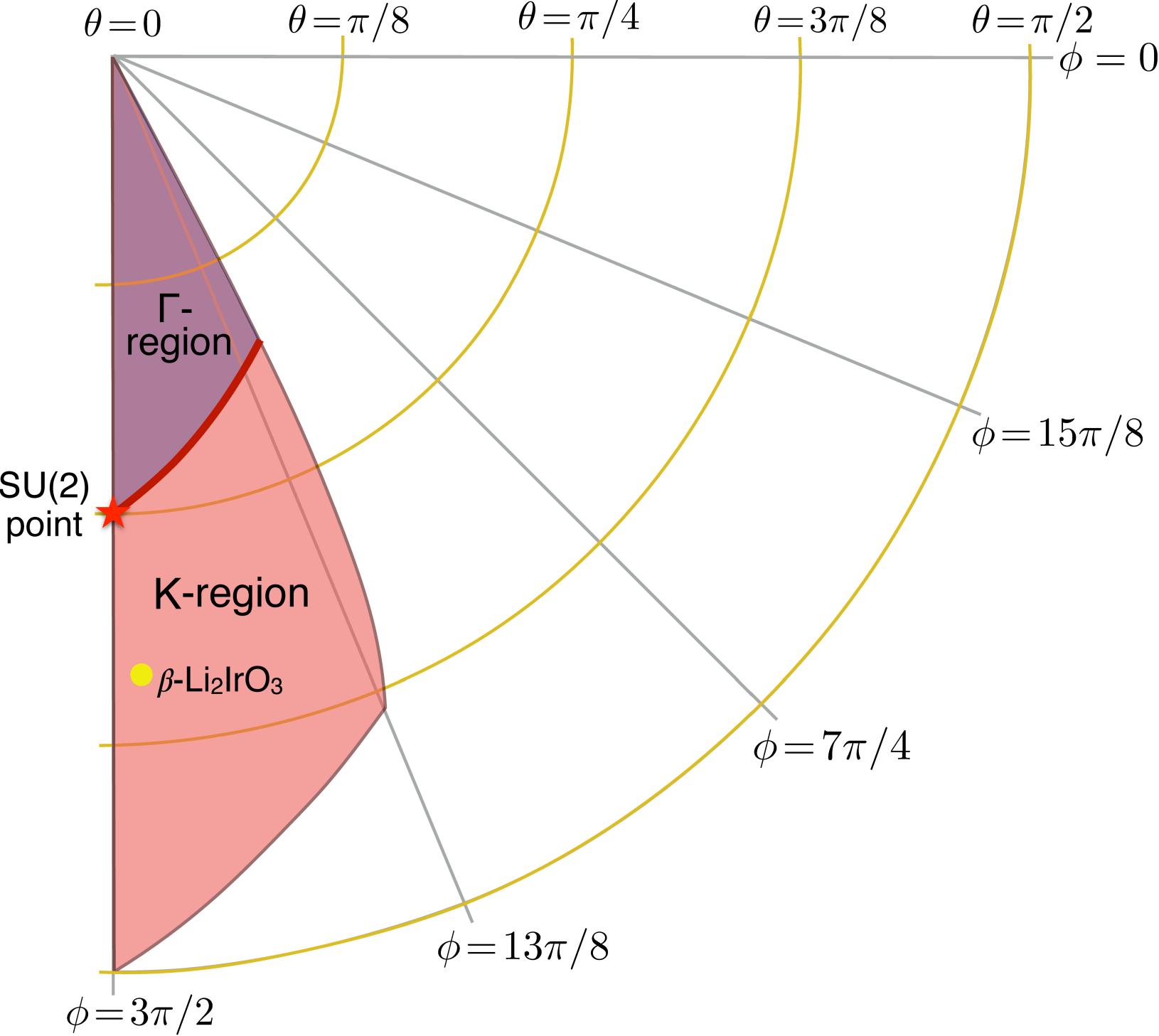}
\caption{The shaded region shows the portion of the parameter space $(\theta,\phi)$ [see Eqs.~(\ref{eq:3d-JKG}) and (\ref{eq:JKGparametrization3D})] that is believed~\cite{Lee2015,Lee2016,Ducatman2018} to be relevant for $\beta$-Li$_2$IrO$_3$. The `$K$-region' and `$\Gamma$-region' host different types of classical states (with varying wavelengths), and their boundary (red line) emerges from the hidden, isotropic SO(3) point (red star)~\cite{Ducatman2018}.  Quantitative estimates of $J$, $K$ and $\Gamma$ (see Table~\ref{tab:NumValues}) place {\bliiro} at the yellow point shown, very close to the special frustrated line $\phi=3\pi/2$. For the phases outside the shaded region (including the white region shown), see detailed analysis in Refs.~\cite{Lee2015,Lee2016}. Adapted from Ref.~\cite{Ducatman2018}.}\label{fig:bLi213-phasediagram}
\end{figure}

{\it Hidden SU(2) points.} In analogy to the 2D honeycomb case, the above model has hidden SU(2) symmetry points at $K\!=\!\Gamma$ and $J\!=\!0$, as these points can be mapped to dual points with $K'\!=\!\Gamma'\!=\!0$ and $J'\!=\!-K$~\cite{Ducatman2018,Stavropoulos2018}. For $K\!=\!\Gamma\!<\!0$ and $J\!=\!0$ the system then maps to an antifferomagnetic Heisenberg model, whereas for $K\!=\!\Gamma\!>\!0$ and $J\!=\!0$, the system maps to a FM Heisenberg model. 
The corresponding duality transformation is similar to the transformation $\mc{T}_6$ of Sec.~\ref{Sec:HiddenSU2KG}, and involves six sublattices on isolated $xy$ chains, and 12  sublattices for the hyperhoneycomb lattice (24 for the stripyhoneycomb). 
The hidden SU(2) point that is most relevant for {\bliiro}, where $K$ and $\Gamma$ are both negative, is shown in Fig.~\ref{fig:bLi213-phasediagram}, where we use the parametrization
\be\label{eq:JKGparametrization3D}
J=\sin\theta\cos\phi,~~
K=\sin\theta\sin\phi,~~
\Gamma=-\cos\theta.
\ee
Here, $\phi\!\in\![0,2\pi)$ and $\theta\!\in\![0,\frac{\pi}{2}]$, and the minus sign in the last term accounts for the negative sign of $\Gamma$ in {\bliiro}.

{\it Relevant region of the phase diagram.} Remarkably, all experimental data reported so far~\cite{Biffin2014PRB,Takayama2015,Ruiz2017,Ruiz2021,Halloran2022,Yang2022PRB} are consistent with parameters close to  $J\!=\!0.4~\text{meV}$,  $K\!=\!-18~\text{meV}$ and $\Gamma\!=\!-10~\text{meV}$~\cite{Ducatman2018,Rousochatzakis2018,Li2020,Li2020b,Yang2022PRB}. These parameters place {\bliiro} somewhere inside the so-called $K$-region of Fig.~\ref{fig:bLi213-phasediagram} \cite{Ducatman2018}. 
This material is therefore extremely close to the special $K$-$\Gamma$ line, which hosts a two-parameter family of degenerate classical ground states, similar to `period-3 chain' family of states with counter-rotating spin sublattices, discussed in Sec.~\ref{sec:KGclassI}. The weak and positive Heisenberg coupling $J$ lifts this degeneracy and selects configurations with the 120$^\circ$ structure, in analogy with the selection of the 120$^\circ$ state of Fig.~\ref{fig:JKPhaseDiag2013} by a negative $J$ in the 2D honeycomb case. 
With increasing $J$, this configuration gets distorted, leading to an IC state with propagation wavevector ${\bf Q}\!=\!(0.574,0,0)$ in orthorhombic units~\cite{Biffin2014PRB,Takayama2015} at zero-field and below $T_I\!=\!38$~K.

The IC order of {\bliiro} can be treated as a long-distance twisting of the closest commensurate period-3 state, with ${\bf Q}=\frac{2}{3}\hat{\bf a}$, in orthorhombic units of $\frac{2\pi}{a}$~\cite{Ducatman2018}. This ansatz is amenable to a semi-analytical treatment, and delivers results that are consistent with almost all experimental findings so far~\cite{Biffin2014PRB,Takayama2015,Ruiz2017,Ruiz2021,Halloran2022,Yang2022PRB}, both in zero field and at finite fields. 
The quantitative success of this ansatz is another evidence of the close proximity of the IC order to the line $\phi\!=\!3\pi/2$.

{\it Physics under a magnetic field.} The application of a magnetic field along ${\bf b}$ axis (see Fig.\ref{fig:hyperhoneycomblattice}) suppresses the zero-field order with a critical field of $H_{\bf b}=2.8$~T~\cite{Ruiz2017}. This rapid decline of the IC order is a signature of the smallness of $J$, since in the ${\bf b}$-direction the critical field only depends on $J$~\cite{Rousochatzakis2018}. For fields along ${\bf a}$ and ${\bf c}$, the system shows a much weaker response, with the IC order remaining robust and the magnetization being linear up to the maximum fields measured~\cite{Ruiz2017,Majumder2019}. Such a strongly anisotropic response signifies a large separation of energy scales between $J$ and $\Gamma$. This large separation also manifests in the structure of the magnetic excitations measured by Resonant Inelastic X-ray Scattering (RIXS)~\cite{Ruiz2021},   inelastic neutron scattering (INS) and time-domain THz spectroscopy~  \cite{Halloran2022}, and in the magnetic Raman response~\cite{Yang2022PRB}.

\section{Quasi-1D $K$-$\Gamma$ models}\label{sec:1D}
The challenges associated with the 2D $K$-$\Gamma$ Honeycomb model (see Sec.~\ref{sec:KGmodel}) have motivated a number of studies which approach the 2D lattice as a collection of 1D chains. This idea has worked quite well for the Heisenberg-Kitaev model, see e.g., Ref.~\cite{CatuneanuPRB2019}. 
This has lead to several quasi-1D versions of the model which are numerically and analytically more accessible than the 2D model and, as it turns out, have very rich physics on their own. 

\subsection{Kitaev-$\Gamma$ chain}
We begin with the geometry of an isolated $xy$ chain (the case of $yz$ and $zx$ chains can be defined analogously), depicted as 
\be\label{eqfig:XYchain}
\includegraphics[width=2.65in]{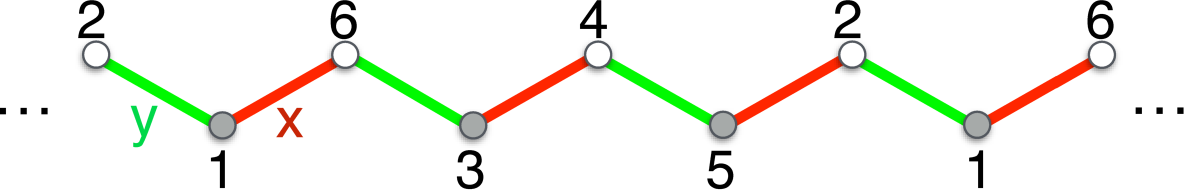}
\ee 
where, for later purposes, we have also labeled the sites according to the six-sublattice decomposition of (\ref{eqfig:T6transfhex}) along $xy$ chains.

\subsubsection{Symmetries and dualities}\label{sec:1Dsec:sym_and_dual}

Before we discuss the phase diagram of the 1D $K$-$\Gamma$ chain model, it is worth emphasizing that the symmetries and dualities of the 2D honeycomb model that we examined in Sec.~\ref{sec:sym_and_dual} (and in particular, the ones involving the $K$-$\Gamma$ model) carry over to the 1D chain as well. To see this, one simply needs to isolate an $xy$ chain from the honeycomb lattice, and, for each given transformation of Sec.~\ref{sec:sym_and_dual} (symmetry or duality),  remove any parts of the operation that pertain to sites not belonging to the given chain. For example, the local symmetries $W_h$ of Eq.~(\ref{eq:WhS1o2}) reduce to three-body operations involving three consecutive sites of the chain. More explicitly, for the first three consecutive sites of (\ref{eqfig:XYchain}), labeled by 2, 1 and 6, the local symmetry is
\be\label{eq:localsymm3sites}
\mathsf{C}_{2x}(2)\mathsf{C}_{2z}(1)\mathsf{C}_{2y}(6)\,.
\ee
Likewise, the four-sublattice transformation $\mc{T}_4'$ discussed in Sec.~\ref{sec:dualityKtomK} for the 2D model, which maps the FM to the AFM Kitaev point, is still present in the chain model, and its form is that of Eq.~(\ref{eq:T4ptransf}) if we keep the 4-sublattice labelling of (\ref{eqfig:T4transf}) for the chain.
Finally, the points $K\!=\!\Gamma$ are hidden SU(2) points with $\widetilde{J}\!=\!-K$, as in the 2D model. The associated transformation $\mc{T}_6$ has again the form of Eq.~(\ref{eq:T6transfhex}) when using the labelling of the six-site decomposition of (\ref{eqfig:XYchain}).

\begin{figure*}[t]
\centering
\includegraphics[width=0.97\textwidth]{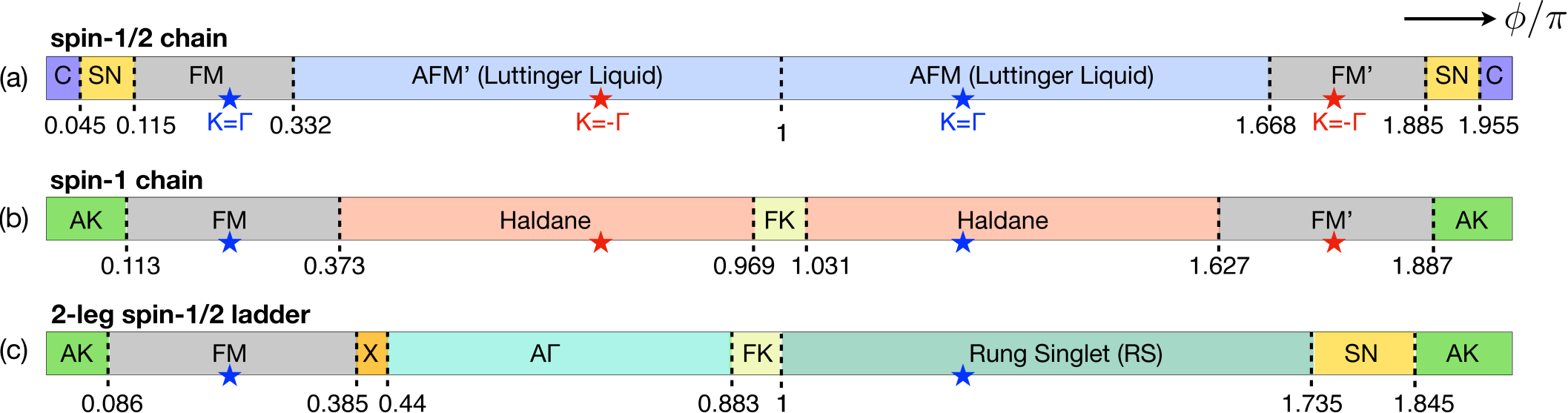}
\caption{Zero-temperature quantum  phase diagram of the  $K$-$\Gamma$ model on (a) spin-$1/2$ chain~\cite{Yang2020PRL,Yang2021PRB,Luo2021PRB}, (b) spin-1 chain~\cite{Luo2021PRR}, and (c) two-leg spin-1/2 ladder~\cite{Sorensen2021PRX}, with the parametrization $K=\cos\phi$ and $\Gamma=\sin\phi$. Blue stars show the positions of the hidden SU(2) symmetry points with $K\!=\!\Gamma$, whereas red stars in (a-b) show the positions of the hidden SU(2) points with $K\!=\!-\Gamma$, see text.}\label{fig:KGChnLdr}
\end{figure*}

Altogether then, the dualities and symmetries of the planar 2D model carry over to the 1D chain model in a straightforward way. 
What is not immediately evident, however, is that the chain model has, in fact, some additional symmetries and dualities compared to the 2D model. We will discuss the three most notable examples: 

i) At the pure Kitaev point, the chain has local symmetries defined on bonds~\cite{Sen2010}. These correspond to $\pi$-rotations (of the two spins of the bond) around the ${\bf x}$-axis (${\bf y}$-axis) for a $y$-type (respectively, $x$-type) of bond. For the bonds (1,2) and (1, 6) of (\ref{eqfig:XYchain}), for example, the corresponding local symmetries take the form 
\be\label{eq:localsymmbonds}
\mathsf{C}_{2x}(2)\mathsf{C}_{2x}(1)~~\text{and}~~
\mathsf{C}_{2y}(1)\mathsf{C}_{2y}(6)\,,
\ee
respectively. We can now recognize that the local 3-site symmetries of Eq.~(\ref{eq:localsymm3sites}) are the product of the bond-symmetries of (\ref{eq:localsymmbonds}).
As shown by Sen {\it et al.}~\cite{Sen2010}, the presence of these bond symmetries gives rise to a qualitative difference between integer and half-odd-integer values of $S$, which stems from the fact that for integer $S$, all the bond symmetries commute with each other, but for half-odd-integer $S$ this is not true any longer. One of the immediate consequences is that all the eigenstates of the chain with half-odd-integer $S$ are $2^{N/2}$-fold degenerate.

ii) Compared to the 2D model, the chain model has an additional pair of hidden SU(2) points, at $K\!=\!-\Gamma$ (with $\widetilde{J}\!=\!-K$)~\cite{Ducatman2018,Yang2020PRL}. For the site-labelling of (\ref{eqfig:XYchain}), the associated transformation $\mc{T}_6'$ reads:  
\be\label{eq:6subM}
\begin{array}{c}
\text{6-sublattice}\\
\text{transf.}~\mc{T}_6'
\end{array}:~~
\renewcommand{\arraystretch}{1.5}
\begin{array}{c |  rrr}
\text{label}~j & S_j^x & S_j^y & S_j^z \\
\hline
1 & \widetilde{S}_1^x& \widetilde{S}_1^y&\widetilde{S}_1^z \\
2 &\widetilde{S}_2^z&-\widetilde{S}_2^y& \widetilde{S}_2^x\\
3 &\widetilde{S}_3^y&-\widetilde{S}_3^z&-\widetilde{S}_3^x \\
4 &-\widetilde{S}_4^y&-\widetilde{S}_4^x&-\widetilde{S}_4^z\\
5 &-\widetilde{S}_5^z&\widetilde{S}_5^x&-\widetilde{S}_5^y\\
6 &-\widetilde{S}_6^x&\widetilde{S}_6^z&\widetilde{S}_6^y\\
\hline
\end{array}
\ee
Indeed, taking, e.g., the $y$-bond $(1,2)$ of (\ref{eqfig:XYchain}) and using (\ref{eq:6subM}) with $J\!=\!\Gamma'\!=\!0$ and $K\!=\!-\Gamma$ gives
\be
K S_1^y S_2^y + \Gamma (S_1^x S_2^z+S_1^z S_2^x) \to
-K\widetilde{{\bf S}}_1\cdot\widetilde{{\bf S}}_2 \,,
\ee
which is again a Heisenberg coupling with $\widetilde{J}\!=\!-K$. 
Altogether then, it follows that when $K$ is positive, the points $K\!=\!\Gamma$ and $K\!=\!-\Gamma$ both map to a FM Heisenberg model, and must therefore show FM ordering at low $T$.
By contrast, when $K$ is negative, the points $K\!=\!\Gamma$ and $K\!=\!-\Gamma$ both map to the AFM Heisenberg model, and we therefore get a gapless critical phase~\cite{Bethe1931}.

iii) The chain model has an additional duality that maps $(K,\Gamma)\to(K,-\Gamma)$. The associated operation is the global $\pi$-rotation around the $z$ axis in spin space alone. This operation maps $(S_i^x,S_i^y,S_i^z)\!\mapsto\!(-S_i^x,-S_i^y,S_i^z)$ for all sites $i$, which effectively reverses the sign of $\Gamma$. Thus, the model with $K$ and $\Gamma$ of opposite signs can be mapped to the model with the same signs. This duality, which can also result from the three-sublattice transformation $\mc{T}_3$ of Ref.~\cite{Yang2020PRL}, is not present in the 2D model.

\subsubsection{Phase diagram for $S=1/2$}
We begin with the physics of the pure Kitaev points, which goes back to the study of Ref.~\cite{Feng2007PRL}. As for the 2D and 3D cases, these points can be mapped to free fermion models via a Jordan-Wigner transformation~\cite{Feng2007PRL}, and it has been shown that the ground states are critical (gapless) phases. In the anisotropic version of the model with different couplings on the two types of bonds (i.e., with $K_x\!\neq\!K_y$), a nonzero gap opens up as soon as we depart from the special point $K_x\!=\!K_y$. The transition between the two gapped phases on the two sides of this critical point does not involve any change of symmetry, but is an example of a change from one topological order to another, each associated with a separate hidden nonlocal order parameter~\cite{Feng2007PRL}.

When the $\Gamma$ coupling is introduced, the Jordan-Wigner transformation is no longer useful, but the model has been studied using non-Abelian bosonization and DMRG techniques~\cite{Yang2020PRL,Yang2021PRB,Luo2021PRB}. The resulting phase diagram is shown in Fig.~\ref{fig:KGChnLdr}\,(a), with the parametrization $K\!=\!\cos\phi$ and $\Gamma\!=\!\sin\phi$. Note that due to the duality which maps $(K,\Gamma)\mapsto (K,-\Gamma)$, the points $\phi$ and $2\pi-\phi$ are dual to each other. 
In total, there are seven extended phases:

i) There are two (dual) FM orders, denoted by FM and FM' in Fig.~\ref{fig:KGChnLdr}\,(a), surrounding the two hidden SU(2) points $K\!=\!\pm \Gamma$ with positive $K$. These orders are gapped everywhere except at the hidden points themselves, where the ordering breaks a continuous symmetry, leading to Goldstone modes. The ordered phase is characterized by the O$_h \rightarrow$ D$_4$ symmetry breaking, where O$_h$ is the full octahedral group and D$_n$ represents the dihedral group of order 2$n$. This phase has a six-fold degenerate ground state, with moments along the $\pm\hat{x}$, $\pm\hat{y}$, and $\pm\hat{z}$~\cite{Yang2020PRL}.

ii) The phases surrounding the two hidden AFM SU(2) points ($K\!=\!\pm \Gamma$, with negative $K$), remain gapless in extended regions of parameter space [denoted by AFM and AFM' in Fig.~\ref{fig:KGChnLdr}\,(a)] and feature an emergent SU(2)$_1$ symmetry. The pure $\Gamma$ points are special members of these extended (dual) phases, and can be described as Luttinger liquids with central charge 1.  

iii) While the FM and AFM mapping is valid in the Kitaev-$\Gamma$ chain, their stability under a finite $\Gamma$ term is very different. As shown in the phase diagram, the FM Kitaev point is immediately unstable and becomes a critical point as soon as $\Gamma$ is introduced. On the other hand, the AFM Kitaev phase (C) remains stable for weak enough $\Gamma$. 

iv) Finally, there are two (dual) magnetic orders, denoted by SN in Fig.~\ref{fig:KGChnLdr}\,(a), which are sandwiched between the AFM Kitaev phase (C) and the FM (FM') phases. These orders are characterized by two types of modulations, with 2-site and 6-site periodicities~\cite{Yang2021PRB}. A finite spin nematicity (SN) on a NN bond is another signature that differentiates these phases from their surrounding AFM Kitaev and FM orders.

\subsubsection{Integer spins: Spin-1 $K$-$\Gamma$ model} 

Quasi-1D models of higher spin have been also studied~\cite{Sen2010,Luo2021PRR}. While the spin-1 Kitaev chain is not exactly solvable, numerical works show that the ground state is translationally invariant and lives in the `flux-free' sector, where all mutually-commuting bond symmetries of Eq.~(\ref{eq:localsymmbonds}) equal to $+1$~\cite{Sen2010,Luo2021PRR}. Compared to the ground state, the first excited states feature one bond with flux $-1$, leading to an $N$-fold degeneracy, where $N$ is the number of bonds~\cite{Luo2021PRR}. 
Due to the unusual excitations, the Kitaev phase displays a nontrivial double-peak structure in the specific heat.  
In contrast to the spin-$1/2$ Kitaev honeycomb model, whose low-temperature peak comes from the freezing of vison fluxes,
the low-temperature peak in the spin-1 Kitaev chain is related to the freezing of $\mathbb{Z}_2$ degrees of freedom and is relevant to the highly degenerate low-lying excited states.
Moreover, the thermal entropy is released gradually without generating a plateau in the crossover temperature region, which is again different from the Kitaev honeycomb model.

The introduction of the $\Gamma$ coupling leads to a rich phase diagram which contains six phases, see Fig.~\ref{fig:KGChnLdr}~(b). These phases include: i) a gapped FM Kitaev (FK) phase and a gapped AFM Kitaev phase (AK), ii) two FM phases (FM and FM') surrounding the two hidden FM SU(2) points, and iii) two Haldane phases surrounding the two hidden AFM SU(2) points, which are gapped, unlike the spin-1/2 case.

\subsection{Spin-1/2 Kitaev-$\Gamma$ ladder}
While the Kitaev-$\Gamma$ chain highlights the interplay of these two interactions leading to both ordered 
and disordered phases, the link to the physics of the 2D Kitaev-$\Gamma$ model is weak. The difference  between the Kitaev-$\Gamma$ chain  behavior and numerical results  obtained for the 2D Kitaev-$\Gamma$ model
is obviously from the missing $z$-bond, and one may wonder if the minimal geometry such as that of the 2-leg Kitaev-$\Gamma$ ladder, depicted as
\be\label{eqfig:Ladder}
\includegraphics[width=3.00in]{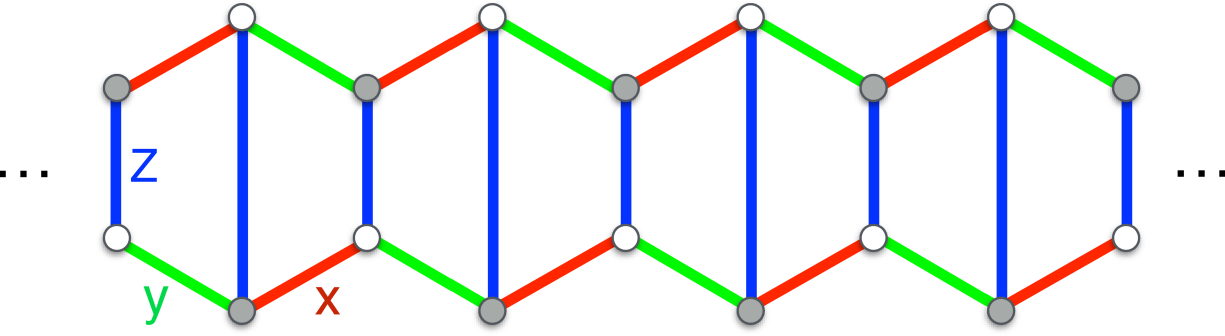}
\ee
may capture the phases found in the 2D model.
This geometry has been the subject of several studies with Kitaev-like models~\cite{Feng2007PRL,WU2012, CatuneanuPRB2019,Gordon2019NC,Agrapidis2019,Metavitsiadis2021,Sorensen2021PRX,YangAffleck2022,ChenPRB2023}.

\subsubsection{Symmetries and dualities}

At the pure Kitaev point, the local plaquette symmetries of the 2D model [see Eq.~(\ref{eq:WhS1o2}) for spin-1/2 or its extension in Eq.~(\ref{eq:WhhighS}) for higher spins] are still present in the 2-leg ladder. Furthermore, the hidden SU(2) points at $K\!=\! \Gamma$ remain present in the parameter space of the ladder, see blue stars in the phase diagram  presented in Fig.~\ref{fig:KGChnLdr}\,(c).  Contrary to the chain, however, the $T_6^\prime$ mapping of Eq.~(\ref{eq:6subM}) is lost in the ladder geometry due to the presence of the $z$-bonds that link the two chains of the ladder. As a result, the points $K\!=\!-\Gamma$ are {\it not} hidden SU(2) points. The same situation  is encountered in the $K$-$\Gamma$ honeycomb model. Similarly, the duality that maps $(K,\Gamma)$ to $(K,-\Gamma)$ in the chain model is {\it not} present in the 2-leg ladder. Therefore the points $\phi$ and $2\pi-\phi$ are not dual.

\subsubsection{Phase diagram} 
The phase diagram of the spin-1/2 $K$-$\Gamma$ ladder is shown in Fig.~\ref{fig:KGChnLdr}\,(c), as obtained by numerical studies~\cite{Gordon2019NC,Sorensen2021PRX}. In total, there are seven extended phases, with the following main features: 

i) The pure Kitaev ladder at $\phi\!=\!0$ and $\pi$ is exactly solvable, and at both points it is in a disordered gapped phase~\cite{Feng2007PRL}, featuring a long-range nonlocal string order parameter~\cite{CatuneanuPRB2019}.
The Kitaev phases survive for small enough $\Gamma$, leading to the extended phases denoted by AK (AFM Kitaev) near $K = 1$ and $\Gamma=0$ and FK (FM Kiteav) near $K=-1$ and $\Gamma=0$ in Fig.~\ref{fig:KGChnLdr}\,(c).

ii) The hidden SU(2) point at $\phi\!=\!\pi/4$ maps to the spin-1/2 Heisenberg ladder with $\widetilde{J}\!=\!-K\!<\!0$, which explains the FM order surrounding this point. 
Likewise, the second hidden SU(2) point at $\phi=5\pi/4$ maps to the AFM spin-1/2 Heisenberg ladder, which gives rise to a rung singlet (RS) phase surrounding this point. This phase is gapped with short-range correlations~\cite{Dagotto1992,Barnes1993}. 
Note that, in the 2D limit, the FM and RS phases will eventually give way to the FM and N\'eel phases, respectively, whose stability regions are outside the parameter space of Fig.~\ref{fig:KGphasediagram}.

iii) Near $\Gamma\!=\!1$ and $K\!=\!0$, there is a disordered phase denoted by A$\Gamma$ (AFM $\Gamma$) in Fig.~\ref{fig:KGChnLdr}~(c). This phase  shows similarities to the nematic paramagnet that has been predicted by iDMRG~\cite{Gohlke2020PRR} for the 2D model, see Fig.~\ref{fig:KGphasediagram}\,(e). 

iv) There is a spin-nematic phase denoted by SN in Fig.~\ref{fig:KGChnLdr}\,(c), which is sandwiched between the RS and the AFM Kitaev phase. This phases features local quadrupolar order parameters and is likely gapless~\cite{Sorensen2021PRX}. 

v) Finally, there is a narrow phase, denoted by X in Fig.~\ref{fig:KGChnLdr}\,(c), which is sandwiched between the FM and A$\Gamma$ phases. A recent DMRG study has reported two symmetry-protected topological (SPT) phases inside X~\cite{Sorensen2023arXiv}.

\subsection{Other developments and phenomena in 1D}
Given the numerical accessibility of 1D models, a variety of modified 1D models have been explored. Examples include an extension of the pure Kitaev chain to bond-alternating Kitaev-$\Gamma$ chain for $S=1/2$ and $S=1$~\cite{Luo2021PRB,Luo2021PRR}. Similar to the spin-1/2 and spin-1 chains, these studies show  clear differences in the ground states between half-integer and integer spin chain systems. 
Another line of investigations concerns the  response of these 1D models to various perturbations, such as external magnetic fields~\cite{Sun2009,Wang2010,You2018,Wu2019,Sorensen2021PRX,Sorensen2023PRRl}. In the following, we briefly review two notable examples.

\subsubsection{Field-induced chiral solitons in Kitaev chains}
 
The influence of magnetic fields on the spin-1/2 Kitaev chain has been explored in several works~\cite{Sun2009,Wang2010,You2018,Wu2019,Sorensen2023PRRl}.
Here we shall briefly expand on the field-induced chiral soliton phase reported in Ref.~\cite{Sorensen2023PRRl}.
Starting from the Kitaev limit at zero field, a phase with finite vector spin chirality can be induced by an in-plane magnetic field along the [110] direction. For $S\!=\!1/2$, the ground state evolves continuously and  remains gapless up to a critical field $h_{c_1}$, but  enters into the polarized phase only above a critical field $h_{c_2}$, leaving an intermediate phase in between. This phase  is characterised by a nonzero vector chirality of the staggered form  $(-1)^j\langle\left({\bf S}_j\times{\bf S}_{j+1}\right)^z\rangle$, similar to the phase found in the context of the anisotropic $J_1$-$J_2$ model with $J_1\!<\!0$ and $J_2\!>\!0$~\cite{Furukawa2008,Furukawa2010,Furukawa2012} and observed recently in the spin-$1/2$ chain compound LiCuVO$_4$~\cite{Grams2022}. For periodic boundary conditions, this phase features a finite excitation gap above the two-fold degenerate ground state. For open boundary conditions, the ground state exhibits a single soliton, lowering the energy, and in-gap excitations. Such a soliton can also be revealed by a twist in the very middle of the magnetization~\cite{Sorensen2023PRRl}.

It is noteworthy that the chiral soliton phase can also appear in the integer-spin Kitaev chains~ \cite{Sorensen2023PRRa}. The ground states of the integer-spin Kitaev chains, at least for  $S\leq4$, are gapped. By adding an in-plane magnetic field, these excitation gaps vanish at certain magnetic fields that are smaller than $\sqrt{2}S K$. Hence, the chiral soliton phase is stabilized in these regions around $h=\sqrt{2}S K$. Finally, we mention that although the chiral soliton phase is easily accessible when the magnetic field is along the [110] direction, it can also exist if the in-plane azimuth angle is slightly away from $45^{\circ}$, and/or the field slightly deviate from the $xy$ plane~ \cite{Sorensen2023PRRa}.

\begin{figure}[!t]
\includegraphics[width=0.995\linewidth]{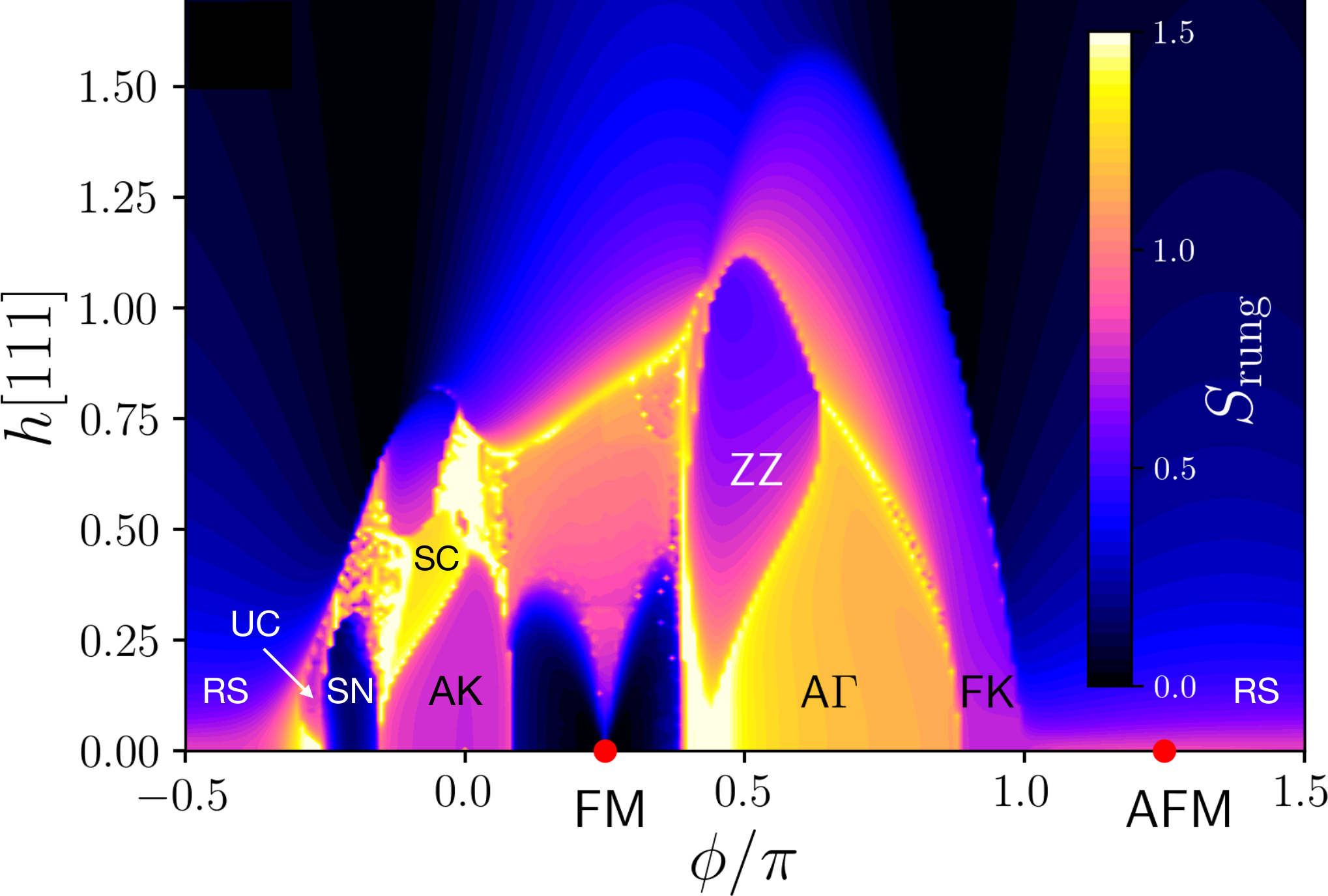}
\caption{Field-induced phase diagram of the spin-1/2 two-leg $K\Gamma$ ladder versus ${\bf h}\parallel[111]$ (where $h=H/(g\mu_B)$ and $H$ is the applied field) and $\phi/\pi$, where $K\!=\!\cos\phi$ and $\Gamma\!=\!\sin\phi$, as obtained  from iDMRG calculations~\cite{Sorensen2021PRX}. The colour-coded intensity corresponds to the bipartite von Neumann entanglement entropy $S_{\text{rung}}$ on the rung. Adapted from Ref.~\cite{Sorensen2021PRX}.}\label{fig:KG_ladder_field}
\end{figure}

\subsubsection{Field-induced scalar chiral orders in $K\Gamma$ ladders}

The entire phase diagram of the spin-1/2 two-leg $K\Gamma$ ladder in the [111] magnetic field has been mapped out by the iDMRG and DMRG methods~\cite{Sorensen2021PRX}, motivated by the rich magnetic field effects in Kitaev materials. The system features a multitude of distinct orders with multicritical points, incommensurate phases, as well as disordered states characterized by various scalar chiralities, see Fig.~\ref{fig:KG_ladder_field} and Figs.~4 and 5 of Ref.~\cite{Sorensen2021PRX}. Throughout the phase diagram, the most exciting parts are the dominant positive $\Gamma$ regime (A$\Gamma$) perturbed by FM Kitaev, and the dominant AFM Kitaev (AK) regime perturbed by a negative $\Gamma$ interaction. The competition between FK and A$\Gamma$ under the [111] magnetic field leads to FK phase sitting above A$\Gamma$ phase leading to a field induced FK phase. However, such field-induced FK phase region shrinks as the number of legs grows~\cite{Gohlke2020PRR}. While A$\Gamma$ is robust under the [111] field, it is replaced by the ZZ order when a small negative $\Gamma^\prime$ is introduced, which has been also found in the honeycomb model~\cite{Rau2014-arxiv}.

\begin{figure}[!t]
\centering
\includegraphics[width=0.95\columnwidth, clip]{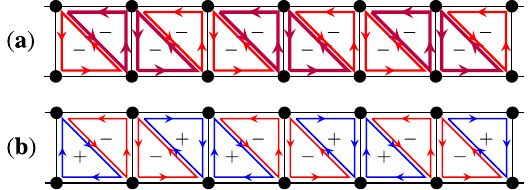}\\
\caption{{Sketch of the scalar chirality in the (a) uniform chirality (UC) case and the (b) staggered chirality (SC) case. The width of the line indicates the stength of the scalar chirality, while the color represents the sign. }
} \label{FIG-Bond}
\end{figure}

Among the phases identified in Ref.~\cite{Sorensen2021PRX}, two scalar chirality orders are intriguing, as they may be connected to the magnetic vortex liquid-like phases found in Ref.~\cite{Chern2021npj} around the similar phase space in 2D classical Kitaev-$\Gamma$ model. This is near the AFM Kitaev regime, where the gapless quantum spin liquid under the magnetic field was also suggested~\cite{Hickey2019NC}. By increasing the magnetic field, there appear two scalar chirality orders. The scalar spin chirality  is defined on triangular plaquettes in anticlockwise direction, as shown in Fig.~\ref{FIG-Bond}. More explicitly, for the lower and upper triangles involving the rungs $r$ and $r+1$, the corresponding chiralities $\kappa_1(r)$ and $\kappa_2(r)$ take the form 
\be\label{EQ:ChiDef}
\renewcommand{\arraystretch}{1.5}
\begin{array}{l}
\kappa_1(r) = \bs{\hat\sigma}_{r,2} \cdot \big(\bs{\hat\sigma}_{r,1} \times \bs{\hat\sigma}_{r+1,1} \big),\\
\kappa_2(r) = \bs{\hat\sigma}_{r,2}\cdot \big(\bs{\hat\sigma}_{r+1,1} \times \bs{\hat\sigma}_{r+1,2}\big)\,,
\end{array}
\ee
where $\mathbf{\hat{S}}_{r,\ell} \!=\!\boldsymbol{\hat\sigma}_{r,\ell}/2$ corresponds to the spin residing at the $r$-th rung and $\ell$-th leg, and where $\ell\!=\!1$ (2) denoting the lower (upper) leg. 
To measure the intensity of the chirality, one introduces the chiral-chiral correlation function between two triangles separated by distance $d$,  $C_{\kappa_\upsilon}(d) = \langle\kappa_{\upsilon}(r)\kappa_{\upsilon}(r+d)\rangle$, such that $C_{\kappa_\upsilon}(d)\!\to\!\kappa_\upsilon^2$ when $|d|\!\gg\!1$. The magnitude of the chirality $\kappa_\upsilon$ is extracted by calculating $C_{\kappa_\upsilon}(d)$ in the large distance limit and its sign is given by the corresponding local expectation value. The resulting pattern of $\kappa_\upsilon$ at two representative points of the phase diagram, $(\phi,h)=(1.73\pi, 0.125)$ and $(1.92\pi, 0.42)$, are shown in Fig.~\ref{FIG-Bond}. In the two panels, the width of the arrows in each triangle indicates the magnitude of the chirality, and the color of the triangle represents its sign. In the upper panel, the scalar chirality is negative in each triangle, and the magnitude shows period-2 modulation, reflecting the period of the Hamiltonian. By contrast, in the lower panel, the sign of chirality shows an alternating staggered pattern but the magnitude is uniform. Since the summation of the chirality within each unit cell is zero,
the net flux in the ladder is zero, despite the presence of the external field.

In summary, the broader picture emerging from the above body of results on quasi-1D geometries is that the interplay of Kitaev and $\Gamma$ interactions on these geometries gives rise to a wide range of new possibilities. Some of the phases are generic to 1D, like the critical Luttinger-Liquids for spin-1/2, or the Haldane-like phases for spin-1. Other phases, seem to be the precursors of particular orders in the 2D model. Examples include the FM and RS phases of Fig.~\ref{fig:KGChnLdr}~(c), which give way to the FM and N\'eel phases of the 2D model; the disordered A$\Gamma$ phase of Fig.~\ref{fig:KGChnLdr}~(c) which shows some similarities to the nematic paramagnet of  Fig.~\ref{fig:KGphasediagram}\,(e)~\cite{Gohlke2020PRR}; and the two field-induced scalar chirality orders of Fig.~\ref{FIG-Bond} (denoted by UC (uniformed chirality) and SC (staggered chirality) in Fig.~\ref{fig:KG_ladder_field}),  which seem to be connected to the magnetic vortex liquid-like phases found in Ref.~\cite{Chern2021npj} in the 2D model. 
Finally, there are phases whose connection to the 2D lattice model is not clear, including the SN phase of Fig.~\ref{fig:KGChnLdr}~(c), and the two SPT phases inside the region denoted by X in Fig.~\ref{fig:KGChnLdr}~(c). 
Further systematic work on $n$-leg ladders with $n>2$ can shed more light into this possible correspondence and eventually to the nature of the enigmatic phases of the 2D $K\Gamma$ model.  
In terms of materials, 1D Kitaev systems are still lacking, apart from the recently suggested twisted Kitaev chain~\cite{morris2021np}. Developing material platforms to explore the physics of quasi-1D models is a promising endeavour.

\section{Summary and discussion}\label{sec:Discussion}
The present review aims to complement previous review articles on various aspects of the broader activity in the field of Kitaev materials~\cite{Rau2016ARCMP,Winter2017,Knolle2017ARCMP,Takagi2019NRP,Motome2019JPSJ,Janssen2019,Trebst2022,Tsirlin2022}, and focuses primarily on the theoretical understanding of how other interactions intertwine with the Kitaev coupling $K$ and what physics to expect beyond the Kitaev limit. 

To expose the relevance of these additional interactions, we begun with a comprehensive overview of the microscopic exchange mechanisms that are generically present in strong spin-orbit coupled honeycomb Mott insulators. 
This led us to the $J$-$K$-$\Gamma$-$\Gamma'$ model as a minimal description of many Kitaev materials, and the identification of the $\Gamma$ coupling as the main source of additional frustration besides the Kitaev coupling. 

We then set out to explore the various phases in this extended  four-dimensional parameter space. To that end, we provided a detailed discussion of a number of special symmetries and duality transformations, which play a crucial role in mapping out the most interesting (and oftentimes hidden) regions of the parameter space. 
We have organized our exploration of this space by reviewing the physics along cuts that connect these special regions. This led us to a number of `simpler' models that have attracted a lot of interest in the community, either due to their close relevance to existing materials or due to their rich physics.

One such region that stands out is the line in the parameter space connecting a pure Kitaev point to a pure $\Gamma$ point, where one encounters a novel interplay between the Kitaev quantum spin liquid and the classical $\Gamma$ spin liquid. 
In particular, we highlighted the region corresponding to negative $K$ and positive $\Gamma$ as the one that is most relevant to existing materials, but also the one that is most challenging to tackle from the theory and numerical side, as this is where the two types of spin liquids compete the most. 
We reviewed the current debate on the nature of the ground state and the phases proposed by various state-of-the-art methods. 
Altogether, despite intensive studies, the ground state for $K\!<\!0$ and $\Gamma\!>\!0$ is currently unknown and remains a challenging question. Although several numerical studies have suggested a magnetically disordered state, most of them
are limited by small cluster sizes or wave function ans{\"a}tze. Developing new numerical techniques could potentially expand our current understanding of this region. 

We have also collected complementary insights from studies of the classical $K\Gamma$ model and highlighted the emergence of `period-three chain' orders with counter-rotating spin sublattices, or long-distance modulations thereof, as one of the key ramifications of the interplay between $K$ and $\Gamma$. 
These orders  have been observed in the three members of the {\liiro} family of 3D polymorphs, for which we have provided a brief overview of the most relevant results and phenomenology. 

Finally, we have briefly reviewed some of the recent exciting results on quasi-1D versions of Kitaev models, which are motivated by the idea that one can approach the 2D lattice by bridging together 1D chains or ladders. Part of the emerging phenomenology is specific to 1D, such as the presence of critical Luttinger-Liquid phases in spin-1/2 chains,  Haldane-like phases in spin-1 chains, or the rung singlet phase in spin-1/2 ladders. Other exciting results in these quasi-1D models include the stabilization of field-induced chiral soliton phases in spin-1/2 Kitaev chains and the scalar chiral orders reported for the spin-1/2 $K$-$\Gamma$ ladder.

The theoretical exploration of the phase diagram of Kitaev materials continues to be strongly driven by discoveries of new materials and the application of state-of-the-art experimental techniques that have been designed for detecting signatures of Kitaev physics.
On the material front, {\rucl} stands center stage due to the report of the half-integer thermal Hall conductivity under an in-plane magnetic field~\cite{kasahara2018thermal,Yamashita2020PRB,Bruin2021,yokoi2021half,kasahara2022quantized,imamura2023majorana}. However, there are debates on whether the measured thermal Hall conductivity can be attributed to chiral magnons~\cite{Zhang2021PRB,Czajka2023NM} or phonons~\cite{Hentrich2019PRB,Lefran2022PRX}. 
The experiments of the longitudinal thermal conductivity, which have shown oscillations as a function of in-plane magnetic field have generated new excitement on {\rucl}~\cite{Czajka2021}. While these oscillations were initially interpreted as quantum oscillations of a gapless spin liquid, akin to Landau levels of gapless metals~\cite{Czajka2021}, later experiments attributed the oscillations to anomalies associated with magnetic transitions~\cite{Lefran2023PRB}. 
Various numerical studies have uncovered an intermediate phase between the magnetic ordered ground state and the polarized paramagnetic state when the field is out of the honeycomb plane~\cite{Gordon2019NC,Lee2020NC}. However, when the field is within the plane, a direct transition from the magnetically ordered phase to the polarized state was reported~\cite{Winter2018PRL,Gordon2019NC}. Further experimental and theoretical studies are needed to confirm or disprove the existence of a field-induced spin liquid in {\rucl}.

Beyond {\rucl} and $d^5$ Iridates, there have been intense studies on $d^7$ honeycomb cobaltates~\cite{Liu2018PRB,Liu2020PRL,Sano2018PRB,Songvilay2020PRB,Janssen2023,Yao2020PRB,Vivanco2020PRB,Ruidan2020SA,Kim2021,Chen2021PRB,Zhang2021arXiv,Samarakoon2021PRB,Sanders2022PRB,Kim2022JPCM,Yang2022PRB,Halloran2023,miao2023persistent,Regnault1977}. There are also several other candidate materials proposed, ranging from quasi-1D systems and twisted Kitaev chains~\cite{morris2021np} to f-electron honeycomb materials\cite{Jang2019PRB,Jang2020PRM,Motome2020}. Further experiments and microscopic theories on these candidates need to be explored.

In parallel to the rapid developments in bulk materials, the experimental community has been actively pursuing the development of various two-dimensional (2D) single layer compounds. Among the Kitaev candidate materials, in Na$_2$IrO$_3$ and Li$_2$IrO$_3$, the presence of Na$^+$ and Li$^+$ ions in the interlayer space introduced unwanted interlayer interactions, which could potentially disturb the desired Kitaev physics in 2D systems. In the case of RuCl$_3$, the interlayer interactions are relatively smaller compared to honeycomb iridates due to the absence of charged ions between the layers, but they are still present and not negligible. Consequently, significant efforts have been made to control and mitigate these interlayer couplings. One approach to achieve a 2D system involves the exfoliation of 2D honeycomb compounds into monolayers~\cite{Weber2016Nano}, allowing for better isolation and manipulation of the desired properties.  Among several measurements, Raman spectroscopy performed to study the evolution of structure with thickness and temperature revealed a symmetry-forbidden mode appearing in only the thinnest flakes at low temperature~\cite{Zhou2019JPCS}. This evidences a possible structural transformation. It is likely that the exfoliation process or substrates generate significant strain leading to the distorted structure. To advance our understanding of the effects of interlayer couplings and various substrates in such van der Waals Kitaev candidates, further studies the role of the structure and the substrate remain to be carried out in future.

On the computational side, exploring the `physics beyond the Kitaev point' that we discuss in this review have been a testing ground for a  broad range of state-of-the-art computational techniques, including {\it ab initio} methods. 
Indeed, the inherent complexity of the materials, the interplay of several degrees of freedom (including correlations, crystal field, spin-orbit coupling, and their sensitivity to structural details), as well as the presence of strong frustration, and the necessity to treat quantum and thermal fluctuations on equal footing renders the modelling of these materials a challenging task.  

As discussed in the introduction, the determination of the right starting minimal model requires using {\it ab initio} methods in tandem with careful analysis of the experimental data based on models, which is quite nontrivial given the large number of microscopic parameters. 

Characterizing the physics of simpler generic models (such as the $K\Gamma$ model) plays a crucial role in mapping out distinctive regions of the broader multi-dimensional phase diagram and building a phenomenology that can then be contrasted with experiment.
To that end, a number of many-body techniques have been employed over the years. Large-scale exact diagonalizations have the advantage of exploiting symmetries to analyze the low-energy part of the many-body spectrum, but the exponential growth of the basis with system size (and the generic lack of continuous spin rotational symmetry in spin-orbit models) imposes severe constraints on the size of finite clusters, typically up to 36 sites. The DMRG method is a powerful tool to target the ground state and the low-lying excited states by mapping a long cylinder into a snake-like chain with more than two hundred sites. However, the width of the cylinder is typically less than ten lattice sites, and the convergence is slow when the ground state is gapless or approaching a quantum critical point. Although the VMC method can achieve a balance of the lattice sites along two sides, it is intrinsically a variational iteration over a mean-field Hamiltonian and relies highly on the proposed trial wave functions. In contrast to these methods mentioned above, the tensor network approach is claimed to solve the model with an infinite number of lattice sites via wave function ansatz like infinite projected entangled pair state. Nevertheless, it may suffer from an artificial optimization route towards the ground state and a fast-growing computational cost when considering further-neighbor interactions and/or large unit cell.
Finally, classical minimizations and semiclassical $1/S$ expansions are often the natural starting point for the description of the ground state properties and low-energy excitations, as most (if not all) materials exhibit magnetic ordering at low enough temperatures. However, these approaches may break down at intermediate and higher energy scales due to strong magnon-magnon interactions and decay processes, which can be strongly amplified when these orders are selected by weak perturbations out of a nearby highly frustrated point.  
While every numerical technique comes with their own weaknesses and limitations, a convincing conclusion can be reached when results from different methods are contrasted and analyzed.

We hope this review will provide a valuable contribution to the current literature on Kitaev materials, and set the basic phenomenology and some useful baselines for researchers working in this flourishing field. We have also exposed the current challenges in the field but also the rich opportunities for exploring novel physics in these materials.

\acknowledgements 
We would like to acknowledge our theory and experimental colleagues from the broader field of Kitaev magnetism all over the world, whose work and insights have been inspirational over the years. In particular, we would like to acknowledge J. G. Analytis, C. Batista, D. Bogdanov, J. van den Brink, C. Broholm, B. B\"uchner, K. S. Burch, A. Catuneanu, G.-W. Chern, A. L. Chernyshev, M. Daghofer, S. Ducatman,  A. Frano, J. Gordon, T. Halloran, L. Hozoi, G. Jackeli, V. Katukuri, G. Khaliullin,  Y. B. Kim, Y. J. Kim, J. Knolle, S. Kourtis, M. Li, R. Moessner, S. Rachel, J. Rau, J. Reuther, U. K. R\"o{\ss}ler, A. Ruiz, Y. Sizyuk, E. S{\o}rensen, P. P. Stavropoulos, H. Takagi, R. Thomale, A. A. Tsirlin, M. Vojta, X. Wang, Y. Yang, and J. Zhao. Finally, we also like to thank our funding agencies: 
IR acknowledges the support by the Engineering and Physical Sciences Research Council, Grant No. EP/V038281/1. NBP acknowledges
the support  by the U.S. Department of Energy, Office of Science, Basic Energy Sciences under Award No. DE-SC0018056. NBP also acknowledge the support of the Alexander von Humboldt Foundation and the hospitality of the Technical University of Munich – Institute for Advanced Study. Additionally, IR and NBP acknowledge the support by the National Science Foundation under Grant No. NSF PHY-1748958 and the hospitality of KITP, UC Santa Barbara, where this work has been finalized. 
QL acknowledges the support from the Natural Science Foundation of Jiangsu Province Grant No. BK20220876
and the National Natural Science Foundation of China Grants No. 12247183 and No. 12304176.
HYK acknowledges the support from the Natural Sciences and Engineering Research Council of Canada Discovery Grant No. 2022-04601, the Canadian Institute for Advanced Research, and the Canada Research Chairs Program. HYK also acknowledges the Aspen Center for Physics, supported by National Science Foundation grant PHY-2210452, where part of this work was carried out.

\appendix
\section*{Appendix}

\subsection{Sign structure of off-diagonal couplings}\label{app:GGpSignStructure} 
Here we return to the issue raised at the end of Sec.~\ref{sec:EffSuperHam} regarding the sign structures of $\Gamma$ and $\Gamma'$ interactions and their dependence on the convention used for the directions of the cubic axes $x$, $y$ and $z$. We know that these axes are perpendicular to the corresponding type of bonds, however their orientation with respect to the crystallographic axes is defined only up to an overall $\pm1$ sign. In Fig.~\ref{fig:geometry}, for example, once we have identified the $x$ and $y$ type of bonds with respect to the crystallographic axes $(a,b,c^\ast)$ of the monoclinic cell, we can choose $(\tilde{x},\tilde{y},\tilde{z})\!=\!(\epsilon_x x,\epsilon_y y,\epsilon_x\epsilon_y z)$, where $\epsilon_{x,y}\!=\!\pm1$ are arbitrary, as the cubic frame instead of the $(xyz)$ frame shown. 

While the choice of $\epsilon_x$ and $\epsilon_y$ may seem innocuous, it can lead to quite different sign structure for the $\Gamma$ and $\Gamma'$ couplings across different bonds. 
Taking the three-fold symmetric honeycomb case as an example, the cubic frame shown in Fig.~\ref{fig:geometry} ($\epsilon_{x}\!=\!\epsilon_y\!=\!1$) features uniform signs of $\Gamma$ across all bonds, 
\be
\Gamma_x\!=\!\Gamma_y\!=\Gamma_z\!\equiv\!\Gamma,
\ee 
which is quite naturally desirable for this three-fold symmetric case. However, the choice $(\tilde{x},\tilde{y},\tilde{z})\!=\!(-x,y,-z)$, which corresponds to  $\epsilon_x\!=\!-1$ and $\epsilon_y\!=\!1$, features 
\be
\Gamma_{\tilde{x}}\!=\!-\Gamma_{\tilde{y}}\!=\!\Gamma_{\tilde{z}}\!=\!-\Gamma,
\ee 
i.e., a modulation of the sign of $\Gamma$ across different types of bonds, despite the fact that the model is threefold symmetric. (Incidentally, this example also shows that such a sign modulation in $\Gamma_\gamma$, that could appear in some model, can be `gauged away' by a twofold rotation around one of the cubic axes.)

The case of $\beta$-{\liiro} offers another striking example. Here, there exist two different sign structures of $\Gamma$, depending on the convention used for the cubic frame. 
Specifically: 

i) the cubic frame of Ref.~\cite{Ducatman2018},
\be\label{eq:bLi213frame}
\hat{\bf x}=\frac{\hat{\bf c}_o+\hat{\bf a}_o}{\sqrt{2}},~~
\hat{\bf y}=\frac{\hat{\bf c}_o-\hat{\bf a}_o}{\sqrt{2}},~~
\hat{\bf z}=-\hat{\bf b}_o,
\ee
where $({\bf a}_o,{\bf b}_o,{\bf c}_o)$ are the orthorhombic axes, see   Fig.~\ref{fig:TriCoord}\,(b-c), gives, by virtue of the crystal symmetries $C_{2{\bf c}}$ and $C_{2{\bf a}}$, 
\be\label{eq:b213-Gmodulation1}
\Gamma_x=-\Gamma_y=-\Gamma_{x'}=\Gamma_{y'},
\ee
where the bonds with $\gamma\!=\!x$ and $y$ are the ones on the $xy$-chains running along ${\bf a}_o$+${\bf b}_o$, whereas $\gamma\!=\!x'$ and $y'$ correspond to the $xy$-chains along ${\bf a}_o$-${\bf b}_o$. 
We note in passing that flipping the directions of two of the orthorhombic axes (without altering the {\it actual} directions of $x$ and $y$), does not change this sign structure. For example, flipping $\hat{\bf a}_o\!\to\!-\hat{\bf a}_o$ and $\hat{\bf b}_o\!\to\!-\hat{\bf b}_o$, would change Eq.~(\ref{eq:bLi213frame}) to $\hat{\bf x}\!=\!(\hat{\bf c}_o-\hat{\bf a}_o)/\sqrt{2}$, $\hat{\bf y}\!=\!(\hat{\bf c}_o+\hat{\bf a}_o)/\sqrt{2}$, and $\hat{\bf z}\!=\!\hat{\bf b}_o$ (which is the frame used in Refs.~\cite{Lee2015,Lee2016}), but would not alter Eq.~(\ref{eq:b213-Gmodulation1}), because the {\it actual} directions of $x$ and $y$ are the same as before.

ii) By contrast, the frame used in Ref.~\cite{Biffin2014PRB}, 
\be\label{eq:bLi213frame2}
\hat{\tilde{\bf x}}=\frac{\hat{\bf c}_o-\hat{\bf a}_o}{\sqrt{2}},~~
\hat{\tilde{\bf y}}=\frac{\hat{\bf c}_o+\hat{\bf a}_o}{\sqrt{2}},~~
\hat{\tilde{\bf z}}=\hat{\bf b}_o,
\ee
which corresponds to the choice $\epsilon_x\!=\!1$ and $\epsilon_y\!=\!-1$, gives, again by virtue of the symmetries $C_{2{\bf c}}$ and $C_{2{\bf a}}$, 
\be\label{eq:b213-Gmodulation2}
\Gamma_{\tilde{x}}=\Gamma_{\tilde{y}}=-\Gamma_{\tilde{x}'}=-\Gamma_{\tilde{y}'},~~~
\Gamma_{\tilde{z}}=-\Gamma_{z}\,,
\ee
which is a drastically different modulation of the sign of $\Gamma$ compared to that of Eq.~(\ref{eq:b213-Gmodulation1}). 
In particular, the $xy$- and $x'y'$-chains now each feature a uniform (instead of alternating) sign of $\Gamma$, opposite to each other. Importantly, the sign of $\Gamma$ on the $z$-bond also depends on the choice of the cubic frame.
Similar modifications in the sign structure of $\Gamma'$ exist as well, and can be found in a straightforward manner. 

The above examples demonstrate that it does not make sense of stating the sign of the off-diagonal couplings without specifying the choice made for the directions of the cubic axes  with respect to the crystalographic frame. This is important both when comparing experimental data to theoretical models, but also when extracting numerical values of the couplings from {\it ab initio} calculations.


%

\end{document}